\documentclass[twocolumn,amssymb,nobibnotes,aps,prd,superscriptaddress,nofootinbib]{revtex4-2}

\usepackage[utf8]{inputenc}
\usepackage{graphicx,amssymb,amsmath,amsthm,amsfonts,epsfig,epsf,environ,float,bbold,upgreek}
\usepackage[usenames,dvipsnames,svgnames,table]{xcolor}
\usepackage{epstopdf}
\definecolor{darkred}{rgb}{0.5,0,0}
\usepackage{bm}
\usepackage{dcolumn}
\usepackage{latexsym}
\usepackage{rotating}
\usepackage[normalem]{ulem}
\usepackage{longtable,tabularx,booktabs,adjustbox}
\usepackage{comment}

\usepackage{enumerate}
\usepackage{tensor,multirow}
\usepackage{url}
\usepackage{hyperref}
\usepackage{aas_macros}

\makeatletter
\NewEnviron{nofleqn}[1]{\@fleqnfalse\begin{#1}\BODY\end{#1}}
\makeatother

\hypersetup{
	colorlinks,
	linkcolor={red!60!black},
	citecolor={green!50!black},
	urlcolor={blue!80!black}
}

\newcolumntype{C}{>{\centering\arraybackslash}X}
\newcommand{\orcid}[1]{\href{https://orcid.org/#1}{\includegraphics[width=8pt]{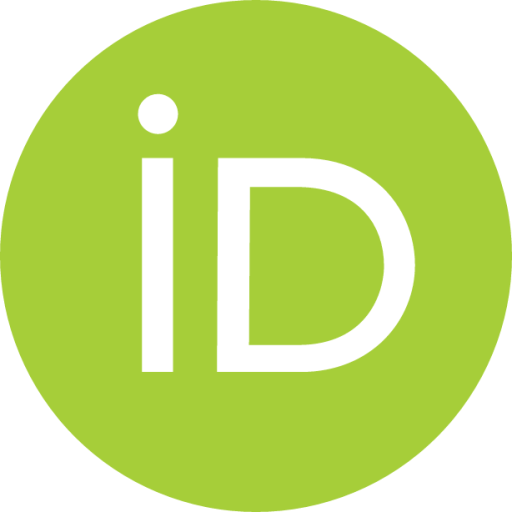}}}

\def\nn{\nonumber}

\def\be{\begin{equation}}
\def\ee{\end{equation}}
\newcommand{\beq}{\begin{eqnarray}}
\newcommand{\eeq}{\end{eqnarray}}
\def\ba{\begin{align}}
\def\ea{\end{align}}

\newcommand{\dd}{\mathrm{d}}
\newcommand{\mb}{m}
\newcommand{\re}{{\mathrm{\,Re}}}

\begin{document}
	
\title{Searching for ultra-light dark matter through frequency modulation of \\ gravitational waves}

\author{Diego Blas \orcid{0000-0003-2646-0112}}
%\email{}
\affiliation{Institut de F\'isica d’Altes Energies (IFAE), The Barcelona Institute of Science and Technology, Campus UAB, 08193 Bellaterra (Barcelona), Spain}
\affiliation{Instituci\'o Catalana de Recerca i Estudis Avan\c cats (ICREA), Passeig Llu\'is Companys 23, 08010 Barcelona, Spain}

\author{Silvia Gasparotto \orcid{0000-0001-7586-1786}}
%\email{}	
\affiliation{Institut de F\'isica d’Altes Energies (IFAE), The Barcelona Institute of Science and Technology, Campus UAB, 08193 Bellaterra (Barcelona), Spain}
\affiliation{Grup de F\'{i}sica Te\`{o}rica, Departament de F\'{i}sica, Universitat Aut\`{o}noma de Barcelona, 08193 Bellaterra (Barcelona), Spain}{}

\author{Rodrigo Vicente \orcid{0000-0002-3150-9315}}
%\email{}
\affiliation{Institut de F\'isica d’Altes Energies (IFAE), The Barcelona Institute of Science and Technology, Campus UAB, 08193 Bellaterra (Barcelona), Spain}
\affiliation{Gravitation Astroparticle Physics Amsterdam (GRAPPA),\\ University of Amsterdam, Amsterdam, 1098 XH, Netherlands}
\date{\today}%

\begin{abstract}
Ultra-light bosons, naturally appearing in well-motivated extensions to the Standard Model, 
can constitute all the dark matter.
Models with particle mass close to the smallest phenomenologically allowed 
exhibit coherent field configurations at (sub)galactic scales, oscillating at a frequency corresponding to the fundamental mass of the dark matter particle. 
The gravitational field of these structures inherits the dark matter field's coherent oscillations, leaving an imprint on gravitational (and electromagnetic) waves sourced close to (or in) such overdensities.
This happens via a heterodyning frequency modulation, which can later be decoded in a gravitational-wave detector. An analogous effect occurs in models with universal (conformal) couplings of ultra-light bosons with ordinary matter, generated by the direct interaction with the oscillating field.
In this work, we explore this phenomenon in detail and assess the capability of near-future interferometers to probe ultra-light dark matter and its potential conformal couplings to matter. 
Using astrophysical population models, together with results from cosmological simulations, we find that the observation of gravitational waves from spinning neutron stars at the Galactic Centre with the Einstein Telescope/Cosmic Explorer would be particularly effective in constraining ultra-light dark matter.
\end{abstract}

%	\keywords{}
\maketitle
%\tableofcontents

%%%%%%%%%%%%%%%%%%%%%%%%%%%%%%%%%
\section{Introduction}
\label{sec:intro}
%%%%%%%%%%%%%%%%%%%%%%%%%%%%%%%%%%
Elucidating the nature of dark matter (DM) is one of the main directions of current research in particle physics and cosmology~\cite{Bertone:2004pz}. Even its most fundamental properties, like the mass of its constituents, remain elusive.
One of the most fascinating scenarios, whose study has attracted plenty of attention in recent years, is that DM particles are so light that the typical interparticle distance in galaxies is much smaller than their de Broglie wavelength. For the Milky Way (MW), this is the case of DM particles with masses smaller than $\sim 1\,\mathrm{eV}$~\cite{Hui:2021tkt}.

These large occupation numbers imply that DM in this ULDM regime is effectively described by \emph{classical wave} equations of a field $\phi$, corresponding to massive particles interacting through gravity (and, maybe, some additional interactions). This results in several new phenomena compared to heavier DM particle candidates. Among these, two are of particular relevance to this work. First, in a virialized DM halo, there exist (transient) regions where the field oscillates coherently with frequency 
\begin{equation}
\label{eq:coh_m}
\omega\sim m\approx \frac{m_{22}}{76 \, \rm{days}} \,,    
\end{equation}
where~$m_{22}\equiv m/(10^{-22}\,\mathrm{eV})$, with~$m$ the ULDM particle mass. These halo patches remain coherent over~$ v_0^{-2}\sim 10^6\,\mathrm{cycles}$ and a length $(m v_0)^{-1}\sim (0.1/m_{22})\,{\rm kpc}$, where $v_0\sim 10^{-3}$ corresponds to the velocity dispersion of the DM particles in the halo~\cite{Hui:2016ltb}. 
Second, there exist dense self-gravitating ULDM structures sustained by wave pressure. These long-lived structures are called ``solitons'', and their associated DM field also oscillates coherently at frequency $\omega\sim m$. At least two categories are possible, originating from different channels. For the lightest bosons,~$m \lesssim 10^{-20}\, \mathrm{eV}$, these solitons form at the centre of galaxies through the gravitational relaxation of transient coherent patches in the halo, equipping galaxies with large DM cores~\cite{Schive:2014dra}.  Another possibility is that they are generated in the early universe through, e.g., a large misalignment mechanism~\cite{Arvanitaki:2019rax,Thompson:2023mfx}. Here, they could exist for heavier bosons~$m \gtrsim 10^{-20}\, \mathrm{eV}$, come with different sizes/densities, and be spread throughout the entire halo.

The gravitational field sourced by the ULDM fields inherits some of its coherent features. In particular, at first post-Newtonian order, it contains an oscillating component with frequency~$\omega \sim 2 m$ (i.e., twice the ULDM field frequency).
Such a peculiar time-dependent phenomenon opens up new ways to probe ULDM. For instance, one can now search for the modulation of electromagnetic signals of known frequency spectrum induced by the aforementioned oscillating background. This was originally proposed in Ref.~\cite{Khmelnitsky:2013lxt}, using the precisely timed rotational period of pulsars immersed in the ULDM halo (it was later extended to ULDM solitons in Ref.~\cite{DeMartino:2017qsa}). Still, despite being one of Nature's most precise clocks, there are practical challenges in timing pulsars at the Galactic Centre, where DM is more abundant, because of the high plasma density and irregularities in the interstellar medium.

Gravitational waves (GWs) with well-understood frequency content are a natural alternative because their sources may be detected even if they reside in the Galactic Centre (GC). These GWs range from (quasi-) monochromatic, such as those sourced by double white dwarfs (DWDs), extremely large mass-ratio inspirals (X-MRIs), or spinning neutron stars (NSs), to ``chirping'' signals, like the ones originating from the coalescence of black hole (BH) or NS binaries. In this work, we explore the possibility of encoding the low-frequency information of a ULDM field in high-frequency \emph{carrier} GWs from these sources via a type of heterodyning modulation. The modulation occurs during propagation through a Sachs-Wolfe effect~\cite{Khmelnitsky:2013lxt}, and can (in principle) be decoded at reception in a GW detector. The same idea was applied in Refs.~\cite{Stegmann:2023glt,Stegmann:2023wzy,Bustamante-Rosell:2021daj} to search for low-frequency GWs, and in Refs.~\cite{Wang:2023phr, Brax:2024yqh} to search for a ULDM halo. Here, we extend it to dense ULDM solitons, including the effect of attractive (quartic) ULDM self-interactions and allowing for universal conformal couplings to ordinary matter, using realistic population models of the sources to provide a first estimate of the prospects of using this effect to detect ULDM.

This work is organised as follows. In Sec.~\ref{sec:formalism}, we describe the ULDM models and coherent structures considered. In Sec.~\ref{sec:heterodyning}, we present the heterodyning effect of ULDM on null geodesics. In Sec.~\ref{sec:MWC},
we discuss the constraints of dynamic tracers on the density profile of the GC and the population of GW sources in the MW, and we conclude with observational prospects of the effect on GWs from the GC. We dedicate Sec.~\ref{sec:othergal} to GWs from other galaxies and Sec.~\ref{sec:halo} to DM halo substructures other than the central core.
Section~\ref{sec:discu} contains our discussion. Some technical material is provided in the appendices.
We use natural units $(c=\hbar=G=1)$ and the mostly positive metric signature~$(-+++)$.

%%%%%%%%%%%%%%%%%%%%%%%%%%%%%%%%%
\section{Coherent structures of ULDM} 
\label{sec:formalism}
%%%%%%%%%%%%%%%%%%%%%%%%%%%%%%%%%
In this section, we present the ULDM models that we consider, together with the relevant equations describing coherent structures of ULDM and their gravitational potentials. We also review the characteristics of such structures, which, depending on their production mechanism, can either condense at the centre of DM halos or be scattered through them.  

%%%%%%%%%%%%%%%%%%%%%%%%%%%%%%%%%
\subsection{ULDM models}
%%%%%%%%%%%%%%%%%%%%%%%%%%%%%%%%%
%

As mentioned, we consider scenarios where DM is made of an ultra-light boson. We will focus on the case of a massive (pseudo)scalar~$\phi$ with attractive quartic self-interactions, although similar results may be derived for ULDM of higher spin. The DM field interacts with ordinary matter fields~$\bm{\upchi}$ via the minimal coupling to the spacetime metric~$\bm{g}$, and, possibly, through direct couplings. 
The effective action we will consider is then
\begin{nofleqn}{align}\label{eq:action}
   S=\int \dd^4x\Big\{\sqrt{-g}\big(\tfrac{1}{16 \pi}R-\tfrac{1}{2}\partial_\mu \phi\partial^\mu \phi-\mathcal{V}[\phi]\big)\nn\\+\mathcal{L}_{\rm m}[\bm{\upchi},\phi,\bm{g}]\Big\},
\end{nofleqn}
with potential
\begin{equation}
\mathcal{V}[\phi]= \tfrac{1}{2}(\mb \phi)^2[1-\tfrac{1}{12}(\phi/F)^2],
\end{equation}
motivated by the low-energy limit ($\phi/F\ll 1$) of shift-symmetric axion-like potentials.\footnote{To generate the observed density of DM via misalignment mechanism would require  $F\sim 10^{17}{\rm GeV}\,(10^{-22} {\rm \,eV}/\mb)^{1/4}$~\cite{Arvanitaki:2009fg, Marsh:2015xka, Hui:2016ltb}.}

The energy-momentum tensor reads~$T_{\mu \nu}=T^\phi_{\mu \nu}+T^{\rm m}_{\mu \nu}$, where the contribution of~$\phi$ is
\begin{equation}\label{eq:enmomtenscalar}
T^\phi_{\;\mu \nu}=\nabla_{\mu} \phi \nabla_{\nu} \phi-g_{\mu \nu}(\tfrac{1}{2}\nabla^\alpha \phi\nabla_\alpha \phi+\mathcal{V}),
\end{equation}
and~$T^{\rm m}_{\;\mu \nu}\equiv -(2/\sqrt{-g})\delta\mathcal{L}_{\rm m}/\delta g^{\mu \nu}$. 
In this work, we will consider ULDM phenomena where the weak-field limit of Einstein's equations describes the relevant gravitational effects. Namely, we can write,\footnote{In Newtonian gauge and using global inertial coordinates~$(t,x^i)$.}
\begin{equation} \label{eq:metric}
    g_{\mu \nu}\dd x^\mu \dd x^\nu\approx-(1-2\Phi)\dd t^2+(1+2\Psi) \delta_{i j} \dd x^i \dd x^j, 
\end{equation}
with~$|\Phi|,|\Psi|\sim \epsilon\ll1$. 
Regarding the equations of motion for $\phi$, 
we will only be interested in their non-relativistic limit, where particles of $\phi$ have energy approximately equal to their rest mass, i.e.~$|\partial_t\log\phi|/\mb\sim 1$. 

To find the relevant equations in the limit of interest, we split the potentials and their sources into slowly and rapidly fluctuating pieces [i.e.,~$\Phi= \bar{\Phi}+\delta \Phi$, where $\bar{\Phi}\equiv(\mb/2\pi) \int_{t-\pi/\mb}^{t+\pi/\mb} \dd t'\, \Phi$]. 
We neglect the direct coupling of DM to ordinary matter for now (we will include it later).
Introducing the field redefinition
\begin{equation}
\phi\equiv \psi\, \frac{e^{-i \mb t}}{\sqrt{2 \mb}}  + {\rm c.c.},   
\label{eq:NRfield}
\end{equation}
with $|\partial_t\log\psi|/\mb\sim \epsilon$, one finds the non-relativistic field equations [averaged over high-frequency modes~$\omega\gtrsim \mb$, cf. App.~\ref{app:potentials}]
\begin{subequations} \label{eqs:SP}%
\begin{gather}
     i  \partial_t \psi= -\Big\{\frac{\nabla^2}{2 \mb}  + \mb\Big[\bar{\Phi} + \frac{\bar{\rho}_\phi}{8\mb^{2}F^2}  \Big]\Big\}\psi, \label{eq:S}  \\
    \nabla^2\bar{\Phi}=- 4 \pi  (\bar{\rho}_\phi+\rho_{\rm m}) , \quad \bar{\Psi}\approx\bar{\Phi}, \label{eq:P}
\end{gather}
\end{subequations}
where~$\nabla^2\equiv \delta^{i j}\partial_i \partial_j$,~$\bar{\rho}_\phi \equiv \mb |\psi|^2$, and~$\rho_{\rm m}\equiv T^{\rm m}_{\;tt}$.

At leading order, the rapidly oscillating parts of the potentials have frequencies close to $2 \mb$ and $4 \mb$ (where the latter is sourced exclusively by the~$\phi^4$ self-coupling, cf. App.~\ref{app:potentials}). We hence introduce the decomposition
\begin{equation}
    \delta \Phi\equiv \Phi_2(x^i) \cos(2 \mb t)+\Phi_4(x^i) \cos(4 \mb t), 
\end{equation}
and a similar one for~$\delta \Psi$.  
Assuming a slowly evolving field configuration 
\begin{equation}
\psi=e^{-i\upgamma \mb t}\uppsi(x^i), \label{eq:profile}    
\end{equation}
with $\gamma=|\partial_t\log\psi|/\mb\ll 1$ and~$\uppsi \in \mathbb{R}$, one finally finds (see  App.~\ref{app:potentials})
\begin{subequations} \label{eqs:delta}
   \begin{alignat}{2}
       & \nabla^2 \big[\Psi_2+\tfrac{\pi}{\mb^2}\bar{\rho}_\phi\big]=-\tfrac{\pi}{6 F^2 \mb^2}  \bar{\rho}_\phi^2, \label{eq:deltaPsi}\\
     &   \nabla^2 \Phi_2=8\pi  \big[5\,\bar{\Phi} +\upgamma -\tfrac{\bar{\rho}_\phi}{12 F^2 \mb^2}\big] \bar{\rho}_\phi \label{eq:osci_eq}\\
      &  \nabla^2 \Psi_4=\tfrac{\pi}{12 F^2 \mb^2} \bar{\rho}_\phi^2, \qquad \nabla^2 \Phi_4= -\tfrac{\pi}{6 F^2 \mb^2} \bar{\rho}_\phi^2.
    \end{alignat}
\end{subequations}
The case without self-interactions is recovered in the limit $F\xrightarrow{}\infty$, and no $\Psi_4$ and $\Phi_4$ are sourced.

The possibility of direct couplings of DM to ordinary matter fields in~$ \mathcal{L}_{\rm m}$ broadens the landscape of ULDM models that we can test with our methods considerably.
A simple option that still encapsulates relevant phenomenological aspects of many models is to consider the case of a \emph{universal} conformal coupling of the ULDM field~$\phi$ to the ordinary matter. This can be summarized through the effective Lagrangian density~\cite{Blas:2016ddr, Blas:2019hxz}
\begin{equation}
    \mathcal{L}_{\rm m}=\mathcal{L}_{\rm m}[\bm{\chi}^i,A^2(\phi)\bm{g}], \label{eq:direct}
\end{equation}
where the function $A(\phi)$ is model dependent.
This theory can be recast into a new form by performing a conformal transformation to the Jordan-Fierz metric
\begin{equation}
    \widetilde{g}_{\mu\nu}=A^2(\phi) g_{\mu\nu}, \label{eq:direct_A}
\end{equation}
where test (ordinary) matter fields fall freely.  In this work, we will restrict to linear  and quadratic couplings
\begin{equation}
\label{eq:Lambdas}
    A\approx1+\phi/\Lambda_1 ~~\mathrm{and} ~~ A\approx1+\phi^2/\Lambda^2_2\,,
\end{equation} 
with the high-energy scales~$\Lambda_{1}$ and~$\Lambda_2$ parameterizing the strength of the interaction. These arise, for instance, in the low-energy limit~$|\phi|\ll \Lambda$ of the Fierz-Jordan-Brans-Dicke~\cite{Fierz:1956zz, Jordan:1959eg, Brans:1961sx, Dicke:1961gz} and Damour-Esposito-Farèse~\cite{Damour:1992we, Damour:1993hw} theories, respectively (see also \cite{Damour:2010rp} for the related dilatonic couplings).

The coupling of DM with matter in Eq.~\eqref{eq:direct_A} adds a term to the right-hand side of Eq.~\eqref{eq:S} that averages to zero in the case of linear couplings, and is~$\sim \rho_{\rm m} \psi/(\mb \Lambda_2^2)$ for quadratic ones. Equation~\eqref{eq:P} is modified by replacing~$\rho_{\rm m}\to A^2\rho_{\rm m}$ on the right-hand side. These corrections are always negligible for the values of~$\Lambda_{1,2}$ considered in this work, and the only impact of Eq.~\eqref{eq:direct} is to change the effective metric over which the matter fields move. Indeed, as clear from Eq.~\eqref{eq:direct_A}, 
the universal direct coupling of~$\phi$ to matter in the Einstein frame, causes the gravitational field to rapidly oscillate with frequency~$\mb$ for the linear coupling, and~$2\mb$ for the quadratic, in the (physical) Jordan frame (see the discussion in Ref.~\cite{Smarra:2024kvv}).

%%%%%%%%%%%%%%%%%%%%%%%%%%%%%%%%%
\subsection{ULDM solitons as DM cores}\label{subsec:soliton}
%%%%%%%%%%%%%%%%%%%%%%%%%%%%%%%%%
%
Self-gravitating ULDM \emph{solitons}---also known as boson stars~\cite{Kaup:1968,Ruffini:1969qy,Liebling:2012fv} (for complex scalar fields) or oscillatons~\cite{Seidel:1991zh,Guzman:2004wj,Visinelli:2021uve} (for real ones)---play an important role in the phenomenology of ULDM, in particular, due to their presence in the central regions of galaxies \cite{Schive:2014dra, Schive:2014hza, Mocz:2017wlg, Veltmaat:2018dfz, Chan:2018, Mina:2020eik, Nori:2020jzx, Chen:2020cef, Chan:2021bja, Mocz:2023adf, Painter:2024rnc}. 

The solitons under consideration are stationary, spherically symmetric, non-relativistic solutions of Eqs.~\eqref{eqs:SP}, minimising the total energy for a fixed number of particles. 
These equations are invariant under the scaling
\begin{equation}
    \{t,x^i, \psi,\bar{\Phi}, F\}\to \{ t/\lambda^2,x^i/\lambda,\lambda^2\psi,\lambda^4 \bar{\Phi},\lambda F\},  \label{eq:scaling}
\end{equation}
which leaves the parameter
\begin{equation}
    \beta \equiv \frac{\sqrt{\bar{\rho}_0/\pi}}{16(F^2\mb)} \approx \frac{0.024}{m_{22}} \Big(\frac{10^{16} {\rm \,GeV}}{F}\Big)^{2} \sqrt{\frac{\bar{\rho}_0}{10^3 M_\odot  \mathrm{pc^{-3}}}}\,. \label{eq:beta}
\end{equation} 
invariant, where
 $\bar{\rho}_0\equiv \mb |\psi(0)|^2$.
Thus, each~$\beta$ defines a one-parameter family of solutions (across different values of~$F$)
\begin{subequations}
\begin{gather}
    \psi_{\lambda}(t,x^i;\beta)=\lambda^2\psi_1(\lambda t,\lambda x^i;\beta),\\
    \bar{\Phi}_{\lambda}(t,x^i;\beta)=\lambda^2\bar{\Phi}_1(\lambda t,\lambda x^i;\beta).
\end{gather}
\end{subequations}
The parameter~$\lambda$ is related to the soliton's central density through
\begin{equation*}
    \lambda=[m_{22}^{-2} \bar{\rho}_0/(10^3 M_\odot \mathrm{pc^{-3}})]^{1/4}.
\end{equation*}

For attractive self-interaction, solitons have a critical mass~\cite{Chavanis:2011a, Chavanis:2011b, Eby:2015hsq}~$M_{\rm s,*}\approx 1.1 \times 10^{10}M_\odot\, F/(10^{16} {\rm \,GeV}) m_{22}^{-1}$, above which they are unstable and collapse (to a more dense soliton~\cite{Braaten:2015eeu, Chavanis:2017loo, Painter:2024rnc}, or a BH~\cite{Chavanis:2016dab, Helfer:2016ljl}). This happens for~\cite{Painter:2024rnc} $\beta_*\approx0.55$.  In this work, we restrict to~$\beta\leq \beta_*$, where the soliton's density profile is well approximated by the fit~\cite{Painter:2024rnc}
\begin{gather}\label{eq:density}
    \bar{\rho}\approx \bar{\rho}_0 \Big[1+0.091\Big(\frac{r}{R_{\rm c}}\Big)^{2-\beta/b} \Big]^{-8},
    \end{gather}
with
\begin{gather}
    \bar{\rho}_0 \approx 4.9 M_\odot \mathrm{pc^{-3}}\, a^{-4\beta/(2b-\beta)} \,m_{22}^{-2}\Big(\frac{0.25\mathrm{kpc}}{R_{\rm c}}\Big)^4, \nn
\end{gather}
and~$a=11.2$ and $b=4.2$.
There is a mass-radius relation~\cite{Painter:2024rnc}
\begin{equation}
    M_{\rm sol}\approx 0.88\,\mathcal{G}(\beta) \times 10^9M_\odot m_{22}^{-2} \left(\frac{0.25 \,\mathrm{kpc}}{R_{\rm c}}\right)\,,
\end{equation}
with $\mathcal{G}\sim \mathcal{O}(1)$ a decreasing function of~$\beta$ with~$\mathcal{G}(0)=1$. We define the core radius to be such that~$\bar{\rho}(R_{\rm c})=\bar{\rho}_0/2$.

The first DM-only cosmological simulations found that a soliton of mass~$M_{\rm sol}$ forms dynamically through relaxation at the centre of DM halos according to the empirical relation~\cite{Schive:2014hza}
\begin{equation}\label{eq:Schive}
    M_{\rm sol}\approx 1.4 \times 10^9 M_\odot \, m_{22}^{-1}\Big(\frac{M_{\rm h}}{10^{12}M_\odot}\Big)^{1/3},
\end{equation}
where~$M_{\rm h}$ is the halo mass.
A significant dispersion on the numerical prefactor and exponent of~$M_{\rm h}$ was found in later simulations~\cite{Mocz:2017wlg, Mina:2020eik, Nori:2020jzx, Chan:2021bja} (probably due to tidal stripping~\cite{Chan:2021bja}), which indicates that cores and halos may coexist in different configurations. Thus, in this work we will take the phenomenological fit of~\cite{Chan:2021bja},\footnote{Note that in~\cite{Chan:2021bja} the core mass is used, which is related to the total soliton mass by~$M_{\rm c}\approx 0.236 M_{\rm sol}$~\cite{Bar:2018acw}.}
\begin{gather}\label{eq:scatter}
   \frac{1.1}{m_{22}} \Big(\frac{M_{\rm h}}{10^{12}M_\odot}\Big)^{1/3} \lesssim  \frac{M_{\rm sol}}{10^9 M_\odot} \lesssim \frac{30}{\sqrt{m_{22}}} \Big(\frac{M_{\rm h}}{10^{12}M_\odot}\Big)^{2/3} ,
  %\\
   % M_{\rm sol} \lesssim 3 \times 10^{10} M_\odot \,m_{22}^{-1/2}\Big(\frac{M_{\rm h}}{10^{12}M_\odot}\Big)^{2/3},
\end{gather}
encompassing the scatter of the soliton-halo mass relation in the literature. 
Baryonic matter~\cite{Veltmaat:2018dfz, Chan:2018} and self-interactions~\cite{Chen:2020cef, Mocz:2023adf, Painter:2024rnc} affect this relation, but we expect the above interval to be sufficiently broad to cover also these effects. 

Equation~\eqref{eq:scatter} is based on simulations of a single-ULDM field. Studies analysing the effects of a single ULDM field on rotation curves~\cite{Bar:2018acw} and the Lyman-$\alpha$ forest~\cite{Rogers:2020ltq} disfavour ULDM with masses $m\lesssim 10^{-21}$~eV as the dominant component of DM. 
The constraints from rotation curves rely on Eq.~\eqref{eq:Schive}, which does not hold for multi-ULDM field scenarios. There we expect the DM halo to relax to a ULDM soliton that will likely differ from the single-field one, and whose mass will depend on the relative abundance and particle mass of the different ULDM fields (as indicated by the results of Refs.~\cite{vanDissel:2023vhu,Luu:2024gnk}. See also \cite{Mirasola:2024pmw}). However, as the final configuration in such scenarios is still not generally known, we focus on the single-field scenario in this paper.  

In addition, one should consider the effect of a supermassive black hole (SMBH), which is likely to be at the centre of all large galaxies~\cite{Kormendy:1995, Kormendy:2013dxa, Reines:2022}, which could potentially `swallow' the soliton~\cite{Bar:2019pnz, Davies:2019wgi, Cardoso:2022nzc}. If the soliton mass is larger than that of the SMBH - the focus of this work - the soliton's gravity dominates the accretion process, which happens on a timescale~\cite{Cardoso:2022nzc}.
\begin{nofleqn}{gather}
    \frac{T_{\rm accr}}{10\, {\rm Gyr}}\approx 3 \times 10^5 \mathcal{F}(\nu) m_{22}^{-6}\Big(\frac{10^9\,M_\odot}{M_{\rm sol}} \Big)^5,
\end{nofleqn}
with~$\nu\equiv M_{\rm bh}/M_{\rm sol}$, and~$\mathcal{F}(\nu)\lesssim 2$ a strictly decreasing function given explicitly in Ref.~\cite{Cardoso:2022nzc}. 
For the MW, we note that Sgr~A*, with~$M_{\rm bh}\approx 4.4 \times 10^6 M_\odot$~\cite{Gillessen:2008qv, Ghez:2008ms, Genzel:2010zy}, would take~$T_{\rm accr}\gtrsim 7 \times 10^8 \, {\rm Gyr}\, m_{22}^{-6} (10^9 M_\odot/M_{\rm sol})^5$ to accrete a considerable amount of mass of a soliton with~$M_{\rm sol}\gtrsim 10^9 M_\odot$. So, we can safely neglect Sgr A* accretion of the soliton in the MW. The same may not be true for larger galaxies~\cite{Davies:2019wgi}.

%%%%%%%%%%%%%%%%%%%%%%%%%%%%%%
\subsection{ULDM overdensities in the halo}
\label{sec:halo}
%%%%%%%%%%%%%%%%%%%%%%%%%%%%%%

We are also interested in other coherent ULDM structures, like the transient coherent patches in the virialized halo (that result in its characteristic granular structure). In the standard halo for ULDM, these can be (roughly) described by assuming that the 
the density profile in Eq.~\eqref{eq:profile} satisfies $\uppsi\approx \uppsi_0 $, with~$\uppsi_0$ sampled from a Rayleigh distribution
\begin{equation}
    {\rm P}(\uppsi_0)=(2 \uppsi_0/\uppsi_{\rm DM}^2)e^{-\uppsi_0^2/\uppsi_{\rm DM}^2},\, \label{eq:Rayleigh}
\end{equation}
with~$\uppsi_{\rm DM}\equiv\sqrt{2\bar{\rho}_{\rm DM}}/\mb$ and $\bar{\rho}_\phi$ the average local DM density (e.g., \cite{Derevianko:2016vpm, Centers:2019dyn}) following a Navarro-Frenk-White (NFW) profile~\cite{Navarro:1995iw}. Their coherence length and time is, respectively,~$L_{\rm c}\sim 1/(m v_0)$ and $T_{\rm c}\sim  1/(m v_0^2)$, with $v_0$ being the velocity dispersion of the DM halo.

As will become clear in the following sections, the average density of the ULDM halo and corresponding~${\mathcal O}(1)$ perturbations (from Eq.~\eqref{eq:Rayleigh}) are too low to yield an observable frequency modulation of GWs. However, the chances of detection could be improved in scenarios where the ULDM halo contains highly overdense regions. These could be produced in the early universe by a ``large misalignment mechanism''~\cite{Arvanitaki:2019rax,Thompson:2023mfx} (see also~\cite{Chatrchyan:2023cmz,Eroncel:2022efc,Eroncel:2022vjg}) in the form of denser and more numerous (sub)halos with a characteristic mass
\begin{gather}\label{eq:massradiusLMM}
     M_{\rm s}\sim 5 \times 10^7 M_\odot \left(\frac{100}{m_{22}}\right)^{3/2}, 
\end{gather}
and typical scale radius
\begin{gather}\label{eq:radiusLMM}
     R_{\rm s}\approx 87{\rm pc}  \left(\frac{M_{\rm s}}{ 5 \times 10^7 M_\odot}\right)^{1/3} \left(\frac{10^5}{\mathcal{B}}\right)^{1/3},
\end{gather}
where $\mathcal{B}\equiv\rho_{\rm s}/\rho_{\rm CDM}$ gives the enhancement in DM halo density scale $\rho_{\rm s}$ compared to the standard cold DM case. 
In Ref.~\cite{Arvanitaki:2019rax} it was found that for large enough misalignment angles, corresponding to~$10^3\lesssim\mathcal{B}\lesssim 10^6$, a soliton is formed at the centre of the subhalo. 
Its density profile would still be given by Eq.~\eqref{eq:density}, now with 
\begin{equation}\label{eq:overdensityLMM}
    \bar{\rho}_0\sim 5 \times 10^2 M_\odot \mathrm{pc^{-3}} \left(\frac{\mathcal{B}}{10^5} \right) \mathcal{G}(\beta)^4 a^{-4\beta/(2b-\beta)},
\end{equation}
which is independent of the ULDM particle mass~$m$.

For even larger misalignment angles, density scales $\rho_{\rm s}$  higher than the critical one for solitons with attractive self-interactions~(cf. Sec.~\ref{subsec:soliton}) can be achieved~\cite{Arvanitaki:2019rax}. This gives rise to \emph{oscillons}---denser (metastable) configurations sustained by attractive self-interactions balanced with the outward wave pressure. These objects are relativistic, meaning that Eqs.~\eqref{eqs:delta} (describing the profile of the oscillating gravitational potentials) are not applicable here. Nevertheless, oscillons will still source periodic gravitational potentials with frequency components close to multiples of~$m$. We leave its investigation for the future, and focus here on the case of gravitational solitons. 

Besides the frequency modulation of GWs studied here, these dense (sub)halos can be probed by other observations, via their gravitational field (e.g., through the motion of nearby stars, and gravitational lensing~\cite{Arvanitaki:2019rax, Thompson:2023mfx, Chatrchyan:2023cmz}), but also via direct couplings to the SM (e.g., through the electromagnetic dispersion introduced by the axion coupling to the photon~\cite{Prabhu:2020pzm}).

%%%%%%%%%%%%%%%%%%%%%%%%%%%%%%%%%
\section{Heterodyning of null geodesics}\label{sec:heterodyning}
%%%%%%%%%%%%%%%%%%%%%%%%%%%%%%%%%

The propagation of high-frequency waves of massless fields (\emph{carriers}) can be described by null geodesics of the background spacetime.
When signals are sourced/detected in or close to a region where a ULDM structure oscillates coherently, they are subjected to heterodyning modulation~\cite{Khmelnitsky:2013lxt}. 
This is similar to what happens in the presence of low-frequency GW backgrounds to: electromagnetic carriers used in pulsar timing array (PTA) experiments (see, e.g., Ref.~\cite{2010CQGra..27h4013H}), or carrier GWs~\cite{Bustamante-Rosell:2021daj,Stegmann:2023glt}.

More concretely, \emph{any} periodically oscillating gravitational background induces a modulation of both the signal's amplitude (through lensing and a difference in gravitational potential between emission and reception events) and frequency (via the ordinary and integrated Sachs-Wolfe effects, and the induced relative motion between emitter and receiver).
An oscillation with frequency~$\omega_\delta=2 m$ is expected for ULDM, while oscillations with frequency $\omega_\delta=4\mb$ appear from the quartic self-coupling. In both cases, the background 
acts effectively as a heterodyne, transferring power from the carrier signal at frequency~$\omega_{\rm e}$ to the frequencies~$\omega_{\rm e} \pm \omega_\delta$.

When a monochromatic GW of amplitude~$h$ is emitted with frequency~$\omega_{\rm e}$ much higher than the characteristic frequency scale of the gravitational background,~$\omega_\delta$, its propagation can be understood using the WKB approximation. The (leading) correction to~$h\approx(\mathcal{Q}_{\rm e}/d_{\rm L})e^{i \omega_{\rm e} t}$, with~$\mathcal{Q}_{\rm e}$ the carrier amplitude at emission and~$d_{\rm L}$ the luminosity distance to the emitter, can be written as (for a derivation see App.~\ref{app:mod}) 
\begin{gather}\label{eq:modsignal}
     \frac{\delta h}{h} \approx i\left( \frac{\omega_{\rm e}}{\omega_{\rm \delta}}\right)\,\Upsilon(\bar{\rho}_0,\mb,x_{\rm e}^i)  \sin(\omega_\delta u),
\end{gather}
where the all-important factor~$\Upsilon$ is defined as
\begin{gather}
     \Upsilon\equiv\begin{cases}
        \big[ \Psi_{2}-\frac{2}{\omega_\delta}n^i\partial_i  \Phi_{2} \big]_{x^i_{\rm e}}, \quad \textrm{(minimal)}\\
        \frac{\sqrt{2}}{\Lambda_1}\left(\frac{\bar{\rho}_\phi(x^i_{\rm e})}{\mb^2}\right)^{1/2}, \quad \textrm{(direct linear)}\\
        \frac{1}{\Lambda_2^2}\frac{\bar{\rho}_\phi(x^i_{\rm e})}{\mb^2}, \quad\qquad \textrm{(direct quadratic)}
    \end{cases} \label{eq:upps}
\end{gather}
with~$\omega_\delta=2 \mb$ for the minimal and the quadratic couplings,~$\omega_\delta=\mb$ for the linear one, and 
where~$(u\equiv t-d_{\rm L},x^i_{\rm e})$ is the emission event and $n^i$ a unit-vector pointing to the sky direction of GW arrival.  Note that the general formula for the phase modification includes an additional term evaluated at the reception event [cf.~Eq.~\eqref{eq:chidef}].
In the minimal scenario, we disregard the contribution of the self-interaction term $\Psi_4$, which oscillates at a frequency $\omega_\delta = 4m$ and is typically subdominant by a factor of $1/2$. A detailed comparison of the amplitude profiles is provided in App.~\ref{app:self-interaction}.
Since the chances of having a close-by dense DM structure will be higher for the source (as there are many sources, some of them may be located at the DM overdensities required to produce a strong enough effect), we neglect here the receptor (or Earth) term.
Note that when the carrier GW is not monochromatic, the perturbation~$\delta h$ has a more complicated form (see App.~\ref{app:Sachs-Wolfe}); we will also consider that case in Sec.~\ref{sec:othergal}, for chirping GW sources.

For a carrier GW observed over a time~$T_{\rm obs}\gtrsim 2\pi/\omega_\delta\approx 0.7 m_{22}^{-1}\, {\rm yr}$, the signal-to-noise ratio (SNR) of the modulation is (see the derivation in App.~\ref{app:snr})
\begin{equation}\label{eq:Gformula}
    {\rm SNR}_\delta\sim \frac{1}{\sqrt{2}}\left(\frac{\omega_{\rm e}}{\omega_\delta}\right)\Upsilon(\bar{\rho}_0,\mb,x_{\rm e}^i) {\rm\, SNR}_{h}, 
\end{equation}
where~${\rm SNR}_{h}$ is the SNR of the unperturbed waveform~$h$. The SNR$_\delta$ gets further enhanced by a factor~$\sim \sqrt{N}$ if multiple~$N$~($\gg1$) uncorrelated sources are detected close to or within the ULDM soliton.

\begin{figure}[t]
    \centering
    \includegraphics[width = \linewidth]{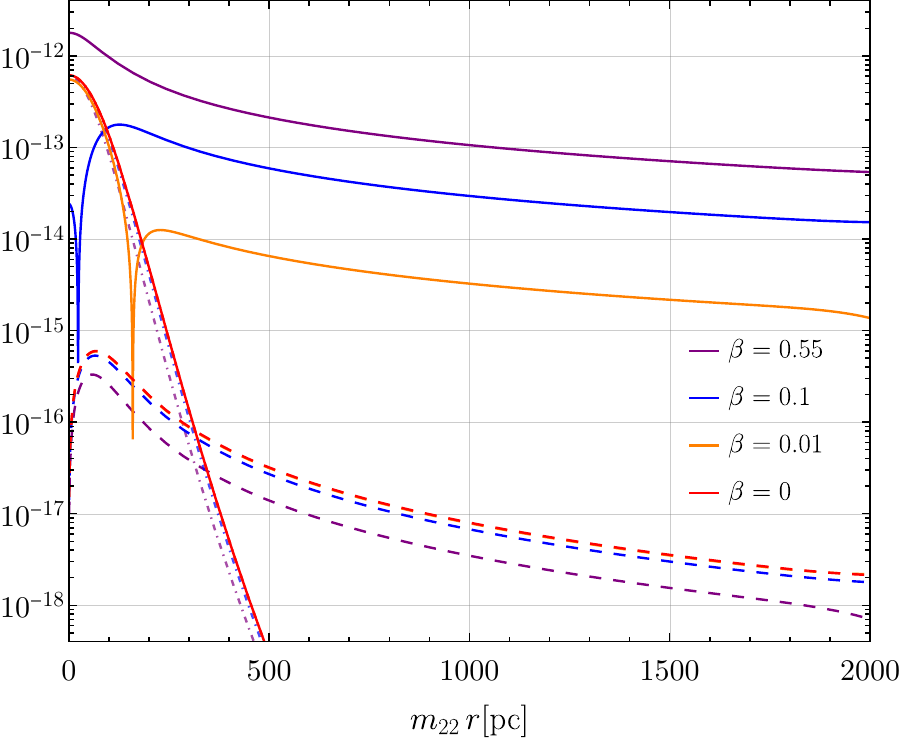}
    \caption{Results for $|\Psi_2|$ (solid) and $m^{-1}\partial_r \Phi_2$ (dashed), compared against the contribution $\pi \bar{\rho}/m^2$ (dot-dashed), for different values of~$\beta$ (defined in Eq.~\eqref{eq:beta}) and fixed~$m_{22}^{-2}\bar{\rho}_0 = 10^{3} M_{\odot}\mathrm{pc^{-3}}$. The curves are left unchanged under  the transformation $r \to r/\lambda$, $\Psi_2\to \lambda^4 \Psi_2$, $\partial_r \Phi_2 \to \lambda^5 \partial_r \Phi_2$ and $\bar{\rho}\to \lambda^4 \bar{\rho}$, with~$\lambda=[m_{22}^{-2} \bar{\rho}_0/(10^3 M_\odot \mathrm{pc^{-3}})]^{1/4}$.}
    \label{fig:potentials}
\end{figure}

In Fig.~\ref{fig:potentials} we present the spatial distributions of relevance for the different contributions to the factor~$\Upsilon$ in Eq.~\eqref{eq:upps}, for different values of the self-coupling~$\beta$ (cf. Eq.~\eqref{eq:beta}) and fixed~$\bar{\rho}_0 m_{22}^{-2}= 10^{3} M_{\odot}\mathrm{pc^{-3}}$. The gravitational potentials of the soliton are obtained from solving numerically Eqs.~\eqref{eqs:delta}, after first solving the Eqs.~\eqref{eqs:SP}. These are the terms entering $\Upsilon$ for the case of minimal coupling. The term~$\Psi_2$ is the dominant contribution close to the soliton in all different cases, while the term~$m^{-1}\partial_r \Phi_2$ becomes dominant at larger distances when ULDM self-interactions are weak (though it is generally very small). 
It is interesting to note that self-interactions have been overlooked in this context. However, Fig.~\ref{fig:potentials} shows that ULDM self-interactions 
extend the region of influence of the soliton to much larger volumes (even for mild self-interactions), and so potentially to more carrier signals. 
Figure~\ref{fig:potentials} also displays $\pi \bar{\rho}/m^2$, from which the contributions for the direct coupling
contribution in Eqs.~\eqref{eq:upps} can be derived (recall that we use units $G=1$).

We can now compare the magnitude of the effect on GWs to the, more conventional one, on electromagnetic signals from pulsars. 
The ordinary Sachs-Wolfe effect of coherent patches of ULDM on pulsar timing signals was first studied in Ref.~\cite{Khmelnitsky:2013lxt} and subsequently applied to ULDM solitons in Ref.~\cite{DeMartino:2017qsa}, and to direct coupling in Refs.~\cite{Kus:2024vpa,Smarra:2024kvv}.
The time-dependent part of the timing residuals is~\cite{Khmelnitsky:2013lxt}
\begin{equation}
    \Delta t(t)= \left(\frac{1}{\omega_\delta}\right) \Upsilon(\bar{\rho}_0,\mb,x_{\rm e}^i)\sin{(\omega_\delta u)}.
\end{equation}
The rms amplitude of the timing signal is then
\begin{equation}
    \sqrt{\langle\Delta t^2}\rangle= \frac{1}{\sqrt{2}}\left(\frac{1}{\omega_\delta}\right) \Upsilon(\bar{\rho}_0,\mb,x_{\rm e}^i), \label{eq:deltat}
\end{equation}
which correspond to the effect of a GW with characteristic strain~$h_c=2\sqrt{3}\Upsilon$~\cite{Khmelnitsky:2013lxt}.
Current PTA measurements are sensitive to a minimum characteristic strain 
\begin{equation}
    h_{\rm c,min}
    %\equiv \omega_\delta\sqrt{3\langle\Delta t^2}\rangle \sqrt{N_p}
    \sim 10^{-14}\left ( \frac{\omega_\delta}{1\,\mathrm{ nHz}}\right )\sqrt{\frac{20}{N_{\rm p}}},
\end{equation} 
with~$N_{\rm p}$ the number of pulsars in the array, while the forthcoming SKA may reach even~$h_{\rm c}\sim 10^{-16}$ at~$\omega_\delta \sim 1\, \mathrm{nHz}$~\cite{Sesana:2008mz,Khmelnitsky:2013lxt}. The maximum sensitivity occurs at the inverse observation time $T_{\rm obs}$, $f_{\rm low}\sim1/T_{\rm obs}$, and the maximum frequency probed is given by the inverse of the observation cadence of each pulsar $f_{\rm high}\gtrsim (2 {\rm weeks})^{-1} \sim 8.3 \times 10^{-7}\,{\rm Hz}$. Using Eq.~\eqref{eq:Gformula} we see that an equivalent sensitivity at~$\omega_\delta \sim 1\,{\rm nHz}$ using $N$ carriers of GW signals with frequency $\omega_e \sim1\, {\rm kHz}$ is achieved for a carrier SNR
\begin{equation}
    {\rm SNR}_h\sim 3\times 10^2 \left( \frac{\omega_{\rm e}}{1\,{\rm kHz}}\right )\sqrt{\frac{N_{\rm p}}{N}},
\end{equation}
using a threshold of SNR$_\delta=3$, corresponding to a~$3\sigma$ detection. 
This shows that PTA data seems better than GW observations in probing ambient oscillating gravitational fields. 
However, we can only radio-time pulsars from our galaxy, and this timing is challenging in the GC, where the DM density is expected to be larger (high plasma density and irregularities in the interstellar medium cause pulse smearing and dispersion in the time of arrivals).\footnote{Note, however, the very recent observation of a millisecond pulsar binary at the GC~\cite{Lower:2024sdi}.} 
GWs interact much more weakly with baryons and can be easily observed even when sourced at the GC or in other galaxies. Furthermore, the number of expected sources of GWs may outnumber that of radio-pulsars in several situations. As we will see, these facts will make of GW emitters to be more relevant to probe ULDM models generating overdensities at the GC.

%%%%%%%%%%%%%%%%%%%%%%%%%%%%%%%%%
\section{The inner Milky Way: constraints and prospects}%
\label{sec:MWC}
%%%%%%%%%%%%%%%%%%%%%%%%%%%%%%%%%
In this section, we discuss some observational constraints on the presence of a ULDM core in the inner MW. Then, we describe the populations of the astrophysical GW sources we consider, providing the numbers to compute the expected magnitude of the ULDM heterodyning effect. Finally, we present some prospects for the observability of such an effect using near-future GW observations.

\subsection{Dynamical constraints on a DM core}

Finding the dark matter component of the MW is challenging~\cite{Read:2014qva, Bland-Hawthorn:2016} (in particular, in its inner region). For Galactocentric radii~$r\gtrsim 2\,{\rm kpc}$, the rotation curve inferred from combining gas kinematics (HI and CO terminal velocities, HI thickness, HII regions, giant molecular clouds), star kinematics (open clusters, planetary nebulae, cepheids, carbon stars), and masers is compatible with a NFW profile
\begin{equation}
    \rho(r)=\frac{4\rho_{\rm s}}{\frac{r}{r_{\rm s}}\big(1+\frac{r}{r_{\rm s}}\big)^2},
\end{equation}
with scale radius~$r_{\rm s}\approx 20\,{\rm kpc}$, normalized to a local DM density~$\rho(r=8{\rm \,kpc})\approx 0.4\,\mathrm{GeV/cm^3}(\sim10^{-2}M_\odot/\mathrm{pc^3})$ (see, e.g., Ref.~\cite{Iocco:2015xga}). Through the dynamical modelling of the Galactic bulge, bar, and inner disk, put together with observations of the red clump giant density, and stellar kinematics in the bulge and bar region, Ref.~\cite{Portail:2017} found evidence for an inner cored profile in the bulge (see also Ref.~\cite{Zoccali:2014}), whose mass in the region~$r\lesssim 0.25{\rm\, kpc}$ could be as large as~$\sim 6 \times 10^9 M_\odot$ (with an average density~$\bar{\rho}\sim 10^3\,\mathrm{M_\odot/pc^3}$). Part of this inferred dynamical mass ($\sim 2\times 10^9 M_\odot$) is thought to be in the form of a (baryonic) nuclear bulge~\cite{Portail:2017, Launhardt:2002, Schonrich:2015, Wegg:2017, Bar:2018acw}, while the rest may be DM. We point out that Refs.~\cite{DeMartino:2018zkx, Li:2020qva} concluded that the inferred dynamics of the GC does not exclude the presence of a central ULDM soliton with total mass given by Eq.~\eqref{eq:Schive} with~$M_{\rm h}\sim 10^{12}M_\odot$, possibly compressed by a nuclear bulge of mass~$\sim 2 \times 10^9 M_\odot$. 

\begin{figure}[t]
    \centering
    \includegraphics[width = \linewidth]{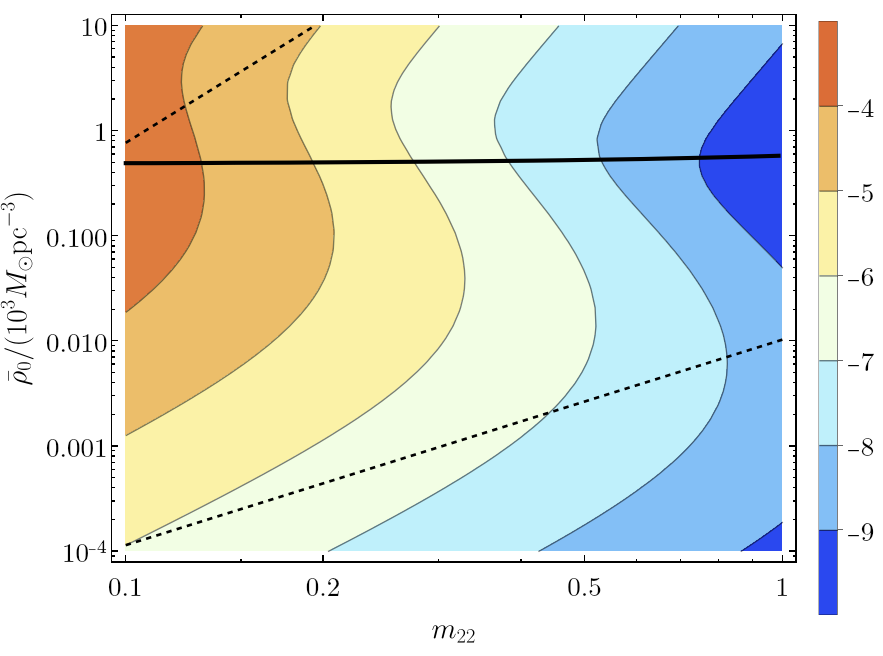}\\
    \includegraphics[width = \linewidth]{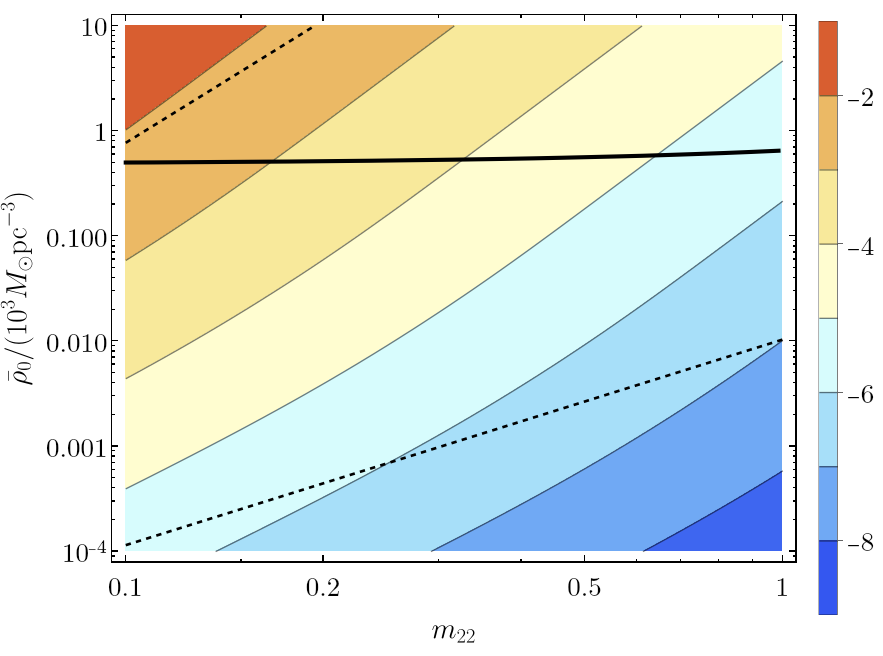}
    \caption{Contour plot of~$\log_{10}\left(\pi |\Upsilon|/(\sqrt{2}\, m_{22})[{\rm s}]\right)$ as a function of~$\bar{\rho}_0$ and~$\mb$, for minimal coupling, fixing~$r_{\rm e}=0.5\,{\rm kpc}$ and~$n^i=x_{\rm e}^i/r_{\rm e}$. Dashed lines show the limits of Eqs.~\eqref{eq:scatter} for~$M_{\rm h}=10^{12} M_\odot$ (a MW soliton is expected to be contained between the two lines). The region above the thick solid curve is disfavored, as the soliton's enclosed mass is in tension with the one inferred by dynamical tracers in the inner~$r\lesssim 0.25\,{\rm kpc}$ of the MW. Top: no self-interactions,~$\beta=0$. Bottom: critical self-interactions, $\beta= 0.55\approx \beta_*$. The final ${\rm SNR}_\delta$ requires a multiplication of $\pi |\Upsilon|/(\sqrt{2}\, m_{22})[{\rm s}]$ by $\sqrt{N}x$, cf. Eq.~\eqref{eq:factor}.} 
    \label{fig:Upsilon_500pc}
\end{figure}

Here, we take an agnostic approach by considering all soliton masses in the interval~\eqref{eq:scatter} with~$M_{\rm h}=10^{12} M_\odot$, imposing that the soliton's enclosed mass is below the one inferred by dynamical tracers in~$r\lesssim 0.25 {\rm\, kpc}$~\cite{Portail:2017} (see Fig.~15 of Ref.~\cite{Bar:2018acw}). Note, however, that observations in this region are subjected to large systematic uncertainties due to the Galactic bar effect on tangent-point velocity measurements of gas flow. A ULDM soliton can also affect the dynamics of baryons at the GC in a non-trivial way (see Ref.~\cite{Li:2020qva} for the impact on the size and kinematics of MW's central molecular zone). A nuclear bulge with mass~$M_{\rm nb}\gtrsim M_{\rm sol}/3$ (as it may be the case for the MW~\cite{Li:2020qva}) will compress the solitonic core by a factor of~$\lesssim 2$~\cite{Bar:2018acw}, but we do not expect this to impact our results considerably. So, in this work, we neglect the interplay of baryons with the ULDM soliton at the GC.

\begin{figure}
    \centering
    \includegraphics[width=\linewidth]{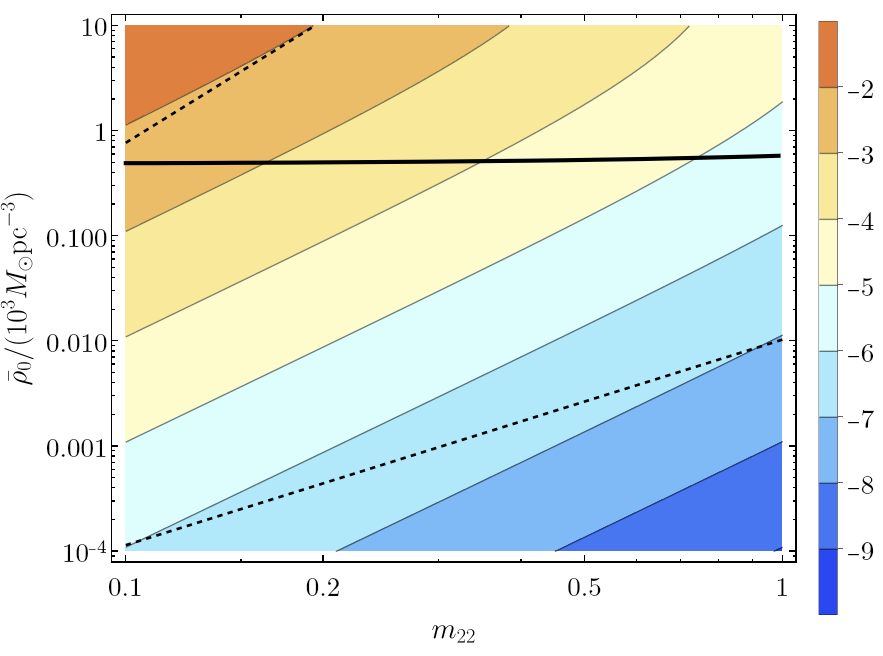}\\
    \includegraphics[width=\linewidth]{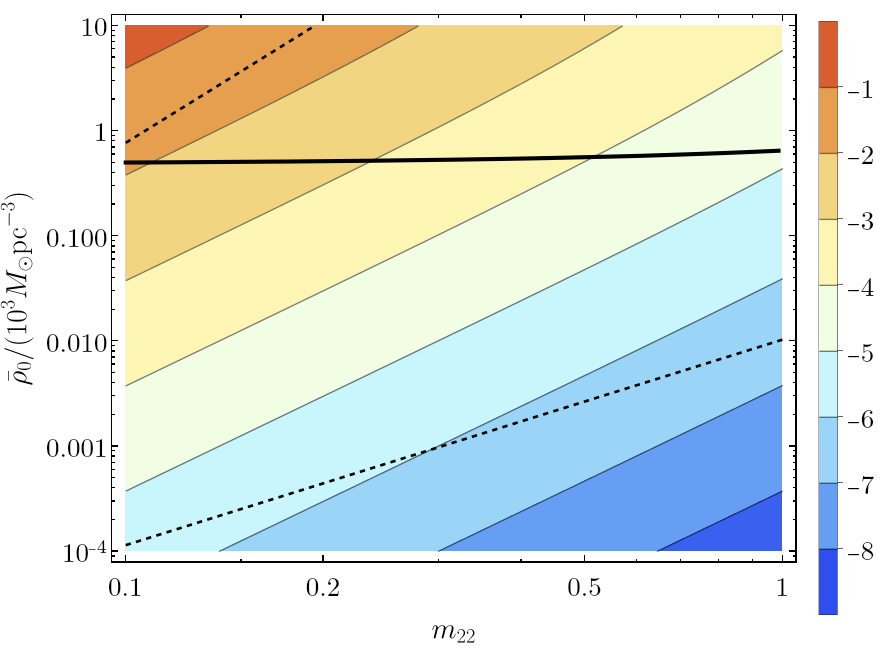}
    \caption{Same as Fig.~\ref{fig:Upsilon_500pc} but with $r_{\rm e}=0.05\,{\rm kpc}$. From the cored profile of the soliton, the above plots (restricted to the parameter space shown) are nearly unchanged for~$r_{\rm e}\lesssim 0.05 \,{\rm kpc}$.} 
    \label{fig:Upsilon_50pc}
\end{figure}

Our results for the heterodyning effect of ULDM cores for two different GW source locations (resp.,~$r_{\rm e}=0.5\,{\rm kpc}$ and~$r_{\rm e}=0.05\,{\rm kpc}$) can be extracted from Figs.~\ref{fig:Upsilon_500pc} and~\ref{fig:Upsilon_50pc}. To find the~${\rm SNR}_{\delta}$ in a single GW observation, one simply multiplies the value of~$\pi |\Upsilon|/(\sqrt{2}\, m_{22})[{\rm s}]$ from the relevant plot by the factor
\begin{equation}
   x\equiv f_{\rm e}[{\rm Hz}] \,{\rm SNR}_h\,, \label{eq:factor}
\end{equation} 
with~$f_{\rm e}$ the frequency of the (quasi)monochromatic GW; combining~$N$ similar
signals (i.e., with similar~$f_{\rm e}$ and~${\rm SNR}_h$) will add an extra factor~$\sqrt{N}$. These numbers depend on the specific sources (cf. Tab.~\ref{tab:DWDs}).

A natural way to improve sensitivity would be if we could coherently combine $N$ sources, which could follow from the coherence of the soliton at the centre of the Galaxy (the phase of the modulation of the GWs at a given time would be the same for every carrier inside the soliton). However, each carrier would have a different phase at the receptor just because of the different distance travelled from emission. The cross-correlation of the signal from different carriers hence requires a precise knowledge of the distance to the sources, or identification of the relative phases.

Analogously, to find the rms amplitude of the timing signal~$\sqrt{\langle\Delta t^2}\rangle[{\rm s}]$ for one pulsar, one simply needs to multiply the plotted quantity ($\pi |\Upsilon|/(\sqrt{2}\, m_{22})[{\rm s}]$) by~$(2 \pi)^{-1}$, cf.  Eq.~\eqref{eq:deltat}. When $N_p$ pulsars are combined, the final sensitivity to the ULDM oscillations also (optimistically) acquires a factor $\sim \sqrt{N_p}$ \cite{Ellis:2012in,Siemens:2013zla,Perrodin:2017bxr}.

%%%%%%%%%%%%%%%%%%%%%%%%%%%%%%%%%%%%%%%%%%%%%%%%%%%%%%%%%%%%
\subsection{Populations of astrophysical sources}
%%%%%%%%%%%%%%%%%%%%%%%%%%%%%%%%%%%%%%%%%%%%%%%%%%%%%%%%%%%%

%%%%%%%%%%%%%%%%%%%%%%%%%%%%%%%%%%%%%%%%%%%%%%%%%%%%%%%%%%%%
\subsubsection{Galactic binaries}\label{sec:population_WD}
%%%%%%%%%%%%%%%%%%%%%%%%%%%%%%%%%%%%%%%%%%%%%%%%%%%%%%%%%%%%
Out of the possible sources of GW carriers, we start by looking at the population of Galactic binaries, which are guaranteed sources of (quasi)monochromatic GWs at millihertz frequencies for space-based interferometers as LISA~\cite{Colpi:2024xhw} or Taiji and TianQin~\cite{Gong:2021gvw}. Those consisting of double WDs (DWDs) will by far outnumber all other (including the extra-galactic) GW sources at $\sim \rm mHz$~\cite{Korol:2018wep, Korol:2020lpq, Korol:2021pun}. Most of their gravitational signals (from millions of DWDs) will be unresolvable, forming a confusion foreground at frequencies $\lesssim {\rm mHz}$. Additionally, we expect around~$\mathcal{O}(10^4$--$10^5)$ DWDs to be resolvable and (quasi)monochromatic~\cite{Strokov:2023ypy}. These has been suggested to be used to detect low-frequency GWs~\cite{Bustamante-Rosell:2021daj, Stegmann:2023glt} (via a heterodyning effect similar to the one studied here) and probe the gravitational potential of the MW~\cite{Korol:2018wep, Ebadi:2024oaq}.

We consider the observational-driven populations of DWDs in the Galactic disk constructed by Ref.~\cite{Korol:2021pun} for LISA (based on the realistic modelling of Ref.~\cite{Maoz:2018}) to quantify the observational prospects of using DWDs to probe the ULDM solitonic core of the MW. We take a number of~$2.6 \times 10^7$ DWDs with frequency and chirp mass distributed according to Fig.~3 of Ref.~\cite{Korol:2021pun} and a spatial distribution following the Galactic disk density profile. We assume an isotropic inclination angle~$\iota$ of the orbital plane (i.e. a uniform distribution of~$\cos \iota$). The relevant numbers, derived from such a population, are shown in Tab.~\ref{tab:DWDs} for LISA, TianQin and Taiji. We also include the prospects for~$\mu$Ares, a concept suggested as the natural continuation of current space-based interferometers~\cite{Sesana:2019vho}.

We are underestimating the number of resolvable binaries in the GC by neglecting the DWDs in the Galactic bulge and bar (see, e.g.,~\cite{Wilhelm_2020,Korol:2023jfz}). However, these are expected to be subdominant populations compared to that of the disc\footnote{We thank Valeryia Korol for pointing this out to us.} so, given the $\sqrt{N}$ dependence of the total SNR, the sensitivity will change by, at most, an order one factor. The same argument applies to other Galactic populations of (quasi)monochromatic binaries in the mHz band, as those involving NSs and BHs that, even if more massive (thus, sourcing carrier GWs with higher SNR), are expected to be orders of magnitude less abundant than DWDs~\cite{Nelemans:2001hp}.
We have also neglected DWDs with~$f_e\lesssim 10^{-4}\,{\rm Hz}$ which could, in principle, be detected by~$\mu$Ares, but are again a small fraction of the total population of DWDs. 
\begin{table}
    \centering
    \noindent
    \begin{tabularx}{\linewidth}{@{} l c c c @{}}
            \toprule
             & $N$ & \hspace{-.3cm}$\langle{\rm SNR}_h\rangle$ & \hspace{-.5cm}$\sqrt{N}\langle{\rm SNR}_h\rangle\langle f_{\rm e}\rangle$[Hz]  \\
             \midrule
            \multicolumn{4}{c}{\textit{Double White Dwarfs}} \\
            LISA & \hspace{-.3cm}$5.5(1.6)\times10^3$ &\hspace{-.3cm} $37(38)$ & \hspace{-.5cm}$7.8(4.3)$ \\
            TianQin & \hspace{-.3cm}$2.5 (0.7)\times10^3$ &\hspace{-.3cm} $37(37)$ & \hspace{-.5cm}$5.1(2.9)$ \\ 
            Taiji & \hspace{-.3cm}$5.8 (1.7)\times10^3$ & \hspace{-.3cm}$59(60)$ & \hspace{-.5cm}$13(6.8)$ \\
            $\mu$Ares & \hspace{-.3cm}$504 (148)\times10^3$ & \hspace{-.3cm} $49(48)$ & \hspace{-.5cm}$97(52)$ \\
            \midrule
            \multicolumn{4}{c}{\textit{X-MRIs}} \\
            LISA & \hspace{-.3cm}$\mathcal{O}(5)$ & \hspace{-.3cm}$\sim10^3$ &\hspace{-.5cm} $\sim 10$ \\
            \midrule
            \multicolumn{4}{c}{\textit{Spinning NSs}} \\
            ET/CE & \hspace{-.3cm}$\mathcal{O}(200)$ & \hspace{-.3cm}$\sim30$ &\hspace{-.5cm} $\sim 10^5$ \\
            \bottomrule
    \end{tabularx}
    
    \caption{Parameters relevant for the sensitivity estimate to the different astrophysical populations considered. 
    The results of DWDs in the Galactic disk come from a simulation based on the population described in Refs.~\cite{Maoz:2018, Korol:2021pun}. $N$ is the number of sources, $\langle{\rm SNR}_h\rangle$ and $\langle f_{\rm e}\rangle$ are their average SNR and frequency, and $\sqrt{N}\langle{\rm SNR}_h\rangle\langle f_{\rm e}\rangle$[Hz] is the multiplicative factor that needs to be applied to the contour plots to find~${\rm SNR}_\delta$. For DWDs,~$N$ is for~${\rm SNR}_{h}\geq 7$ and Galactocentric radii~$<1{\rm kpc}$ ($<0.5 {\rm kpc}$). }
    \label{tab:DWDs}
\end{table}

%%%%%%%%%%%%%%%%%%%%%%%%%%%%%%
\subsubsection{X-MRIs}
%%%%%%%%%%%%%%%%%%%%%%%%%%%%%%

Another population of interest at the GC for  LISA, TianQin, and Taiji is that of X-MRIs---extremely large mass-ratio inspirals of brown dwarfs around Sgr A*~\cite{Amaro-Seoane:2019umn, Gourgoulhon:2019iyu, Vazquez-Aceves:2021xwl}. Ref.~\cite{Amaro-Seoane:2019umn} estimates that at any time there will be~$\gtrsim 20$ such sources in the LISA band: about~$\gtrsim 15$ highly eccentric and at lower frequencies (with SNR$_{h}$ of a few $100$, for a one year observation) and~$\gtrsim 5$ circular and at higher frequencies (with SNR$_h$ from a few $100$ up to~$2 \times 10^4$). So, taking~$f_{\rm e}\sim {\rm mHz}$ and the SNR$_h\gtrsim 10^3$, we find~$\sqrt{N}\langle{\rm SNR}_h\rangle\langle f_{\rm e}\rangle\lesssim 10\,\mathrm{Hz}$. An appealing feature of X-MRIs is that they are always well inside the soliton and may have considerably large SNR allowing for reaching similar prospects as WDs with fewer sources.

%%%%%%%%%%%%%%%%%%%%%%%%%%%%%%%%%%%%%%%%%%%%%%%%%%%%%%%%%%%%
\subsubsection{NS deformations}
%%%%%%%%%%%%%%%%%%%%%%%%%%%%%%%%%%%%%%%%%%%%%%%%%%%%%%%%%%%%

We turn now to the continuous (quasi)monochromatic GWs sourced by rapidly spinning deformed NSs. These (still elusive) signals have high-frequency ($\gtrsim 0.1\,{\rm kHz}$) and may be observed soon by ground-based detectors~\cite{Piccinni:2022vsd}. The GW strain depends crucially on the degree of (non-axially symmetric) deformations of the NS.
Several mechanisms can cause such deformations, from magnetic field distortion to accretion, and their magnitude depends on the NS equation of state (see, e.g., \cite{Gittins:2024zbg} for a review). The frequency of these GWs is associated with the angular velocity of the NS, which is highly correlated with its initial spin and magnetic field.

Prospects for the detection of such signals with current and third-generation ground-based detectors 
were studied in Refs.~\cite{Cieslar:2021viw,Pagliaro:2023bvi,Branchesi:2023mws, Maggiore:2019uih}. For instance, in Ref.~\cite{Pagliaro:2023bvi} they considered eight astrophysical population models differing on their initial distribution of magnetic fields, birth spin periods, and NS deformation (ellipticity), which result in different detection rates. All the models are consistent with the non-detection or the detection of~$\mathcal{O}(1)$ events by current GW interferometers. The situation is greatly improved for the next-generation detectors, such as the Einstein Telescope (ET)~\cite{Maggiore_2020} and the Cosmic Explorer (CE) \cite{Reitze:2019iox}, which will have higher sensitivities and better coverage at lower frequencies.
For the most promising models considered in Ref.~\cite{Pagliaro:2023bvi},\footnote{These are the A2 models in Ref.~\cite{Pagliaro:2023bvi}, we refer to their Tabs.~5 and 6.} one expects between~$200$ and~$500$ detectable events with ET and CE, and more than half of them with frequencies in $[100,500]\mathrm{\, Hz}$.  

Since most of the detectable pulsars are expected to be young, they will most probably be close to the GC, where the formation rate is higher, as they have not had time to migrate away from it~\cite{Pagliaro:2023bvi}. Therefore, considering $\mathcal{O}(200)$ detectable pulsars at the GC with an average SNR of 30 over 8 years of observations, and an average frequency of~$\langle f_e\rangle\simeq 300\, \mathrm{Hz}$, we obtain the multiplicative factor~$\sqrt{N}\langle{\rm SNR}\rangle\langle f_e\rangle\sim 10^5\, \mathrm{Hz}$. Note that we have neglected that some of the GWs from NSs close to the GC could be strongly lensed by Sgr A*, which may further improve the detection prospects~\cite{Savastano:2022jjv}.

\begin{figure}
    \centering
    \includegraphics[scale=0.4]{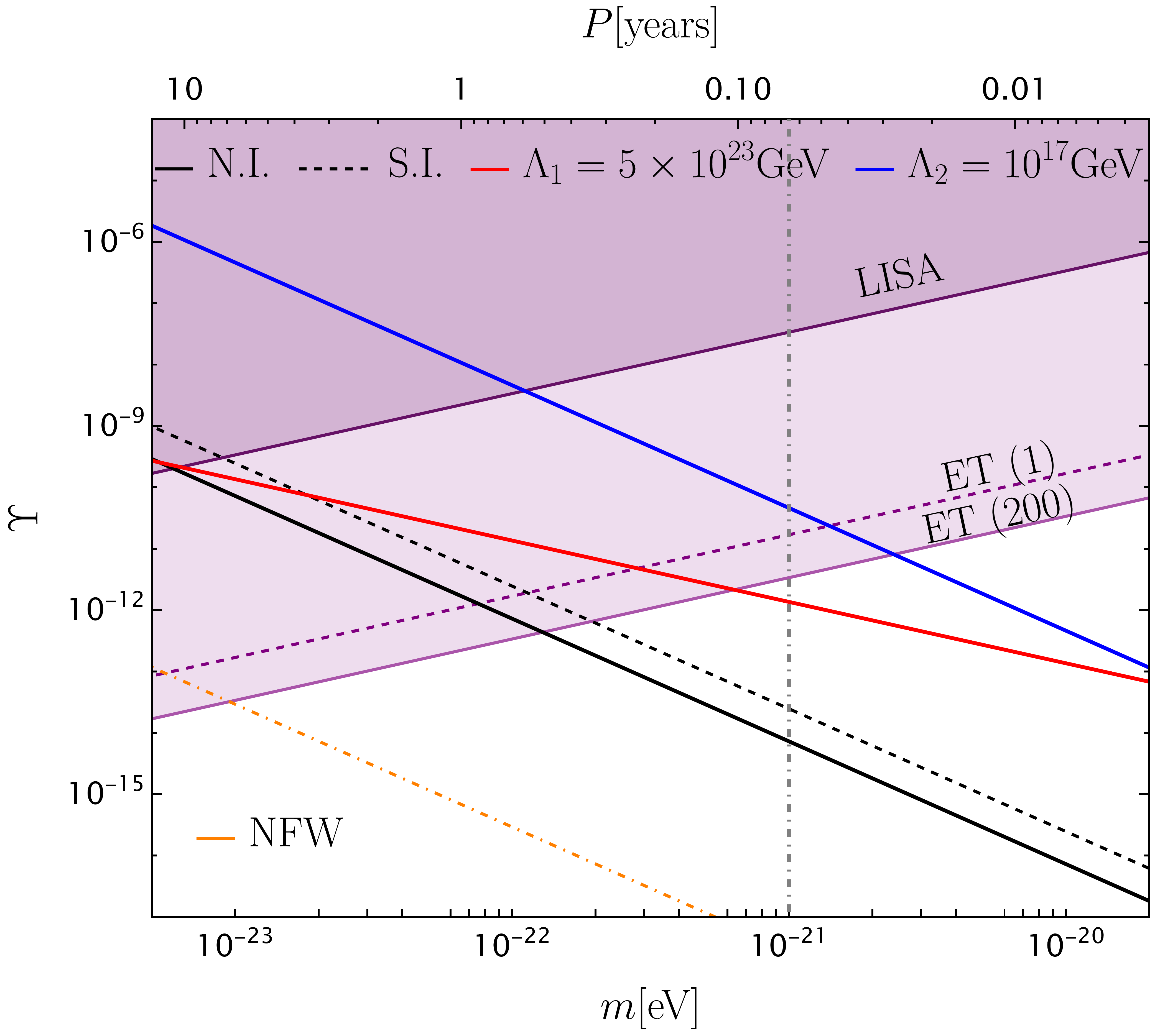}
    \caption{Sensitivities of LISA and ET/CE to~$\Upsilon$ (cf. Eq.~\eqref{eq:upps}). 
    The region in dark (light) purple is probed by LISA (ET/CE) for the values of Tab.~\ref{tab:DWDs} for X-MRIs (spinning NS). 
    The dashed purple line instead represents the sensitivity for a single young NS with $f_{\rm e}=10^3\,{\rm Hz}$ and ${\rm SNR}=20$. The amplitudes of $\Upsilon$ are shown in black-solid for no self-interactions (N.I.) and black-dashed for critical self-interactions (S.I.) in the minimal coupling case. The blue (red) line corresponds to an example case of direct interaction with quadratic (linear) coupling. All curves correspond to a soliton at the GC with central density~$\bar{\rho}_0=10^3 M_\odot {\rm pc}^{-3}$, except the orange which is for the Galactic NFW density profile at $r_{\rm e}=500\, {\rm pc}$. ULDM with masses below the grey dashed line is disfavoured as the dominant component DM, see main text. The top horizontal axis shows the period corresponding to~$\omega =2 m$.}
    \label{fig:Upsilonres}
\end{figure}
\begin{figure}
    \centering
    \includegraphics[width = \linewidth]{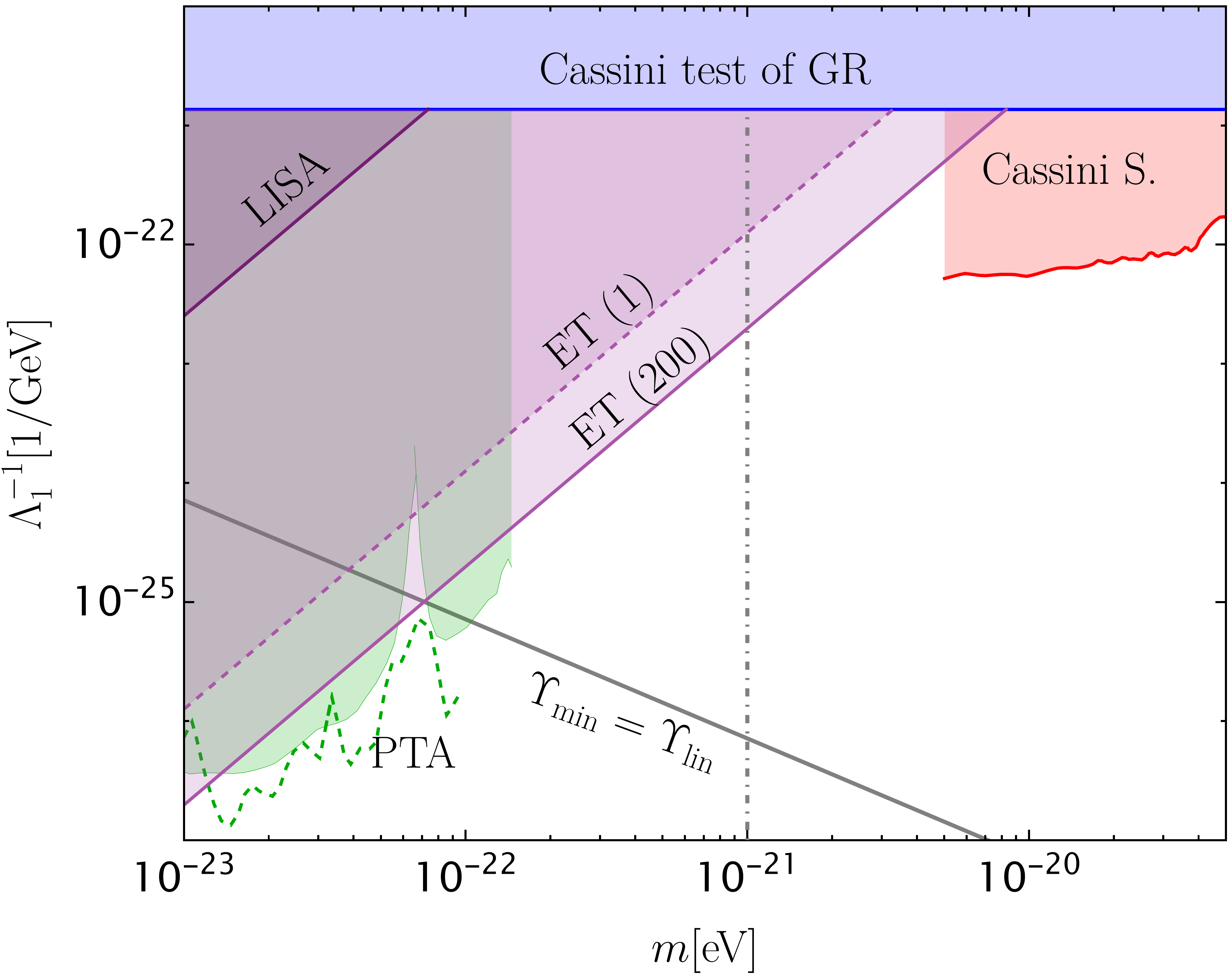}
    \caption{Sensitivities of LISA and ET/CE to the linear direct coupling~$\Lambda_1^{-1}$ from the heterodyning of GWs by a ULDM soliton at the GC with central density~$\bar{\rho}_0=10^3 M_\odot {\rm pc}^{-3}$. The coloured regions are excluded by PTA (olive), Cassini tests of General Relativity (violet), and Cassini bounds on a stochastic GW background (red). As in~Fig. \ref{fig:Upsilonres}, in the case of ET/CE we show lines both for a single observation of a spinning NS and for a population of $\mathcal{O}(200)$ (cf. Table~\ref{tab:DWDs}). Here, we choose a threshold of SNR$_{\rm th}=5$ for a more fair comparison with the other bounds. The gray line indicates the values at which the effect of minimal coupling is equal to the direct interaction considered here (excluding the PTA and Cassini probes). The shaded regions below this line represent scenarios where the effect is observable and predominantly governed by minimal coupling.}
    \label{fig:Lambda1res}
\end{figure}
\begin{figure}
    \centering
    \includegraphics[width = \linewidth]{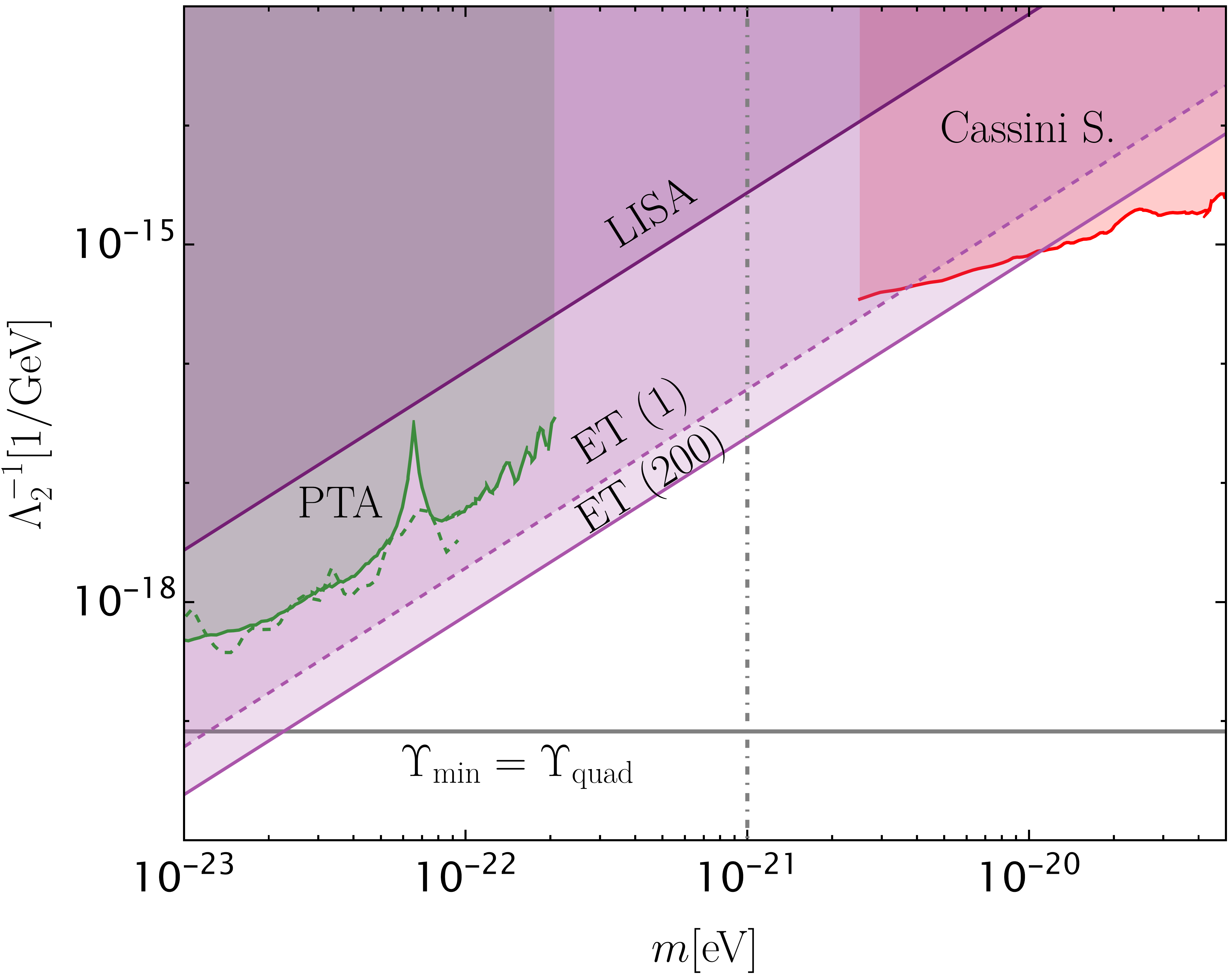}
    \caption{Same as Fig. \ref{fig:Lambda1res}, but for the quadratic coupling $\Lambda_2^{-1}$. }
    \label{fig:Lambda2res}
\end{figure}

%%%%%%%%%%%%%%%%%%%%%%%%%%%%%%%%%%%%%%%%%%%%%%%%%%%%%%%%%%%%%%%%%%%%
\subsection{Observational prospects}
\label{sec:ObsMW}
%%%%%%%%%%%%%%%%%%%%%%%%%%%%%%%%%%%%%%%%%%%%%%%%%%%%%%%%%%%%%%%%%%%%

The sensitivities of near-future detectors to the ULDM heterodyning effect on GWs are presented in Figs.~\ref{fig:Upsilonres}--\ref{fig:Lambda2res} for the source populations discussed in the previous sections (whose relevant parameters are summarized in Tab.~\ref{tab:DWDs}). 
We consider~$\langle{\rm SNR}_{h,\rm tot}\rangle\langle f_{\rm e}\rangle= 10\,\mathrm{Hz}$ for the LISA sources, a representative value valid for both WDs and X-MRIs. TianQin and Taiji lead to similar prospects as LISA (cf. Tab.~\ref{tab:DWDs}), whilst $\mu$Ares improves the sensitivity by one order of magnitude. Also, combining the sources from LISA, TianQin and Taiji would improve the prospects by a factor of $\sim\sqrt{3}$ from the $\sqrt{N_{\rm det}}$ scaling, so it will not substantially change the sensitivity of these millihertz detectors. This is far below the reach of ET/CE (we consider~$\langle{\rm SNR}_{h,\rm tot}\rangle\langle f_{\rm e}\rangle= 10^4$--$10^5\,\mathrm{Hz}$ for the GWs from spinning NSs). Given the uncertain status of the $\mu$Ares mission, we omit it in our plots.

In Fig.~\ref{fig:Upsilonres} we show the reach of LISA and ET/CE computed from Eq.~\eqref{eq:Gformula}, using a threshold of SNR$_{\delta,{\rm th}}=1$ (a different choice of threshold corresponds to a linear rescaling of the curves). For comparison, we show the value of~$\Upsilon$ for a soliton with a central density of~$\rho_0=10^3 M_\odot {\rm pc}^{-3}$ (allowed by dynamical constraints in the inner MW) both for no self-interactions and for critical self-interactions, in the case of minimal coupling. We also include the value of~$\Upsilon$ for an example case of (unconstrained) direct linear and quadratic couplings. Additionally, we show a reference value of~$\Upsilon$ for a coherent halo patch, using the Galactic NFW density profile at~$r_{\rm e}= 500\,{\rm pc}$.

In Figs.~\ref{fig:Lambda1res} and~\ref{fig:Lambda2res} we show the sensitivities of LISA and ET/CE to the linear and quadratic direct couplings of ULDM to the SM, through the heterodyning of GWs by a ULDM soliton with central density~$\bar{\rho}_0=10^3 M_\odot {\rm pc}^{-3}$. For comparison, we show also the excluded regions by PTA~\cite{Smarra:2024kvv}, Cassini tests of General Relativity~\cite{Blas:2016ddr,2003Natur.425..374B} and Cassini bounds on a stochastic GW background~\cite{Blas:2016ddr,Armstrong:2003ay}. We do not show the prospects from binary pulsars \cite{Blas:2016ddr,Blas:2019hxz,Kus:2024vpa}, as they are not competitive for $\Lambda_1$ and are only relevant for a narrow set of frequencies for $\Lambda_2$. 
These constraints assume that a single ULDM field constitutes the whole DM. However, as discussed in Sec.~\ref{subsec:soliton}, studies of the Lyman-$\alpha$ forest~\cite{Rogers:2020ltq} and rotation curves~\cite{Bar:2018acw} disfavour that scenario at masses~$m\lesssim 10^{-21}$~eV. We indicate this mass range with the grey dot-dashed line in Figs.~\ref{fig:Upsilonres},~\ref{fig:Lambda1res} and~\ref{fig:Lambda2res}.  The most stringent constraint, coming from Lyman-$\alpha$, basically disappears if this fraction is below 20$\%$~\cite{Ferreira:2020fam}, and most of the other bounds also deteriorate in that case.

From Fig.~\ref{fig:Upsilonres}, we conclude that LISA will not be able to probe the oscillations of the gravitational potentials of a ULDM soliton at the GC for~$m\gtrsim 10^{-23}$ eV (not even for a soliton with critical self-interactions). The simultaneous use of TianQin and Taiji with LISA does not change this conclusion. The prospects for LISA are more promising for constraining the direct coupling of the ULDM to the SM. Through the observation of (quasi)monochromatic GWs from Galactic binaries, LISA can probe quadratic couplings with unprecedented sensitivity in the mass window~$2\times 10^{-22} \lesssim m[{\rm eV}] \lesssim 3\times 10^{-21}$ (cf. Fig.~\ref{fig:Lambda2res}), where the current stringent constrain is derived from Cassini tests of General Relativity~\cite{Blas:2016ddr,2003Natur.425..374B}. However, in the case of linear direct couplings, we do not expect LISA to do better than current probes (cf. Fig.~\ref{fig:Lambda1res}).

Remarkably, our results in Fig.~\ref{fig:Upsilonres} show that the observation of (quasi)monochromatic GWs at ET/CE from a single (young) spinning NS at the GC with frequency~$f_e\sim 10^3\,{\rm Hz}$ and~${\rm SNR}_h\sim 20$ can be used to probe the oscillations of the gravitational potentials of a ULDM soliton for~$m\lesssim 10^{-22}\,{\rm eV}$. Additionally, this type of observation with ET/CE would outperform LISA in probing direct couplings of ULDM to the SM (cf. Figs.~\ref{fig:Lambda1res} and~\ref{fig:Lambda2res}). In particular, it would not only probe the mass window~$2\times 10^{-22} \lesssim m[{\rm eV}] \lesssim 3\times 10^{-21}$ with unprecedented sensitivities for both linear and quadratic direct couplings, but would even outperform current PTA probes of quadratic couplings at~$m \lesssim 2\times 10^{-22}\,{\rm eV}$, showing the great discovery potential of our method.
The main reason for the improved prospects for (quasi)monochromatic GWs at ET/CE compared to LISA is the linear dependence of ${\rm SNR}_\delta$ on the carrier's frequency~$f_{\rm e}$, which is more than five orders of magnitude higher for spinning NSs in ET/CE than for mHz sources in LISA. We also show the expected improved sensitivity from stacking~$\mathcal{O}(200)$ GW signals from spinning NSs (which is larger than with a single GW by a factor~$\sqrt{200}$).

In App.~\ref{app:snr} we comment on the possibility of coherently combining signals from binary pairs, which could improve the overall SNR by more than the $\sqrt{N}$ factor considered for the case of incoherent combination we considered before. This could be possible if the SNR is high enough to localise the binary precisely, or by performing a global fit to all the relative modulation phases of the different binaries.

%%%%%%%%%%%%%%%%%%%%%%%%%%%%%%
\section{Other galaxies}
\label{sec:othergal}
%%%%%%%%%%%%%%%%%%%%%%%%%%%%%%

The quasi-monochromatic sources discussed in Sec.~\ref{sec:MWC} are expected to be observed only within (or close enough to) the MW. 
But we will also observe GWs from other galaxies: from extreme-mass-ratio inspirals (EMRIs) and intermediate-mass-ratio inspirals (IMRIs) in LISA, to compact binaries with BH/NS components, e.g., in the deci-Hertz band of (B-)DECIGO~\cite{Kawamura:2011zz, Yagi:2011wg, Kawamura:2020pcg}. These sources have a non-negligible chirp, making the computation of the ULDM effect on the waveform considerably more involved than in the quasi-monochromatic case (which we presented in Sec.~\ref{sec:heterodyning}). We discuss that computation in detail in App.~\ref{app:Sachs-Wolfe}.

For chirping signals, the amplitude of the frequency modulation increases with the frequency of the main signal instead, cf. Eq.~\eqref{eq:modsignal}. But, interestingly, as shown in App.~\ref{app:Sachs-Wolfe} (cf.,~Fig.~\ref{fig:eta}), if several ULDM oscillations are observed in-band, the cumulative relative phase difference introduced in the signal is maximum when the time to coalescence,~$\tau$, is close to~$1/\omega_\delta$ (i.e., when the time to coalescence corresponds to one oscillation of the ULDM gravitational field). In the dominant (quadrupole) mode and in frequency domain, this maximum imprint corresponds to frequencies~$f$ close to the $f_{\rm e}$ at~$\tau \sim \/\omega_\delta$ (cf. Eq.~\eqref{eq:Newfreqs}).

For concreteness, we consider four systems: an EMRI with redshifted masses $(m_1,m_2)=(10^6M_\odot,60\,M_\odot)$, an IMRI with $(m_1,m_2)=(10^4M_\odot,10\, M_\odot)$, both observed in LISA, and two stellar-mass compact binaries in (B-)DECIGO with the same parameters as the binary BH (BBH) event \texttt{GW170608}~\cite{LIGOScientific:2017vox} and the binary NS (BNS) event \texttt{GW170817}~\cite{LIGOScientific:2017vwq}. 
We use a Fisher analysis to compute the uncertainty on the estimation of the ULDM modulation amplitude,~$\sigma_\Upsilon$ (see Sec.~\ref{sec:modwave} of App.~\ref{app:Sachs-Wolfe}).
We take the threshold for detectability to be the minimal~$\Upsilon_*$ for which  $\sigma_\Upsilon\lesssim \Upsilon_*$. 
Since the uncertainty~$\sigma_\Upsilon$ depends on the value of~$\Upsilon_*$ where the Fisher matrix is evaluated, we can find the minimal~$\Upsilon_*$ via an iterative procedure (differently than Ref.~\cite{Brax:2024yqh}, which evaluated the Fisher matrix at $\Upsilon_*=0$). 

\begin{figure}
    \centering
    \includegraphics[width = \linewidth]{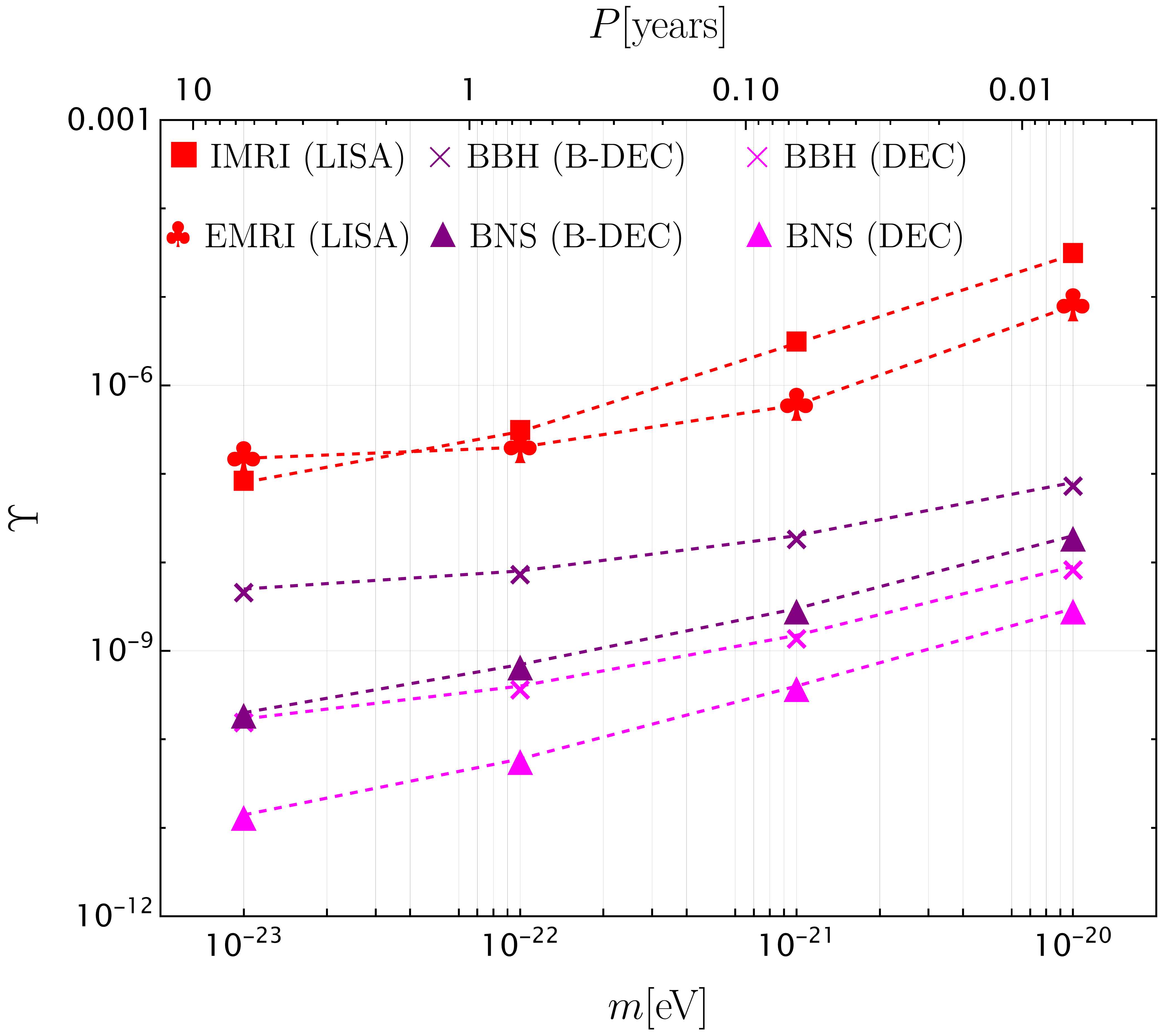}
    \caption{Threshold value~$\Upsilon_*$ for different sources and ULDM particle masses, obtained via a Fisher analysis. The minimum frequency is fixed by the total observation time of~$6$~yrs and the maximum one is chosen as~$f_{\rm max}=\min(f_0,f^{\rm det}_{\rm max})$, with ~$f^{\rm LISA}_{\rm max}=1\,{\rm Hz}$ for LISA and $f^{\rm DECIGO}_{\rm max}=100\,{\rm Hz}$, and~$f_0$ as defined in Ref.~\cite{Robson:2018ifk} (marking the transition between the inspiral and the merger stage).}
    \label{fig:chirpres}
\end{figure}

Our results are shown in Fig.~\ref{fig:chirpres} for different ULDM particle masses~$m$. In this analysis, we considered~$6$~yrs of observations and used a distance of $d_{\rm L}=1\,{\rm Gpc}$ for the LISA sources, and the distances estimated for the real \texttt{GW170608} and \texttt{GW170817} events for (B)DECIGO. Modulo cosmological redshift corrections, the threshold~$\Upsilon_*$ scales simply with~$d_{\rm L}$. We considered only the inspiral waveform, modelled through the impact of the emission from the quadrupole formula into Newtonian dynamics. The total SNR for 6~yrs of observation in LISA of the EMRI and IMRI, and in B-DECIGO and DECIGO of \texttt{GW170608} and
\texttt{GW170817}-like signals is, respectively, ${\rm SNR}_h=(22,305,220,374,2737,4414)$.

The results in Fig.~\ref{fig:chirpres} show that stellar-mass compact binaries are more sensitive to the ULDM effect than EMRIs and IMRIs. This is mainly due to their higher frequencies, as it would hold even if one scales $\sigma_\Upsilon$ to the same ${\rm SNR}_h$. The case is similar to the (quasi)monochromatic signals. Indeed, we find that the sensitivities given by Eq.~\eqref{eq:Gformula} still give the correct order of magnitude for chirping signals. However, the correlations between the different parameters in the waveform have some impact on the sensitivity. We remark that our results agree with the ones of Ref.~\cite{Brax:2024yqh} (cf. their Figs.~1 and~2).

\begin{figure}
    \centering
    \includegraphics[width = \linewidth]{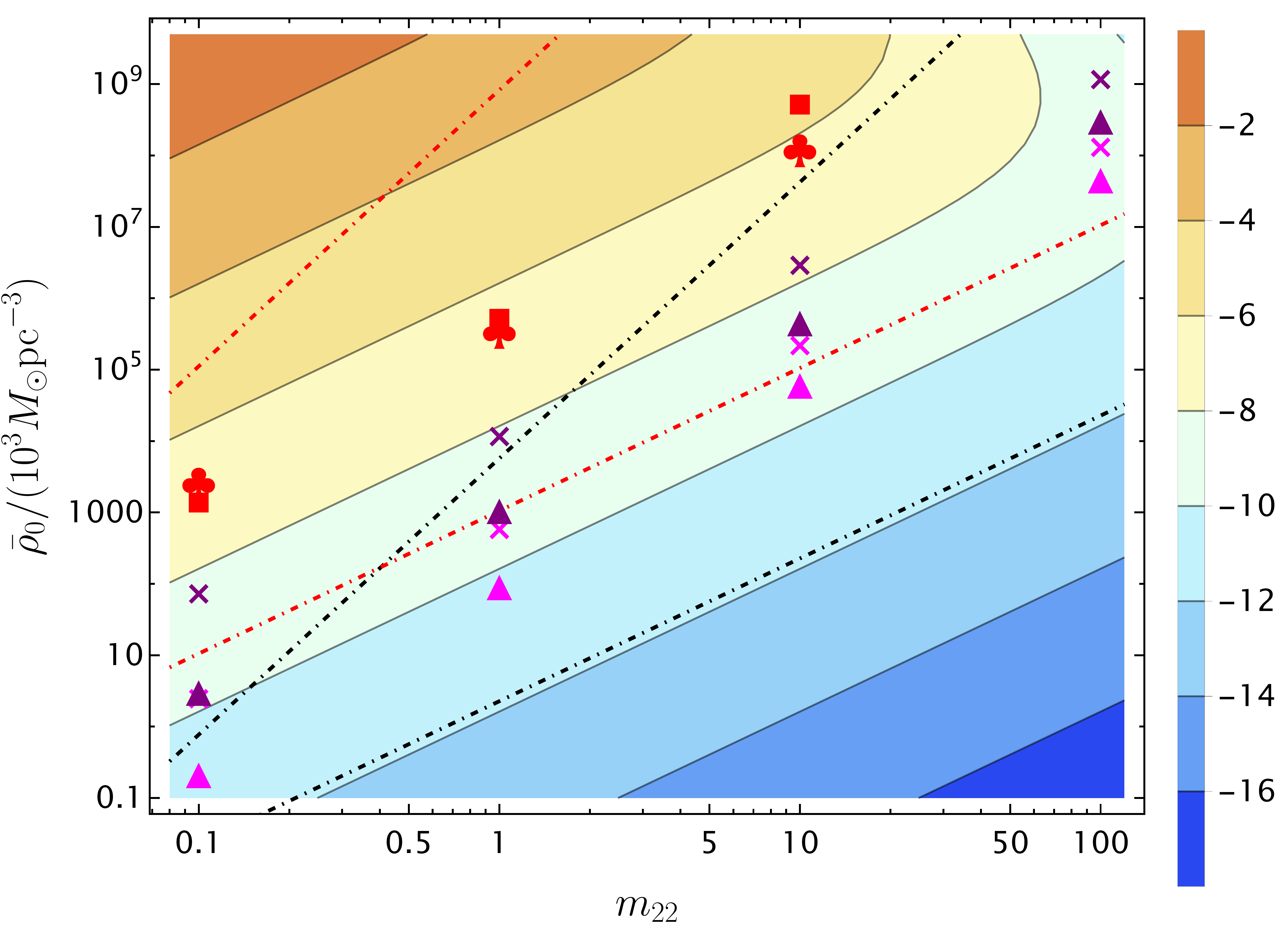}
    \caption{Same as the upper panel of Fig.~\ref{fig:Upsilon_50pc}, now for other galaxies. The region delimited by black (red) dashed dotted lines corresponds to solitons satisfying Eq.~\eqref{eq:scatter} with $M_{\rm h}\sim10^{12}M_\odot\,(10^{14}M_\odot)$. 
    The red, purple, and magenta symbols are the threshold densities for detectability given in Fig.~\ref{fig:chirpres}. To get an estimate of the sensitivity for masses in between symbols, one can simply extrapolate linearly between them, as we showed in Fig.~\ref{fig:chirpres}.}
    \label{fig:othergal}
\end{figure}

When extending the analysis to other galaxies, one of the challenges is estimating the expected value of~$\Upsilon$ at the position of the source, required to connect Fig.~\ref{fig:chirpres} to a specific ULDM model. The most favourable situation is if the source is inside the soliton at the corresponding galactic centre, which is expected, at least, for EMRIs. In Fig.~\ref{fig:othergal} we show~$\pi |\Upsilon|/(\sqrt{2}\, m_{22})$ in that case (restricting to the minimal coupling), together with the sensitivities of the different chirping sources (c.f. Fig.~\ref{fig:chirpres}). 
Note that since $\Upsilon\propto \bar{\rho}_ 0$, the threshold~$\bar{\rho}_0$ corresponding to a detectable effect scales (roughly) with~$d_{\rm L}$ (modulo cosmological redshift corrections). For reference, we also highlight the parameter space region where ULDM solitonic cores of halos of mass~$M_{\rm h}\sim 10^{12}M_\odot$ or~$\sim 10^{14}M_\odot$ are expected to lie (cf. Eqs.~\eqref{eq:scatter}).

From these results, we conclude that stellar-mass compact objects observed in (B-)DECIGO might be able to probe solitons in DM halos similar to or more massive than the MW one ($M_{\rm h}\gtrsim 10^{12}M_\odot$). As a result, one would be able to probe some ULDM models from purely gravitational effects, cf. Fig.~\ref{fig:othergal}. On the other hand, EMRIs and IMRIs in LISA could be used to probe the effect only for more massive halos. Unfortunately, the primary BH of an EMRI in the LISA band has a mass in the range~$M_{\rm BH}\simeq 10^{4.5}\text{--}10^{6.5} M_\odot$, which are unlikely to be observed at the centre of such massive halos. For instance, according to the empirical relation between the central BH mass and the corresponding halo mass of the galaxy of Ref.~\cite{Bandara_2009}, the primary BHs of EMRIs in LISA are expected to be in halos of mass~$M_{\rm h}\approx10^{10}\text{--}10^{12.5}M_\odot$.

Finally, we remark that in this section we have only considered the purely gravitational effect of ULDM (i.e., the minimal coupling), while, as shown in previous sections, these systems could also be used to probe direct couplings with ordinary matter.\footnote{The sensitivity to such coupling can be easily estimated using Fig.~\ref{fig:chirpres} together with Eq.~\eqref{eq:upps}.}
Moreover, by considering other galaxies, with few or no constraints on the density profile in their inner regions, we are free to take the full scatter in Eq.~\eqref{eq:scatter} and explore larger central densities (and more compact) solitons than in the MW [compare the values for~$\bar \rho_0$ of Fig.~\ref{fig:othergal} with those of Fig.~\ref{fig:Upsilon_500pc}]. However, solitons with larger central densities are also smaller and so will affect less sources.

%%%%%%%%%%%%%%%%%%%%%%%%%%%%%%
\section{Other overdensities in DM halos} \label{sec:large_miss}
%%%%%%%%%%%%%%%%%%%%%%%%%%%%%%

We concluded in previous sections that the only possibility to probe ULDM
through the frequency modulation of GWs is if sources are in the presence of large DM overdensities (cf. the NFW line in Fig.~\ref{fig:Upsilonres}). As discussed in Sec.~\ref{sec:halo}, certain ULDM models generate these overdensities not only at the galactic centres (for which we derived observability prospects in Secs.~\ref{sec:MWC} and~\ref{sec:othergal}), but also throughout the DM halo in the form of compact structures. This possibility is relevant as it expands the region where emitters can be found in galaxies.

For concreteness, we discuss briefly the detectability prospects for the effect from (subhalo) overdensities produced via the large misalignment mechanism, already introduced in Sec.~\ref{sec:halo}. Given the prospects derived in the previous sections, we restrict to the most promising scenario: observing the effect from these overdensities in the MW,\footnote{Considering other galaxies will expand the parameter space one can access. We leave such exploration for future work.} and consider the solitons with $\rho_s$ given by Eq.~\eqref{eq:overdensityLMM}. From Refs.~\cite{Arvanitaki:2019rax, Thompson:2023mfx}, we know that a certain fraction~$Y_{\rm s}$ of ULDM in the Galactic halo will be condensed into these subhalo structures. The number of subhalos is 
\begin{equation}
    N_{\rm s}= \frac{Y_{\rm s} M_{\rm h}}{M_{\rm s}}\sim 20 \left(\frac{Y_{\rm s}}{0.1}\right) m_{22}^{3/2},
\end{equation}
where we used Eq.~\eqref{eq:massradiusLMM}.  
We estimate the probability of a Galactic binary being inside such overdensities by the ratio of the total volume in these overdensities to the volume of the Galaxy halo, which from  Eq.~\eqref{eq:radiusLMM} gives 
\begin{equation}
    \mathcal{P}\approx N_{\rm s}\left(\frac{R_{\rm s}}{R_{\rm h}}\right)^3\sim 10^{-6} \left(\frac{Y_{\rm s}}{0.1}\right) \left(\frac{10^3}{\mathcal{B}}\right). \label{eq:prob}
\end{equation}
This is independent of the ULDM particle mass. 

From Eq.~\eqref{eq:prob}, we see that a lower value of~$\mathcal B$ increases the probability of having sources residing within subhalos. However, such subhalos would also have lower densities (cf.~Eq.~\eqref{eq:overdensityLMM}) and lead to smaller effects on the carrier GWs (cf.~Fig.~\ref{fig:Upsilon_500pc}). For an intermediate value~$\mathcal B\sim 10^3$, the corresponding density is $\rho \sim 20 M_\odot/\mathrm{pc^3}$, which could be detectable with ET/CE; however, the probability for one spinning NS to reside in such subhalos seems very small.   
Nevertheless, note that this estimate is very crude, as it does not consider the proper distribution of sources and subhalos in the Galaxy. Moreover, when self-interactions are effective, the oscillating gravitational potential is relevant well beyond the size of the soliton, as shown in Fig.~\ref{fig:potentials}, which would enhance the chances of emitters to be affected by nearby overdensities. We will come back to a more realistic estimate in future work.

%%%%%%%%%%%%%%%%%%%%%%%%%%%%%%%%%
\section{Conclusions}
\label{sec:discu}
%%%%%%%%%%%%%%%%%%%%%%%%%%%%%%%%%

The extension of the landscape of DM models to ultra-light masses brings new opportunities to search for this (still) mysterious component of the Universe. A striking signature of these models is the existence of large dense coherent (sub)structures, whose gravitational potentials oscillate at frequencies proportional to the ULDM particle mass. These result in the frequency modulation of gravitational (and electromagnetic) signals sourced in their surroundings. ULDM conformal couplings to ordinary matter are responsible for the same effect. Stochastic fluctuations of the ULDM field in the halo can also lead to observable effects~\cite{Kim:2023kyy, Kim:2023pkx, yao2024probing}. These effects are complementary to others which are also of \emph{purely gravitational} nature (and hence test models independently of the direct coupling to light or matter), like the change in GW propagation speed~\cite{Dev:2016hxv, Banerjee:2022zii}, the orbital dephasing due to dynamical friction and accretion~\cite{Annulli:2020ilw, Annulli:2020lyc, Baumann:2021fkf, Traykova:2021dua, Vicente:2022ivh, Bamber:2022pbs, Traykova:2023qyv, Tomaselli:2023ysb, Aurrekoetxea:2023jwk, Brito:2023pyl, Duque:2023cac} in ULDM structures or 
modification of orbital motion \cite{Blas:2016ddr,Blas:2019hxz,Kus:2024vpa,Zwick:2024hag,Kim:2024rgf}.

In this work, we explored how the imprint left by coherent DM structures in the phase of GWs may be used to detect ULDM with GW observations. For that, we considered GW (carrier) signals with well-known frequency content that will be observed with near-future detectors. We have included the effect of attractive quartic self-interactions in our analysis, which considerably increases the influence region of ULDM structures. To assess the effect's observability prospects, we used realistic astrophysical population models for the carrier sources, together with the soliton-halo mass relations derived from cosmological simulations~\cite{Schive:2014hza,Chan:2021bja}. We have also used the results from structure formation simulations in the context of the large-misalignment mechanism~\cite{Arvanitaki:2019rax}.
While our results confirm the findings of Refs.~\cite{Wang:2023phr, Brax:2024yqh}, that typical average densities in halos lead to unobservable effects, we show that the DM cores (solitons) seen to form in ULDM cosmological simulations can be observable.

For (quasi)monochromatic GW signals from the MW, our results are summarized in Fig.~\ref{fig:Upsilonres}. We conclude that the effect from a ULDM field minimally coupled to gravity on (quasi)monochromatic GWs can be detected \emph{only} if we observe GWs from a spinning NS at the GC with ET/CE. In that case one could probe a ULDM soliton for masses~$10^{-23}\lesssim m[{\rm eV}]\lesssim 10^{-22}$, outperforming current (and perhaps future) PTA observations~\cite{EuropeanPulsarTimingArray:2023egv}. LISA Galactic sources can be used to constrain quadratic couplings of ULDM to ordinary matter in a relevant region of parameter space. But, again, observing the GW signal from a spinning NS at the GC would be the most effective in constraining ULDM conformal couplings. We also note that, as compare to other constraints, our bounds are based on the existence of ULDM solitons generated by relaxation, which means that they may be less impacted when one considers that the ULDM that generates the relevant soliton constitutes a part of the total DM in the Universe. The concrete impact is model dependent and we leave a detail study for future work.

The case for chirping signals (sourced in other galaxies) is summarized in Fig.~\ref{fig:othergal}, which shows that the effect from a ULDM field minimally coupled to gravity could be observed in the GW signals from stellar-mass compact binaries observed in (B-)DECIGO, if they merge close enough~($\lesssim 250\,{\rm pc}$) to the centre of a galaxy at least as massive as the MW; this could be used to probe ULDM particle masses~$10^{-23}\lesssim m[{\rm eV}]\lesssim 10^{-20}$. The effect on EMRIs and IMRIs signals would only be observed in more massive galaxies ($M_{\rm h}\gtrsim 10^{14}M_\odot$), which are expected to host a central supermassive BH too large (i.e., with orbital frequencies too low) to be in the LISA band. 

On a general note, we remark that while larger overdensities lead to larger effects, these are typically of smaller sizes, affecting fewer or no GW sources at all. Indeed, this was what we found in our estimate for the effect of subhalos originating from a large misalignment mechanism. Similar (but even more extreme) instances are oscillons and superradiant clouds which were not studied here. Such boson structures lead to the same frequency modulation effects. However, our framework does not apply immediately to those, as oscillons are relativistic and superradiant clouds are not spherically symmetric. The attractive quartic self-interactions may increase the influence region of these structures to such an extent that would enhance their discovery potential. We leave a thorough exploration for the future.

\section{\label{acknowledgments} Acknowledgments}
We are grateful to Enrico Barausse, Kfir Blum, Valerya Korol, Michele Maggiore, Angelo Ricciardone, Alberto Sesana  and Fabio Van Dissel for their  insights during the development of this project. D.B. would like to thank the Department of Theoretical Physics at CERN for its hospitality during the last stages of this work. 
S.G. has the support of the predoctoral program AGAUR FI SDUR 2022 from the Departament de Recerca i Universitats from Generalitat de Catalunya and the European Social Plus Fund. R.V. is supported by grant no. FJC2021-046551-I funded by MCIN/AEI/10.13039/501100011033 and by the European Union NextGenerationEU/PRTR. The authors acknowledge the support from the Departament de Recerca i Universitats from Generalitat de Catalunya to the Grup de Recerca 00649 (Codi: 2021 SGR 00649).
The research leading to these results has received funding from the Spanish Ministry of Science and Innovation (PID2020-115845GB-I00/AEI/10.13039/501100011033).
This publication is
part of the grant PID2023-146686NB-C31 funded by MICIU/AEI/10.13039/501100011033/ and by FEDER, UE.
IFAE is partially funded by the CERCA program of the Generalitat de Catalunya.
D.B. and S.G. acknowledge the support from the European Research Area (ERA)
via the UNDARK project of the Widening participation
and spreading excellence programme (project number 101159929).
%\newpage
\appendix
%%%%%%%%%%%%%%%%%%%%%%%%%%%%%%%%%%%%%%%%%%%%
%%%%%%%%%%%%%%%%%%%%%%%%%%%%%%%%%%%%%%%%%%%%
\section{Gravitational potential fluctuations}
\label{app:potentials}

The gravitational fluctuations produced by a non-relativistic, self-gravitating, real scalar field are described by Eqs.~\eqref{eqs:delta}. To show this, let us first consider the perturbed metric
\begin{equation}
    g_{\mu \nu}\approx\eta_{\mu \nu} +{}^{(1)}\gamma_{\mu \nu}+{}^{(2)}\gamma_{\mu \nu},
\end{equation}
where $\eta_{\mu \nu}=\mathrm{diag}(-1,\mathbb{1}_3)$, and~$O[{}^{(j)}\gamma_{\mu \nu}]= \epsilon^j$, with~$\epsilon \ll 1$. The perturbation is assumed to be sourced by a non-relativistic scalar field of the form~\eqref{eq:NRfield}, with
\begin{equation*}
    O[\tfrac{1}{\mb^n}\partial_t^{(n)} \log \psi]=\epsilon^n, \quad O[\tfrac{1}{\mb^n}\partial_i^{(n)}\log \psi]=\epsilon^{n/2}.
\end{equation*}
The field is also assumed to be dilute (so that it sources a weak gravitational field). In particular, the Einstein's equations imply~$O[\psi/\sqrt{\mb}]=\epsilon$ and
\begin{nofleqn}{equation*}
    O\big[\tfrac{1}{\mb^n}\partial_t^{(n)} {[}^{(j)}\gamma_{\mu \nu}]\big]=\epsilon^{n+j}, \quad O\big[\tfrac{1}{\mb^n}\partial_i^{(n)} {[}^{(j)}\gamma_{\mu \nu}]\big]=\epsilon^{n/2+j}.
\end{nofleqn}
The energy-momentum tensor of the scalar field is also expanded as~$T_{\mu \nu}\approx {}^{(1)}T_{\mu \nu}+{}^{(2)}T_{\mu \nu}$, where~$O[{}^{(j)}T_{\mu \nu}]=\epsilon^{2j}$.

The perturbations~${}^{(j)}\gamma_{\mu \nu}$ can be decomposed into irreducible representations of $SO(3)$, which can then be further Helmholtz decomposed into transverse and longitudinal pieces, resulting in~\cite{Flanagan:2005yc}
\begin{gather*}
        \gamma_{tt}=2\Phi, \qquad \gamma_{ti}=\beta_i+\partial_i h,\\
        \gamma_{kl}=2\Psi+\gamma_{kl}^{\rm TT}+\partial_{(k}\varepsilon_{l)}+\{\partial_k\partial_l-\tfrac{1}{3}\delta_{kl}\nabla^2\}\tilde\lambda,
\end{gather*}   
with the four scalars~$(\Phi, \Psi, h, \tilde\lambda)$, the two transverse spatial vectors $(\beta_i,\varepsilon_i)$ satisfying~$\partial^i \beta_i=\partial^i \varepsilon_i=0$, and the transverse-traceless (TT) spatial tensor~$\gamma_{ij}^{\rm TT}$ satisfying~$\eta^{ij}\gamma_{ij}^{\rm TT}=0$ and~$\partial^i \gamma_{ij}^{\rm TT}=0$. 
The same decomposition of the energy-momentum tensor~${}^{(j)}T_{\mu \nu}$ gives
\begin{gather*}
    T_{tt}=\rho, \qquad T_{ti}=S_i+\partial_i S, \\
    T_{ij}=P\delta_{ij}+\sigma_{ij}+\partial_{(i}\sigma_{j)}+\{\partial_i\partial_j-\tfrac{1}{3}\delta_{ij} \nabla^2\}\sigma,
\end{gather*}
where~$\partial^i S_i=\partial^i\sigma_i=0$, $\partial^i\sigma_{ij}=0$ and $\sigma_{ii}=0$.

Defining the variables~$\uline{\Phi}\equiv \Phi-\partial_t h+\partial_t^2\tilde\lambda/2$,~$\uline{\Psi}\equiv\Psi-\nabla^2\tilde\lambda/3$, and~$\uline{\beta_i}\equiv \beta_i-\partial_t \epsilon_i$, the linearized Einstein tensor reads~\cite{Flanagan:2005yc}
\begin{subequations}
\begin{align}
    {}^{(1)}G_{tt}[\bm{\gamma}]&=-2\nabla^2 \uline{\Psi}, \\
    {}^{(1)}G_{ti}[\bm{\gamma}]&=-\tfrac{1}{2}\nabla^2\uline{\beta}_i-2\partial_i\partial_t\uline{\Psi}, \\
    {}^{(1)}G_{ij}[\bm{\gamma}]&=-\tfrac{1}{2} \Box_{\bm{\eta}}\gamma_{ij}^{\rm TT}-\partial_{(i}\partial_t\uline{\beta}_{j)}\nn \\
    &-\partial_i\partial_j[\uline{\Psi}-\uline{\Phi}]+\delta_{ij}\big\{\nabla^2[\uline{\Psi}-\uline{\Phi}]-2\partial_t^2\uline{\Psi}\big\}.
\end{align}
\end{subequations}

At leading order $\epsilon^2$, the Einstein's equations are simply ${}^{(1)}G_{\mu \nu}[{}^{(1)}\bm{\gamma}]={}^{(1)}T_{\mu \nu}$. These are linear in~${}^{(1)}\gamma_{\mu \nu}$, implying that the scalar, vector, and tensor sectors decouple. The $(tt)$, the spatial divergence of~$(ti)$, and the trace of~$(ij)$ components results, respectively, in the equations
\begin{gather}
    \nabla^2 [{{}^{(1)}\uline{\Psi}}]=-4\pi\, {}^{(1)}\rho, \label{eq:appPsi}\\
    \partial_t [{{}^{(1)}\uline{\Psi}}]=-4\pi \,{}^{(1)}S,\\
    3 \partial_t^2 [{}^{(1)}\uline{\Psi}]+\nabla^2[{}^{(1)}\uline{\Phi}-{}^{(1)}\uline{\Psi}]=-12\pi\, {}^{(1)} P ,
\end{gather}
from which,
\begin{equation} \label{eq:appPhi}
    \nabla^2 [{{}^{(1)}\uline{\Phi}}]=-4\pi\big[ {}^{(1)}\rho+3( {}^{(1)}P-\partial_t  {}^{(1)}S) \big].
\end{equation}
Finally, the vector and tensor modes are, respectively, described by
\begin{gather}
    \nabla^2[{}^{(1)}\uline{\beta}_i]=-16\pi\, {}^{(1)}S_i,\\
    \Box_{\bm{\eta}}[{}^{(1)}\gamma_{ij}^{\rm TT}]=-16 \pi \, {}^{(1)}\sigma_{ij}
\end{gather}

\begin{subequations}
Using the stress-energy tensor of the scalar field in Eq.~\eqref{eq:enmomtenscalar}, we note that the only non-vanishing sources, at the leading order~$\epsilon^2$, are the scalars
\begin{gather}
    {}^{(1)}\rho =\mb |\psi|^2,\\
    \partial_t[{}^{(1)}S]=- \mb\re[e^{-2 i \mb t}\psi^2 ], \\
    {}^{(1)}P=- \mb\re[e^{-2 i \mb t}\psi^2 ].
\end{gather}
\end{subequations}
In particular, ${}^{(1)}\sigma=0$. Thus, only scalar modes are sourced at leading order.
In the Newtonian gauge~${}^{(1)}\gamma_{\mu \nu}=2 \mathrm{\,diag}({}^{(1)}\Phi,{}^{(1)}\Psi\,\mathbb{1}_3)$, the Einstein's equations imply 
\begin{equation}
    \nabla^2[{}^{(1)}\Phi]= -4 \pi {}^{(1)}\rho,\qquad {}^{(1)}\Psi={}^{(1)}\Phi,
\end{equation}
where we used that~${}^{(1)}P=\partial_t [{}^{(1)}S]$, i.e. ${}^{(1)}\sigma=0$. Note that, at this order, the gravitational perturbations~${}^{(1)}\gamma_{\mu \nu}$ are slowly varying. The KG equation reduces to
\begin{equation}
    i \partial_t \psi = -\Big\{\tfrac{\nabla^2}{2 \mb}  + \mb\big[{}^{(1)}\Phi + \tfrac{1}{8\mb F^2}|\psi|^2 \big]\Big\}\psi,
\end{equation}
where we averaged out high frequency modes~$\gtrsim \mb$. Formally, in our perturbative scheme we use~$O[F]=\epsilon^{1/2}$. 

At the (next-to-leading) order~$\epsilon^3$, the Einstein's equations are
\begin{equation}
    {}^{(1)}G_{\mu \nu}[{}^{(2)}\bm{\gamma}]=8\pi\, {}^{(2)}T_{\mu \nu}-{}^{(2)}G_{\mu \nu}[{}^{(1)}\bm{\gamma}],
\end{equation}
and so there are source terms quadratic in the gravitational potentials~${}^{(1)}\Phi$ arising from~${}^{(2)}G_{\mu \nu}[{}^{(1)}\bm{\gamma}]$. However, note that these source perturbations are slowly varying in time. Our goal here is to compute (at leading order) the rapidly oscillating gravitational potentials with frequency~$\gtrsim 2\mb$, and not the corrections to the slowly oscillating gravitational potentials. Thus, we can simply use~
\begin{equation}
    {}^{(1)}G_{\mu \nu}[{}^{(2)}\bm{\gamma}]=8\pi\, {}^{(2)}T_{\mu \nu}.
    \end{equation}
Again, the scalar, vector, and tensor modes are decoupled. In this work we focus our attention on the scalar modes. Then, the relevant equations are still Eqs.~\eqref{eq:appPsi} and~\eqref{eq:appPhi}, replacing~$(1)\to(2)$. We restrict now to stationary field configurations,~$\psi=e^{-i \upgamma \mb t} \uppsi(x^i)$. At order~$\epsilon^3$, the rapidly oscillating source terms of the scalar sector are
\begin{nofleqn}{subequations}
\begin{align}
    &\delta\,{}^{(2)} \rho =\re\Big[\{\tfrac{\nabla^2}{4 \mb}+\tfrac{|\psi|^2}{24 F^2} \}\psi^2e^{-2i \mb t}-\tfrac{1}{48 F^2}\psi^4e^{-4i \mb t}\Big], \\
    &\partial_t[\delta\,{}^{(2)} S]= -2\upgamma \mb\re\big[ \psi^2e^{-2i \mb t} \big],\\
    &\delta\,{}^{(2)} P=-\re\Big[\big\{\tfrac{\nabla^2}{12 \mb}+\tfrac{2\mb}{3}(2\upgamma+5\,{}^{(1)}\Phi)-\tfrac{1}{24 F^2}|\psi|^2\big\}\nn\\ 
    &\qquad\quad\times \psi^2 e^{-2i\mb t}-\tfrac{1}{48 F^2}\psi^4 e^{-4i \mb t} \Big].
\end{align}
\end{nofleqn}
The missing scalar $\sigma$ can be derived from the conservation of the $T^{\mu\nu}$. In the Newtonian gauge~${}^{(2)}\gamma_{\mu \nu}=2 \mathrm{\,diag}({}^{(2)}\Phi,{}^{(2)}\Psi\,\mathbb{1}_3)$. Decomposing the gravitational fluctuations as~$\delta\,{}^{(2)}\Phi=\re[e^{-2i \mb t}\Phi_2+e^{-4i \mb t}\Upphi_4]$, and likewise for $\delta\,{}^{(2)}\Psi$, we finally find
\begin{subequations}
    \begin{gather}
        \nabla^2 \big[\Psi_2+\tfrac{\pi}{\mb}\psi^2\big]=-\tfrac{\pi}{6F^2}|\psi|^2 \psi^2,\\
        \nabla^2 \Phi_2=8\pi \mb \big[5\,{}^{(1)}\Phi +\upgamma -\tfrac{|\psi|^2}{12F^2\mb}\big]\psi^2,\\
        \nabla^2 \Uppsi_4=\tfrac{\pi}{12 F^2} \psi^4, \qquad \nabla^2 \Upphi_4= -\tfrac{\pi}{6 F^2} \psi^4.
    \end{gather}
    \label{eq:eqspots}
\end{subequations}
These expressions are the same as those in Eq.~\eqref{eqs:delta}.
%

%%%%%%%%%%%%%%%%%%%%%%%%%%%%%%%%%%%%%
\section{Oscillatory potential from quartic self-interactions}\label{app:self-interaction}
%%%%%%%%%%%%%%%%%%%%%%%%%%%%%%%%%%%%%

\begin{figure}
    \centering
    \includegraphics[width=\linewidth]{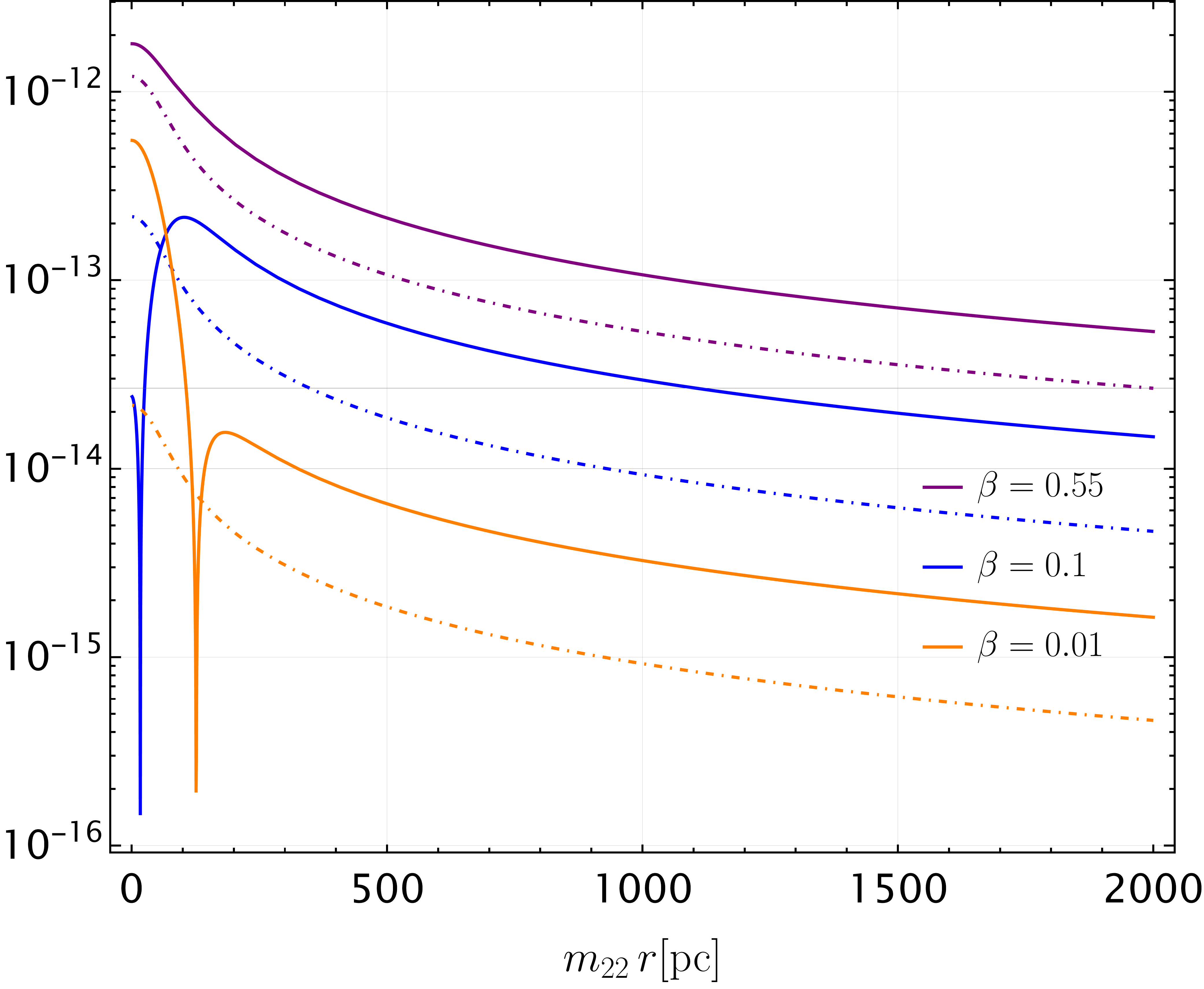}
    \caption{Comparison of $|\Psi_2|$ (solid) with $|\Psi_4|$ (dashed) for the different self-interaction values considered in Fig.~\ref{fig:potentials}.}
    \label{fig:selfinterpots}
\end{figure}

In App.~\ref{app:potentials}, we have seen that quartic self-interactions source a gravitational potential $\Psi_4$, oscillating at a frequency $\omega = 4m$. As demonstrated in Eqs.~\eqref{eq:eqspots}, the amplitude of $\Psi_4$ satisfies the same equation as $[\Psi_2 + (\pi/m)\psi^2]$, with a scaling factor of $(-1/2)$.

Figure~\ref{fig:selfinterpots} compares the profiles of $|\Psi_2|$ and $|\Psi_4|$ for various values of the self-interaction parameter $\beta$ defined in Eq.~\eqref{eq:beta}.
Far from the soliton, the relationship $|\Psi_2| \approx 2|\Psi_4|$ holds. However, near the soliton's core, where the $(\pi/m)\psi^2$ term becomes significant, $|\Psi_4|$ can locally exceed $|\Psi_2|$. This happens in particular for moderately strong self-interaction (e.g., as shown for $\beta = 0.1$). 
Despite this, since $|\Psi_2|$ is generally larger than $|\Psi_4|$, we focused our detectability analysis on solitons with zero or maximal self-interactions, neglecting the contribution of~$\Psi_4$ to the effect. We note that given the larger frequency modulation imparted by~$\Psi_4$ on carrier GWs (twice the one due to~$\Psi_2$), a shorter (half) observation time is necessary to probe the effect, which in some cases may allow the exploration of lower ULDM masses than with~$\Psi_2$.

%%%%%%%%%%%%%%%%%%%%%%%%%%%%%%%%%%%%%
\section{Sachs-Wolfe effects on GWs}\label{app:Sachs-Wolfe}
%%%%%%%%%%%%%%%%%%%%%%%%%%%%%%%%%%%%%

The propagation of GWs in non-trivial backgrounds is widely studied in cosmology and gravitational physics \cite{Maggiore:2007ulw, Dodelson:2003ft}. A key aspect of our background is that it has non-trivial time dynamics of characteristic frequency $\omega_b$ related to the mass of the ULDM candidate, which adds to the possible lensing features. 
For completeness, we now provide a brief derivation of the effects on a GW of frequency $\omega$,~$h_{\mu \nu}$, propagating on a dynamical background~$g_{\mu \nu}=\eta_{\mu \nu}+\gamma_{\mu \nu}$, assuming~$|\gamma_{\mu \nu}|\sim \epsilon_\gamma\ll1$ (i.e., weak gravitational field) and~$|\partial^{(n)} h_{\mu \nu}|= O[\epsilon_h^{1-n}]$, with~$\epsilon_h\sim \omega_b/\omega\ll 1$ (i.e., the background field changes over length and time scales much larger than the GW's). Here, we follow the treatment of Ref.~\cite{Laguna:2009re}.

In a Lorenz traceless gauge with respect to the background metric, $\nabla_\mu h^{\mu \nu}=0$ and $g^{\mu \nu} h_{\mu \nu}=0$, the Einstein's equations at leading order in~$\epsilon_h$ (high-frequency limit) are~\cite{Podolsky:2003bm}
\begin{equation}\label{eq:linear_Einstein}
    \Box_g h_{\mu \nu}+\big[2R_{\gamma \mu \delta \nu}-(g_{\mu \gamma} R_{\nu \delta}+g_{\nu \gamma} R_{\mu \delta}) \big]h^{\gamma \delta}=0,
\end{equation}
where~$R_{\mu \nu \gamma \delta}$ and~$R_{\mu \nu}$ are the Riemman and Ricci tensors of the background scalar field configuration.\footnote{In the Einstein's equations, we neglected the response of the scalar field to the propagating~$h_{\mu \nu}$, since the GW frequency,~$\omega$, is supposed to be much higher than the normal mode frequencies of the background ($\sim |\gamma| \mb$), which characterize the response timescale of the configuration (see, e.g., Ref.~\cite{Annulli:2020lyc}).} 
In this limit, we can use the WKB (\emph{geometrical optics}) approximation~\cite{Isaacson:1968hbi, Podolsky:2003bm}, with $h_{\mu \nu} \equiv%\re [\mathfrak{e}_{\mu \nu}h]=
\re [\mathfrak{e}_{\mu \nu}\mathcal{A} e^{i \alpha}]$, where~$\mathcal{A}$ is a slowly changing real function of the position such that $\partial^{(n)}\mathcal{A}=O[\epsilon_h]$, $\alpha$ is a phase such that~$\partial^{(n)}\alpha=O[\epsilon_h^{-1}]$, and~$\mathfrak{e}_{\mu \nu}$ is a (normalized) polarization tensor. At leading order in~$\epsilon_h$, Eq.~\eqref{eq:linear_Einstein} implies~\cite{Isaacson:1968hbi, Podolsky:2003bm}
\begin{gather*}
    k_\mu k^\mu=0,  \qquad k^\alpha\nabla_\alpha \mathfrak{e}_{\mu \nu}=0, \qquad  \nabla_\alpha (\mathcal{A}^2 k^\alpha)=0,\\
     k^\mu \mathfrak{e}_{\mu \nu}=0, \qquad g^{\mu \nu} \mathfrak{e}_{\mu \nu}=0, \qquad \mathfrak{e}^{\mu \nu}\mathfrak{e}_{\mu \nu}=1,
\end{gather*}
where the null rays~$k_\mu \equiv \nabla_\mu \alpha$ are tangent vectors to null geodesics, i.e., $k^\alpha \nabla_\alpha k^\mu=0$.

The previous equations are solved in an expansion in~$\epsilon_\gamma$. 
At zero order, the null rays are ${}^{(0)}k^\mu=\omega(1,-n^i)$, where~$n^i$ is a unit-vector pointing in the sky-direction of arrival of the GW. The integral null geodesic that is connected to the reception event~$x_{\rm r}^\mu$ is
\begin{equation}
    {}^{(0)}X^\mu(\lambda)=x_{\rm r}^\mu- (\lambda_{\rm r}-\lambda){\,}^{(0)}k^\mu/\omega.
\end{equation}
From the fact that~${}^{(0)}\nabla_\mu {}^{(0)}k^\mu=2 \omega/{}^{(0)}d_{\rm L}$, with~${}^{(0)}d_{\rm L}\equiv\lambda-\lambda_{\rm e}$ the luminosity distance from the emitter (as computed with~$\eta_ {\mu \nu}$), one finds ${}^{(0)}\mathcal{A} {\,}^{(0)}d_{\rm L}=\mathcal{Q}_{\rm e}$, with $\mathcal{Q}_{\rm e}$ a function of the retarded time~$u=t-{}^{(0)}d_{\rm L}$.

The linear order corrections in~$\epsilon_\gamma$ can be found following Ref.~\cite{Laguna:2009re}. In Newtonian gauge~$\gamma_{\mu \nu}\equiv 2 \mathrm{\,diag}(\Phi,\Psi\,\mathbb{1}_3)$, one finds
\begin{gather}
	{}^{(1)}k^0=(\Phi+ \Psi)|_{\lambda_{\rm r}}-2 \Phi|^{\lambda_{\rm r}}_{\lambda}+ I_{\rm i SW}(\lambda),\\
	{}^{(1)}k^i_{||}={}^{(0)}k^i \left[\frac{d}{d \lambda}(\Phi-\Psi)-\partial_t(\Psi+\Psi)\right],\\
	{}^{(1)}k^i_{\perp}=\big(\eta^{ij}-{}^{(0)}k^i {\,}^{(0)}k^j\big) \partial_j (\Phi+\Psi),
\end{gather}
where
\begin{equation}
    I_{\rm i SW}\equiv \int_\lambda^{\lambda_{\rm r}}\partial_t(\Phi+\Psi)\dd\lambda',
\end{equation} 
corresponds to the integrated Sachs-Wolfe (iSW) effect, ${}^{(1)}k^i_{||}\equiv {}^{(0)}k^i {\,}^{(0)}k_j {}^{(1)}k^j$ and ${}^{(1)}k^i_{\perp}\equiv \perp^i_j {}^{(1)}k^i$, with $\perp^i_j \equiv \delta^i_j-{}^{(0)}k^i{\,}^{(0)}k_j$. The GW frequency measured by the receptor (emitter) is~$\omega_{\rm r,e}=k^\mu (u_{\rm r,e})_\mu$, with observers 4-velocities $u_{\rm r, e}^\mu\equiv d x_{\rm r, e}^\mu/d\tau_{\rm r, e}\approx (1+ \Phi_{\rm r,e},v_{\rm r,e}^i)$. It is straightforward to show then
\begin{equation}\label{eq:relrecem}
	\frac{\omega_{\rm r}}{\omega_{\rm e}}\approx 1-\chi, 
 \end{equation}
 where
 \begin{equation}\label{eq:chidef}
      \chi=-\Phi|^{\rm r}_{\rm e}-n^i v_i|^{\rm r}_{\rm e}+I_{\rm iSW}(\lambda_{\rm e}).
\end{equation}

It will be convenient to rewrite
\begin{equation}
    I_{\rm iSW}(\lambda_{\rm e})=(\Phi+\Psi)|^{\rm r}_{\rm e}+n^i\int_{\lambda_{\rm e}}^{\lambda_{\rm r}}\partial_i(\Phi+\Psi)\dd\lambda',
\end{equation}
where we used~$\dd/\dd\lambda=\partial_t- n^i\partial_i$ and integrated by parts. Moreover, note that if the receptor and emitter move along geodesics, one has~$v_{\sf r,e}^i\approx \int \partial_i \Phi_{\rm r, e}\, \dd t$. Thus, at the receptor's worldline, the GW phase is 
\begin{equation}\label{eq:recphase}
	\alpha_{\rm r}= \int \omega_{\rm r} \dd\tau_{\rm r}\approx [1-\eta(\tau_{\rm r})] \int \omega_{\rm e} d\tau_{\rm r},
\end{equation}
with 
\begin{equation}\label{eq:eta}
    \eta(\tau_r)\equiv\frac{\int \omega_{\rm e} \chi \dd\tau_{\rm r}}{\int \omega_{\rm e} \dd\tau_{\rm r}}.
\end{equation}
The luminosity distance also receives corrections at linear order in~$\epsilon_\gamma$ which modifies the amplitude of the GW, which is usually referred to as lensing effect, see~\cite{Laguna:2009re}.
%\footnote{Indeed, it is easy to see that~$\mathcal{A}\, d_{\rm L}= \mathcal{Q}_{\rm e}(u_{\rm e})$ with~$u_{\rm e}=\tau_{\rm e}-d_{\rm L}$ (at all orders), where~$ \ln d_{\rm L}\equiv\frac{1}{2}\int \nabla_\mu k^\mu d\lambda$.} 
However, in the limit~$\omega_\delta \ll \omega_{\rm e}$, lensing effects are subleading w.r.t. the iSW ones.
%

%%%%%%%%%%%%%%%%%%%%%%%%%%%%%%%%%%%%%%%%%%
\section{ULDM effects on the GW phase}\label{app:mod}
%%%%%%%%%%%%%%%%%%%%%%%%%%%%%%%%%%%%%%%%%%

Using Eqs.~\eqref{eq:recphase} and~\eqref{eq:eta}, together with~\eqref{eq:chidef}, we can find the changes in the GW phase induced by the gravitational potentials of non-relativistic ULDM structures. We do it now for two distinct cases: monochromatic and chirping signals.

%%%%%%%%%%%%%%%%%%%%%%%%%%%%%%%%%%%%%%%%%%
\subsection{Monochromatic GW signal}
%%%%%%%%%%%%%%%%%%%%%%%%%%%%%%%%%%%%%%%%%%

For a monochromatic signal: $\omega_{\rm e}(u)=\omega_{\rm e}$. At leading order, the effect from coherent oscillations of the ULDM gravitational potentials on a propagating GW is captured by the de-phasing (at the receptor)\footnote{We take~$u=0$ as the retarded time when the observation starts.}
\begin{equation}
	\label{eq:SW1}
	\delta \alpha \approx \frac{\omega_{\rm e}}{\omega_{ \delta}} \Upsilon(r_{\rm e}) \left[\sin(\omega_\delta u+\varphi_\delta)-\sin\varphi_\delta\right],
\end{equation}
where we defined~$\Upsilon \equiv \big[ \Psi_2-\frac{2}{\omega_\delta}n^i\partial_i \Phi_2 \big]$, and assumed that~$n^i \partial_i \log \Phi_2(x^i_{\rm e})\ll \omega_{\delta}/2\pi$, and a much larger DM density close to the emission event than at reception. 

Using the wave's principal axis to define the (two-dimensional) orthonormal polarization tensor basis, the GW reads~$h_{\mu \nu}=\mathcal{A}_+ \mathfrak{e}^+_{\mu \nu} \cos \alpha+\mathcal{A}_\times \mathfrak{e}^\times_{\mu \nu} \sin \alpha$. The strain in the detector is~\cite{Cutler:1997ta}
\begin{equation}\label{eq:modwaveform}
    h \approx \mathcal{A}\cos\left[\omega_{\rm e}u+\delta \alpha(u)+ \alpha_0\right],
\end{equation}
with~$\mathcal{A}\equiv\sqrt{(\mathcal{A}_{+}F_+)^2 +(\mathcal{A}_{\times}F_{\times})^2}$, where $F_{+}(t)$ and $F_{\times}(t)$ are the detector's beam pattern coefficients.  
Using the trigonometric relation~\cite{Maggiore_2020}\footnote{This can be immediately obtained from the Jacobi-Anger expansion of the complex exponential~$e^{iz\sin{\theta}}=\sum_{-\infty}^{\infty}J_n(z)e^{in\theta}$.}
\begin{equation}
    \cos(x+\beta\sin y)=\sum_{n=-\infty}^{\infty}J_n(\beta)\cos(x+n y),
\end{equation}
where~$J_n$ are Bessel functions of the first kind, we can analyze the signal in different regimes. When $|\beta|\ll1$, one has~$J_n(\beta)\approx[{\rm sign}(n)]^n (\beta/2)^{|n|}/|n|!$. In that case, only the first two sidebands~$n=\pm 1$ are significant. Therefore, the signal is simply 
\begin{multline}\label{eq:hplusdelta}
    h\approx \mathcal{A}
    \Big[\cos(\omega_e u+\alpha'_0)\\
    \pm \frac{\omega_{\rm e}}{2\omega_{ \delta}}\Upsilon|_{r_{\rm e}}\cos[(\omega_e\pm \omega_\delta)u+\alpha'_0\pm\varphi_\delta]\Big],
\end{multline}
with~$\alpha'_0\equiv \alpha_0-(\omega_{\rm e}/\omega_{ \delta})\Upsilon (r_{\rm e}) \sin \varphi_\delta$.
When $|\beta|\gtrsim 1$, the signal gets spread into higher harmonics. This leads to approximately $\mathcal{O}(2|\beta|)$ peaks around the dominant frequency due to the broader spread of the signal components. The cases considered in this work are always restricted to the regime where $|\beta|\ll1$.

%%%%%%%%%%%%%%%%%%%%%%%%%%
\subsection{Chirping GW signal}\label{sec:modwave}
%%%%%%%%%%%%%%%%%%%%%%%%%%

Some of the binary sources we consider evolve due to the emission of GWs. This generates a frequency chirp, which at Newtonian order for the binary system and considering the emission from the quadrupole formula is given by~\cite{Maggiore:2007ulw}
\begin{equation}\label{eq:Newfreqs}
    \omega_{\rm e}=\frac{1}{4}  \mathcal{M}^{-5/8} \left( \frac{5}{ \tau} \right)^{3/8},
\end{equation}
where $\tau\equiv u_{\rm c}-u$ is the time to coalescence, and~$u=0$ corresponds again to the start of the observation. From Eq.~\eqref{eq:eta}, using $\chi=\Upsilon(r_{\rm e}) \cos{(\omega_\delta u+\varphi_\delta)}$, we find 
\begin{widetext}
\begin{equation}\label{eq:etaeq}
\eta(\tau)=\Upsilon(r_{\rm e})  \left[ {}_1F_2\left(\tfrac{5}{16}; \tfrac{1}{2}, \tfrac{21}{16}; -\tfrac{1}{4} \tau^2 \omega_\delta^2\right)\cos\Theta \, + \frac{5}{13} \tau \omega_\delta \, {}_1F_2\left(\tfrac{13}{16}; \tfrac{3}{2}, \tfrac{29}{16}; -\tfrac{1}{4} \tau^2 \omega_\delta^2\right) \sin\Theta \right],
\end{equation}
\end{widetext}
where~$\Theta\equiv \varphi_{\rm \delta}+ \omega_\delta u_{\rm c}$, and the total phasing becomes
\begin{equation}
    \alpha=2[1- \eta(\tau)]\left(\frac{\tau}{5 \mathcal{M}} \right)^{5/8}+\alpha_{\rm c}.
\end{equation}
Our result is consistent with Eq.~(18) of Ref.~\cite{Stegmann:2023wzy}. 

\begin{figure}
    \centering
    \includegraphics[scale=0.4]{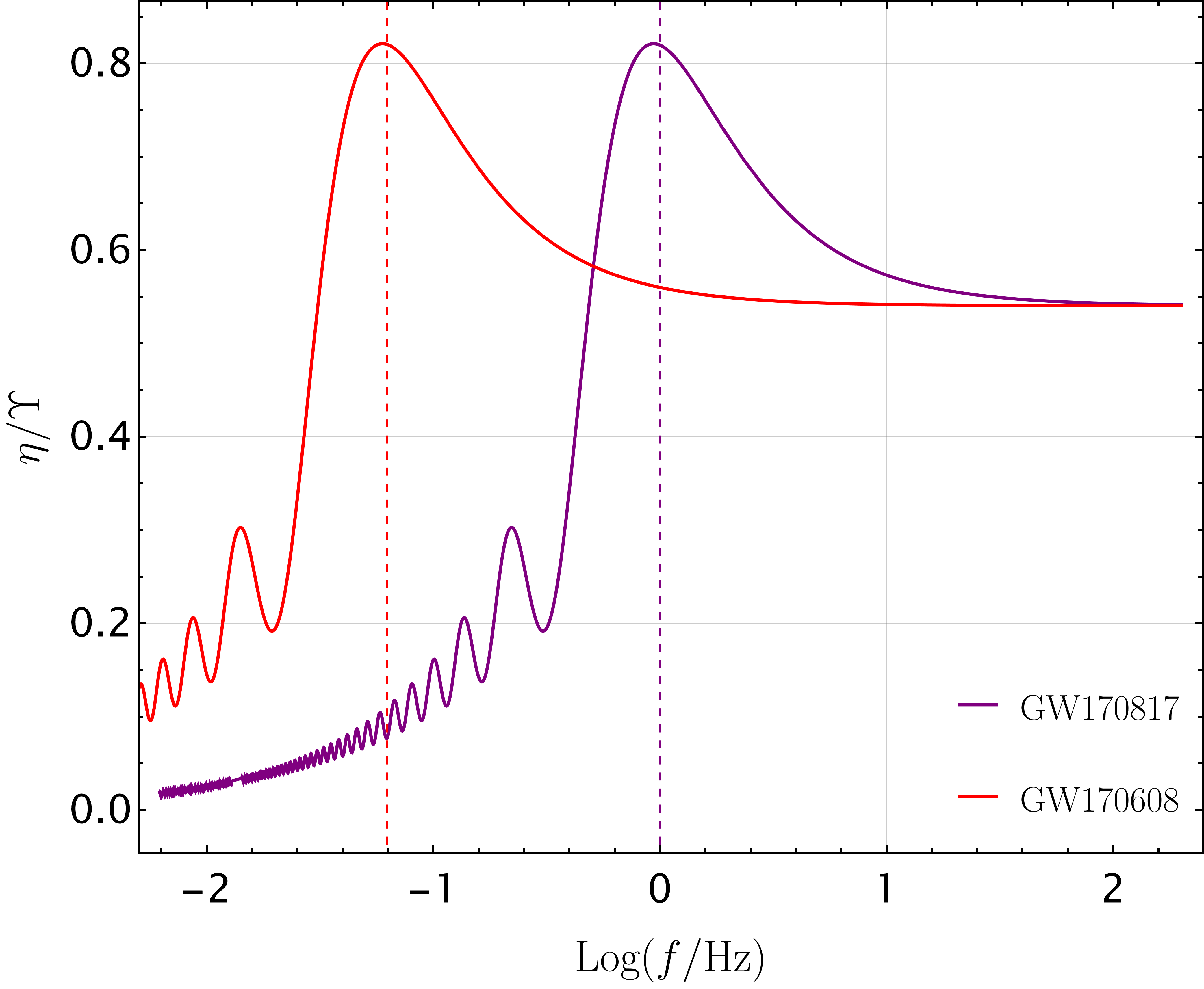}
    \caption{Factor~$\eta/\Upsilon$ in Eq.~\eqref{eq:etaeq} describing the correction to the phase of a chirping GW signal, from an oscillating gravitational field, as a function of~$f_{\rm e}$ (cf. Eq.~\eqref{eq:Newfreqs}). The results are for \texttt{GW170817}~\cite{LIGOScientific:2017vwq} (purple) and \texttt{GW170608}~\cite{LIGOScientific:2017vox} (red) binaries, for a ULDM particle mass~$m=10^{-21}\,{\rm eV}$, and an example phase of~$\Theta=1$. Vertical lines show the frequencies~$f_{\rm e}(\tau=1/\omega_\delta)$ for each of the two binaries. }
    \label{fig:eta}
\end{figure}

\begin{figure}
    \centering
    \includegraphics[scale=0.4]{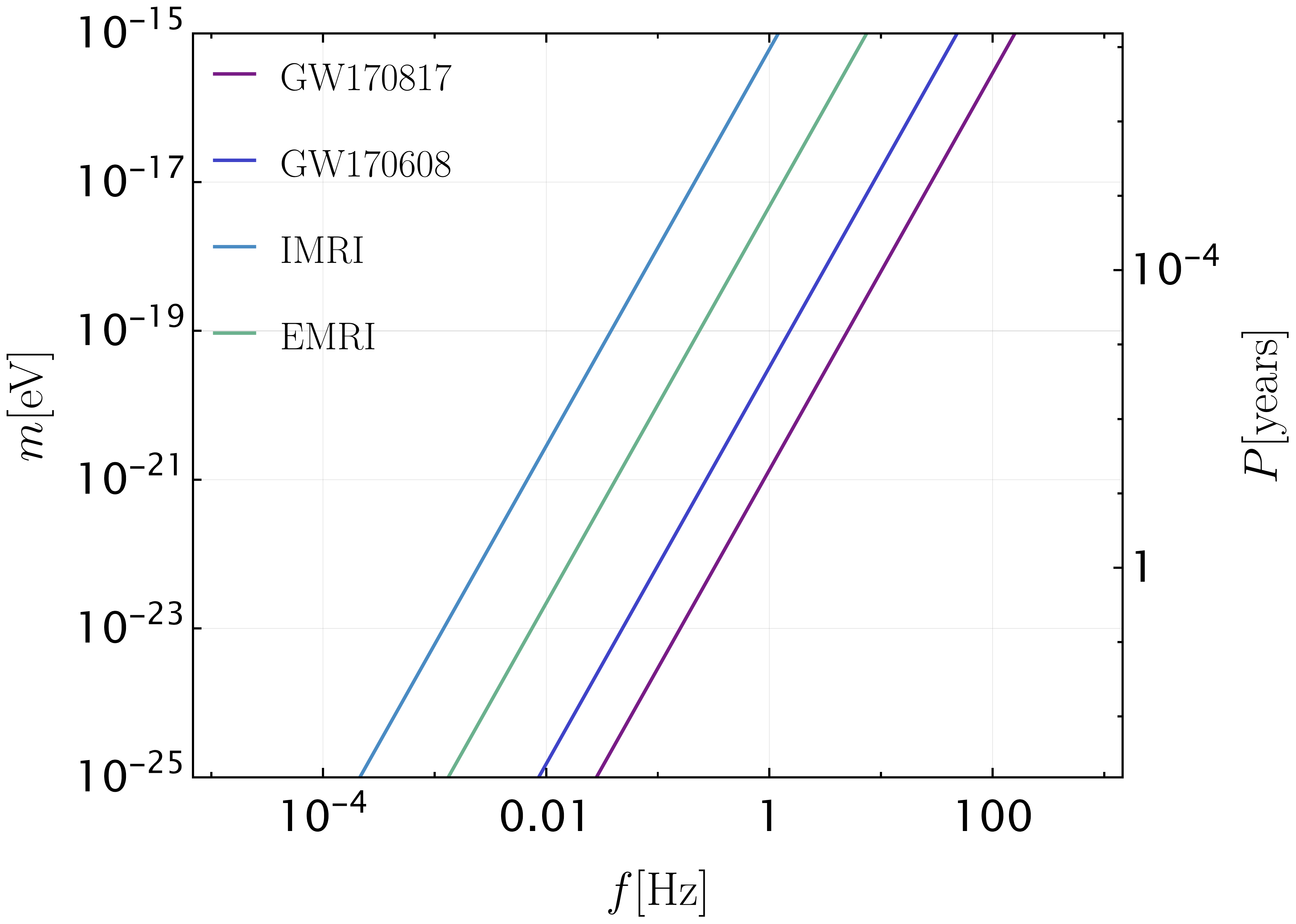}
    \caption{The ULDM particle mass corresponding to the condition~$\omega_\delta \tau(f)= 1$, for different binary systems. For concreteness, we take the minimal coupling case~$\omega_\delta= 2 m$. These illustrate the order of magnitude of the ULDM masses probed by the different chirping sources at given frequencies~$f_{\rm e}$.}
    \label{fig:cond}
\end{figure}

Figure~\ref{fig:eta} shows~$\eta/\Upsilon(r_{\rm e})$ as function of~$f\equiv \omega_{\rm e}(\tau)/(2 \pi)$, for two different binaries and fixed ULDM particle mass. Unlike the monochromatic case, the modulation of chirping GW signals has a time-dependent amplitude that reaches a maximum absolute value at~$\omega_\delta\tau\sim 1$, and whose value depends on the phase~$\Theta$. Thus, we expect the detectability of the ULDM effect to be mainly determined by the sensitivity at~$\omega_\delta\tau(f)\sim 1$, which corresponds to different modulation frequencies (ULDM particle masses) depending on the source/detector, as shown in Fig.~\ref{fig:cond} for the different binaries considered in Sec.~\ref{sec:othergal}. 

The (quasi)monochromatic result is recovered in the limit~$\tau \omega_\delta\gg1$. At leading order in $1/(\tau \omega_\delta)\ll1$, Eq.~\eqref{eq:etaeq} is 
\begin{equation}
    \eta \approx \frac{5 \Upsilon(r_{\rm e})}{8 \tau\omega_\delta}\sin(\omega_\delta u+\varphi_\delta),
\end{equation}
resulting in (for~$u\ll u_{\rm c}$)
\begin{align}
    &\alpha(u_{\rm c}-u)-\alpha(u_{\rm c})\approx\omega_{\rm e}u \nn\\
    &\qquad+ \frac{\omega_{\rm e}}{\omega_{ \delta}} \Upsilon(r_{\rm e}) \left[\sin(\omega_\delta u+\varphi_\delta)-\sin\varphi_\delta\right].
\end{align}
This matches precisely the argument in Eq.~\eqref{eq:modwaveform}.

The GW waveform in the Fourier domain can be found using the stationary phase approximation~\cite{Cutler:1994ys},
\begin{equation}\label{eq:STAwave}
    \Tilde{h}(f)\approx  \sqrt{\frac{5}{24}} \frac{1}{\pi^{2/3}} \frac{1}{d_{\rm L}} \mathcal{M}^{5/6} f^{-7/6} e^{i\bar{\alpha}(f)},
\end{equation}
with
\begin{equation}\label{eq:GWphase}
    \bar{\alpha}=2\pi f u_{\rm c}-\alpha_{\rm c}-\frac{\pi}{4}-\alpha[\tau(f)],
\end{equation}
where~$\tau(f)$ is found by inverting~$f_{\rm e}(\tau)$ in Eq.~\eqref{eq:Newfreqs}, with~$f_{\rm e}\equiv \omega_{\rm e}/(2 \pi)$. In the stationary phase approximation, we have neglected corrections from the ULDM effect to the amplitude in~$|\bar{h}|\propto1/\sqrt{\alpha''[\tau(f)]}$ that are subleading, and the sidebands arising from the non-injectivity of~$f_{\rm e}(\tau)$ due to the ULDM frequency modulation (see, e.g.,~\cite{Strokov:2023ypy}).
The latter, which is the all-important effect for (quasi)monochromatic sources, can be neglected for chirping sources when~$|\omega'_{\rm e}|/\omega_\delta^2\gg 1$ or equivalently $\omega_\delta \tau \ll (\mathcal{M} \omega_\delta)^{-5/11}$, corresponding to the regime where the~$\omega_{\rm e}$ chirps by more than~$\omega_\delta$ during one modulation period.


%%%%%%%%%%%%%%%%%%%%%%%%%%%%%%%%%%%%%%%%%%%%
%%%%%%%%%%%%%%%%%%%%%%%%%%%%%%%%%%%%%%%%%%%%
\section{Matched Filtering and Cross-correlation}
\label{app:snr}

Match filtering is the standard technique used to maximize the signal from a GW~$h$, given the knowledge of the detector noise~$n$. We consider the noise approximately Gaussian and characterized by the two-point function $\langle n^*(f)n(f')\rangle=\delta(f-f')\frac{1}{2}S_n(f)$, where~$S_n(f)$ is the noise spectral density. 
%It's customary to define the following noise-weighted inner product between two signals in the frequency domain $\Tilde{a}, \Tilde{b}$
The SNR of a signal $h$ is defined in terms of the inner product~\cite{Maggiore:2007ulw} 
\begin{equation}\label{eq:Innersp}
    {\rm SNR}=\langle h|h\rangle=4 \text{Re}\int_{0}^{\infty}\frac{\Tilde{h}^*(f)\Tilde{h}(f)}{S_n(f)}{\rm d}f ,
\end{equation}
where $\Tilde{h}$ is the Fourier transform of $h$, defined as
\begin{equation}
    \Tilde{h}\equiv\int_{-\infty}^{\infty} h(t)e^{-2\pi i f t} {\rm d}t,
\end{equation}
and $h$ is the strain at the detector defined as in Eq.~\eqref{eq:modwaveform}.

%%%%%%%%%%%%%%%%%%%%%%%%%%%%%%%%%%%%%%%
\subsection{Monochromatic signal}
%%%%%%%%%%%%%%%%%%%%%%%%%%%%%%%%%%%%%%%

We focus now on the case of a monochromatic signal and derive analytical results for the SNR we use in the analysis of Sec.~\ref{sec:MWC}. 
Since the signal is observed for a finite time~$T$, we introduce the regularized Dirac delta function 
\begin{equation}
    \delta_T(\omega)=\int_{-T/2}^{T/2}e^{i\omega t}{\rm d}t=T{\rm sinc}(\omega T/2),
\end{equation}
with~$\delta_T(0)=T$. For a monochromatic signal 
\begin{equation}
\Tilde{h}=\frac{\mathcal{A}}{2}e^{-i\alpha_0}\delta_T(\omega-\omega_e),    
\end{equation}
we have that
\begin{align}
   &{\rm SNR}^2=\frac{\mathcal{A}^2T^2}{S_n(f_e)}\int_0^\infty{\rm sinc}^2[\pi(f-f_e)T]{\rm d}f \\
   &=\frac{\mathcal{A}^2T}{\pi S_n(f_e)}\int_{-\pi f_e T}^\infty{\rm sinc}^2(x){\rm d}x\approx\frac{\mathcal{A}^2T}{S_n(f_e)},
\end{align}
where we take $\pi f_e T\gg1$.

Similarly, we can compute the SNR of the modulation effect. If there are~$N$ loud signals in band, after extracting the carrier components (found through match-filtering) from the data, we are left with the residuals
\begin{align}
    &\bar{\delta h} =\sum_{i=1}^N \bar{\delta h}_i=\frac{1}{4\omega_\delta}\sum_{i=1}^N \mathcal{A}_i \omega_{i} \Upsilon_i e^{-i\alpha'_{0,i}} \nn\\
    &\times\left[e^{-i\varphi_{\delta,i}}\delta_T(\omega-\omega_i-\omega_\delta)-e^{i\varphi_\delta}\delta_T(\omega-\omega_i+\omega_{\delta,i}) \right].
\end{align}

By applying a narrow-filter~$\mathcal{F}_i[\,\cdot\,]$ centered at~$\omega_i$ with width~$\Gamma_i$ sufficiently small to not pick other carrier components, but large enough that~$\Gamma_i>\omega_\delta$, one could (in principle) isolate each individual~$\bar{\delta h_i}$.\footnote{Note that for ULDM masses~$m\sim 10^{-22}\,{\rm eV}$, the side-band separation is~$\omega_\delta \sim 1\,{\rm nHz} \ll \omega_{i}$.} 
Now, notice that if we whiten each individual residual like~$\bar{\delta h}_i/\sqrt{S_n(\omega_i)}$, shift their frequency by the corresponding~$\omega_i$, and sum them all, we are able to accumulate power at~$\pm\omega_\delta$, i.e.,
\begin{align}
    &\sum_{i=1}^N \frac{\mathcal{F}_i[\bar{\delta h}](\omega'+\omega_i)}{\sqrt{S_n(\omega_i)}}=\nn \\
    &\qquad\pm\frac{\delta_T(\omega'\mp\omega_\delta)}{4\omega_\delta}\sum_{i=1}^N  \frac{\omega_{i} \mathcal{A}_i \Upsilon_i}{\sqrt{S_n(\omega_i)}} e^{-i(\alpha'_{0,i}\pm\varphi_{\delta,i})}.
\end{align}

The SNR of the modulation effect (after applying such procedure) is that
\begin{align}
    &{\rm SNR}_\delta^2\approx \sum_{i=1}^N \left(\frac{\omega_i \Upsilon_i}{\sqrt{2} \omega_\delta}\right)^2 {\rm SNR}_i^2+\sum_{i<j} \frac{\omega_i \omega_j \Upsilon_i \Upsilon_j}{\omega_\delta^2} \nn\\
    &\qquad\times {\rm SNR}_i {\rm SNR}_j\cos\left(\alpha'_{0,i}-\alpha'_{0,j}+\varphi_{\delta,i}-\varphi_{\delta,j}\right),
\end{align}
where we used~$\pi f_\delta T\gg1$ (for~$f_\delta T=1$ we get a 5\% error).

Most certainly, one will not be able to measure the luminosity distances well enough to estimate each~$\varphi_{\delta,i}-\varphi_{\delta,j}$ with uncertainty much smaller than~$\pi$\footnote{If two sources are close to or within the same coherent structure, the difference~$\varphi_{\delta,i}-\varphi_{\delta,j}$ can be estimated from the match-filtering with the carrier waveforms, as it is equal to~$\omega_\delta (d_{{\rm L},j}-d_{{\rm L},i})$}. Indeed, as we show below, this would require estimating each luminosity distance with an accuracy~$\sigma_{d_{{\rm L},i}}\lesssim 0.1/(2 m_{22})\,{\rm pc}$, which for~${\rm mHz}$ sources at the GC would be possible only if~${\rm SNR}_i\gtrsim 10^4 m_{22}$. 
This implies that the second sum will average out to zero for~$N\gg1$ (the same applies to sources close to/ within different ULDM coherent structures), unless we treat the~$\varphi_{\delta,i}-\varphi_{\delta,j}$ as free parameters in the match-filtering of the residuals, which would most probably result in a very expensive analysis. So, if we do not treat~$\varphi_{\delta,i}-\varphi_{\delta,j}$ as free parameters in the match-filtering, the SNR of the modulation effect becomes simply
\begin{align}
    &{\rm SNR}_{\delta}=\sqrt{\sum_{i=1}^{N}{\rm SNR}_{i}^2\left(\frac{  \Upsilon_i\omega_{i} }{\sqrt{2}\omega_\delta}\right)^2}.
\end{align}
If~$\Upsilon_i$ is approximately the same for all the sources, the SNR of the effect scales with~$\sqrt{N}$.

On the other hand, if one is able to estimate each~$\varphi_{\delta,i}-\varphi_{\delta,j}$ with uncertainty much smaller than~$\pi$, or if one takes them as free parameters, the ${\rm SNR}_\delta$ will scale instead with~$N$. However, in the latter case, the false alarm rate (FAR) is also increased, which means that the threshold for detectability at a given FAR should be increased. Noting that~\cite{Takahashi:2002ky}
\begin{equation}
    d=\frac{5}{96\pi^2}\frac{\dot{f}}{f^3\mathcal{A}}.
\end{equation}
and using a Fisher matrix analysis (see Sec.~\ref{app:fisher}) one can show that for LISA
\begin{equation}\label{eq:distsigma}
    \frac{\sigma_{d_{\rm L}}}{d_{\rm L}}\simeq\frac{\sigma_\mathcal{A}}{\mathcal{A}}+\frac{\sigma_{\dot{f}_e}}{\dot{f}_e}\sim 0.0015\left(\frac{900}{\rm SNR}\right).
\end{equation}

%%%%%%%%%%%%%%%%%%%%%%%%%%%%%
\section{Fisher Information Matrix}\label{app:fisher}
%%%%%%%%%%%%%%%%%%%%%%%%%%%%%

The Fisher matrix formalism is an often used simple procedure to estimate the model parameter uncertainties given the sensitivity of a detector (e.g.,~\cite{Vallisneri:2007ev}). For GWs, the Fisher matrix is defined in terms of the derivatives of the waveform with respect the different model parameters $\Lambda_i$, and the inner product defined in Eq.~\eqref{eq:Innersp} 
\begin{equation}
    \Gamma_{ij}\equiv\Big\langle\frac{\partial \Tilde{h}}{\partial \Lambda_i}\Big|\frac{\partial  \Tilde{h}}{\partial \Lambda_j}\Big\rangle\Big |_{\Lambda=\Bar{\Lambda} },\label{eq:FisherFormula}
\end{equation}
with the whole matrix evaluated at the true value of the parameters~$\Lambda=\Bar{\Lambda}$. In the limit of high SNR, the covariance matrix is well approximated by the inverse of the Fisher matrix, thus the correlation coefficient for each pair of parameters, and the uncertainty on the parameters are computed as
\begin{equation}
   c_{ij}=\frac{(\Gamma^{-1})_{ij}}{\sigma_i\sigma_j}, \qquad \text{and}\qquad \sigma^2_i=(\Gamma^{-1})_{ii}.
\end{equation}
%
%%%%%%%%%%%%%%%%%%%%%%%%%%%%%%%%%%%%%%%%%%%%%%%%%%%%%%%%%%
\subsection{(Quasi)Monochromatic Waveform}
%%%%%%%%%%%%%%%%%%%%%%%%%%%%%%%%%%%%%%%%%%%%%%%%%%%%%%%%%%

A monochromatic GW signal of frequency~$f_{\rm e}$, averaged over sky direction and polarization, has the waveform~\cite{Maggiore:2007ulw}
\begin{equation}
    h(t;\Lambda)=\sqrt{\frac{(1+\cos^2\iota)^2}{4}+\cos^2{\iota} }\, \mathcal{A}(f_{\rm e})e^{i(2\pi f_{\rm e} t+\varphi)}, 
\end{equation}
where~$\iota$ is the inclination angle. For a binary source, the amplitude is given in terms of the orbital frequency~$f_e/2$, (red-shifted) chirp mass~$\mathcal{M}$, and luminosity distance $d_{\rm L}$ by~\cite{Maggiore:2007ulw}
\begin{equation}
    \mathcal{A}=\frac{4\mathcal{M}^{5/3}(\pi f_e)^{2/3}}{d_{\rm L}}.
\end{equation}
For LISA we have used the sensitivity curve provided in Ref.~\cite{Robson:2018ifk}, averaged over sky location and polarization, together with the estimated noise contribution from unresolved galactic binaries for a four years observation. For $\mu$Ares we considered the values reported in Ref.~\cite{Sesana:2019vho}. For TianQin we used the sensitivity curve given in Ref.~\cite{Torres-Orjuela:2023hfd}, whereas for Taiji that in Ref.~\cite{Liu:2023qap}. 

To estimate the uncertainty on the luminosity distance of the binary, we need to consider the time derivative of the GW frequency, so that the waveform gets extended to~$h\propto e^{2\pi i(f_e t +\frac{1}{2}\dot{f}_e t^2)+i\varphi}$. Computing the Fisher matrix for the intrinsic parameters $\Lambda=\{\mathcal{A},\ln{f_e},\dot{f}_e\}$, we find the following uncertainties  
\begin{equation*}
    \frac{\sigma_\mathcal{A}}{\mathcal{A}}=\frac{1}{\rm SNR},\quad
    \frac{\sigma_{f_{\rm e}}}{{f_e}}=\frac{2}{\rm SNR}\frac{1}{T_{\rm obs}},\quad
    {\sigma_{\dot{f}_e}}=\frac{4}{\rm SNR}\frac{1}{T^2_{\rm obs}}, 
\end{equation*}
where we averaged over sky location, polarization and inclination angles.

\subsection{Chirping Waveform}

For the Fisher Matrix analysis of Sec.~\ref{sec:othergal}, we consider the waveform of Eq.~\eqref{eq:STAwave}. We computed the Fisher matrix for LISA, B-DECIGO and DECIGO. For the last two, we use the analytical sensitivity curves of Ref.~\cite{Piorkowska-Kurpas:2020rfy}. We consider the derivatives of the waveform $\Tilde{h}(f)$ w.r.t. the parameters $\Lambda=\{\ln{\mathcal{M}},u_c,\alpha_{\rm c},\Upsilon, \Theta\}$, so that the entire Fisher matrix is 5-dimensional. These parameters are~$u_c$ the (initial) time to coalescence,~$\alpha_{\rm c}$ the phase at coalescence, $\Upsilon$ the amplitude of the oscillating gravitational background, and $\Theta$ the phase appearing in Eq.~\eqref{eq:etaeq}. The last two parameters concern only the background modulation, whereas the first three are parameters of the usual (isolated vacuum) waveform. We imposed uniform priors for the parameters~$\{\alpha_{\rm c}, \Theta\}$ in the interval~$[0,2\pi]$.

We find that, depending on the mass of the ULDM particle, there can be a strong correlation between~$\Upsilon$ and~$\Theta$. 
This is to be expected: note that for~$\tau \omega_\delta \ll 1$, the only dependence of the waveform on~$\Upsilon$ and~$\Theta$ is through~$\eta\approx \Upsilon \cos \Theta$.
We remark that for heavier ULDM particle masses the Fisher matrix components involve highly oscillatory integrals, which demand high numerical precision.

\bibliography{References}

\begin{thebibliography}{151}%
\makeatletter
\providecommand \@ifxundefined [1]{%
 \@ifx{#1\undefined}
}%
\providecommand \@ifnum [1]{%
 \ifnum #1\expandafter \@firstoftwo
 \else \expandafter \@secondoftwo
 \fi
}%
\providecommand \@ifx [1]{%
 \ifx #1\expandafter \@firstoftwo
 \else \expandafter \@secondoftwo
 \fi
}%
\providecommand \natexlab [1]{#1}%
\providecommand \enquote  [1]{``#1''}%
\providecommand \bibnamefont  [1]{#1}%
\providecommand \bibfnamefont [1]{#1}%
\providecommand \citenamefont [1]{#1}%
\providecommand \href@noop [0]{\@secondoftwo}%
\providecommand \href [0]{\begingroup \@sanitize@url \@href}%
\providecommand \@href[1]{\@@startlink{#1}\@@href}%
\providecommand \@@href[1]{\endgroup#1\@@endlink}%
\providecommand \@sanitize@url [0]{\catcode `\\12\catcode `\$12\catcode
  `\&12\catcode `\#12\catcode `\^12\catcode `\_12\catcode `\%12\relax}%
\providecommand \@@startlink[1]{}%
\providecommand \@@endlink[0]{}%
\providecommand \url  [0]{\begingroup\@sanitize@url \@url }%
\providecommand \@url [1]{\endgroup\@href {#1}{\urlprefix }}%
\providecommand \urlprefix  [0]{URL }%
\providecommand \Eprint [0]{\href }%
\providecommand \doibase [0]{https://doi.org/}%
\providecommand \selectlanguage [0]{\@gobble}%
\providecommand \bibinfo  [0]{\@secondoftwo}%
\providecommand \bibfield  [0]{\@secondoftwo}%
\providecommand \translation [1]{[#1]}%
\providecommand \BibitemOpen [0]{}%
\providecommand \bibitemStop [0]{}%
\providecommand \bibitemNoStop [0]{.\EOS\space}%
\providecommand \EOS [0]{\spacefactor3000\relax}%
\providecommand \BibitemShut  [1]{\csname bibitem#1\endcsname}%
\let\auto@bib@innerbib\@empty
%</preamble>
\bibitem [{\citenamefont {Bertone}\ \emph {et~al.}(2005)\citenamefont
  {Bertone}, \citenamefont {Hooper},\ and\ \citenamefont
  {Silk}}]{Bertone:2004pz}%
  \BibitemOpen
  \bibfield  {author} {\bibinfo {author} {\bibfnamefont {G.}~\bibnamefont
  {Bertone}}, \bibinfo {author} {\bibfnamefont {D.}~\bibnamefont {Hooper}},\
  and\ \bibinfo {author} {\bibfnamefont {J.}~\bibnamefont {Silk}},\ }\bibfield
  {title} {\bibinfo {title} {{Particle dark matter: Evidence, candidates and
  constraints}},\ }\href {https://doi.org/10.1016/j.physrep.2004.08.031}
  {\bibfield  {journal} {\bibinfo  {journal} {Phys. Rept.}\ }\textbf {\bibinfo
  {volume} {405}},\ \bibinfo {pages} {279} (\bibinfo {year} {2005})},\ \Eprint
  {https://arxiv.org/abs/hep-ph/0404175} {arXiv:hep-ph/0404175} \BibitemShut
  {NoStop}%
\bibitem [{\citenamefont {Hui}(2021)}]{Hui:2021tkt}%
  \BibitemOpen
  \bibfield  {author} {\bibinfo {author} {\bibfnamefont {L.}~\bibnamefont
  {Hui}},\ }\bibfield  {title} {\bibinfo {title} {{Wave Dark Matter}},\ }\href
  {https://doi.org/10.1146/annurev-astro-120920-010024} {\bibfield  {journal}
  {\bibinfo  {journal} {Ann. Rev. Astron. Astrophys.}\ }\textbf {\bibinfo
  {volume} {59}},\ \bibinfo {pages} {247} (\bibinfo {year} {2021})},\ \Eprint
  {https://arxiv.org/abs/2101.11735} {arXiv:2101.11735 [astro-ph.CO]}
  \BibitemShut {NoStop}%
\bibitem [{\citenamefont {Hui}\ \emph {et~al.}(2017)\citenamefont {Hui},
  \citenamefont {Ostriker}, \citenamefont {Tremaine},\ and\ \citenamefont
  {Witten}}]{Hui:2016ltb}%
  \BibitemOpen
  \bibfield  {author} {\bibinfo {author} {\bibfnamefont {L.}~\bibnamefont
  {Hui}}, \bibinfo {author} {\bibfnamefont {J.~P.}\ \bibnamefont {Ostriker}},
  \bibinfo {author} {\bibfnamefont {S.}~\bibnamefont {Tremaine}},\ and\
  \bibinfo {author} {\bibfnamefont {E.}~\bibnamefont {Witten}},\ }\bibfield
  {title} {\bibinfo {title} {{Ultralight scalars as cosmological dark
  matter}},\ }\href {https://doi.org/10.1103/PhysRevD.95.043541} {\bibfield
  {journal} {\bibinfo  {journal} {Phys. Rev. D}\ }\textbf {\bibinfo {volume}
  {95}},\ \bibinfo {pages} {043541} (\bibinfo {year} {2017})},\ \Eprint
  {https://arxiv.org/abs/1610.08297} {arXiv:1610.08297 [astro-ph.CO]}
  \BibitemShut {NoStop}%
\bibitem [{\citenamefont {Schive}\ \emph
  {et~al.}(2014{\natexlab{a}})\citenamefont {Schive}, \citenamefont {Chiueh},\
  and\ \citenamefont {Broadhurst}}]{Schive:2014dra}%
  \BibitemOpen
  \bibfield  {author} {\bibinfo {author} {\bibfnamefont {H.-Y.}\ \bibnamefont
  {Schive}}, \bibinfo {author} {\bibfnamefont {T.}~\bibnamefont {Chiueh}},\
  and\ \bibinfo {author} {\bibfnamefont {T.}~\bibnamefont {Broadhurst}},\
  }\bibfield  {title} {\bibinfo {title} {{Cosmic Structure as the Quantum
  Interference of a Coherent Dark Wave}},\ }\href
  {https://doi.org/10.1038/nphys2996} {\bibfield  {journal} {\bibinfo
  {journal} {Nature Phys.}\ }\textbf {\bibinfo {volume} {10}},\ \bibinfo
  {pages} {496} (\bibinfo {year} {2014}{\natexlab{a}})},\ \Eprint
  {https://arxiv.org/abs/1406.6586} {arXiv:1406.6586 [astro-ph.GA]}
  \BibitemShut {NoStop}%
\bibitem [{\citenamefont {Arvanitaki}\ \emph {et~al.}(2020)\citenamefont
  {Arvanitaki}, \citenamefont {Dimopoulos}, \citenamefont {Galanis},
  \citenamefont {Lehner}, \citenamefont {Thompson},\ and\ \citenamefont
  {Van~Tilburg}}]{Arvanitaki:2019rax}%
  \BibitemOpen
  \bibfield  {author} {\bibinfo {author} {\bibfnamefont {A.}~\bibnamefont
  {Arvanitaki}}, \bibinfo {author} {\bibfnamefont {S.}~\bibnamefont
  {Dimopoulos}}, \bibinfo {author} {\bibfnamefont {M.}~\bibnamefont {Galanis}},
  \bibinfo {author} {\bibfnamefont {L.}~\bibnamefont {Lehner}}, \bibinfo
  {author} {\bibfnamefont {J.~O.}\ \bibnamefont {Thompson}},\ and\ \bibinfo
  {author} {\bibfnamefont {K.}~\bibnamefont {Van~Tilburg}},\ }\bibfield
  {title} {\bibinfo {title} {{Large-misalignment mechanism for the formation of
  compact axion structures: Signatures from the QCD axion to fuzzy dark
  matter}},\ }\href {https://doi.org/10.1103/PhysRevD.101.083014} {\bibfield
  {journal} {\bibinfo  {journal} {Phys. Rev. D}\ }\textbf {\bibinfo {volume}
  {101}},\ \bibinfo {pages} {083014} (\bibinfo {year} {2020})},\ \Eprint
  {https://arxiv.org/abs/1909.11665} {arXiv:1909.11665 [astro-ph.CO]}
  \BibitemShut {NoStop}%
\bibitem [{\citenamefont {Thompson}(2023)}]{Thompson:2023mfx}%
  \BibitemOpen
  \bibfield  {author} {\bibinfo {author} {\bibfnamefont {J.~O.}\ \bibnamefont
  {Thompson}},\ }\emph {\bibinfo {title} {{The large-misalignment mechanism for
  the formation of compact axion structures}}},\ \href@noop {} {Ph.D. thesis},\
  \bibinfo  {school} {Stanford U.} (\bibinfo {year} {2023})\BibitemShut
  {NoStop}%
\bibitem [{\citenamefont {Khmelnitsky}\ and\ \citenamefont
  {Rubakov}(2014)}]{Khmelnitsky:2013lxt}%
  \BibitemOpen
  \bibfield  {author} {\bibinfo {author} {\bibfnamefont {A.}~\bibnamefont
  {Khmelnitsky}}\ and\ \bibinfo {author} {\bibfnamefont {V.}~\bibnamefont
  {Rubakov}},\ }\bibfield  {title} {\bibinfo {title} {{Pulsar timing signal
  from ultralight scalar dark matter}},\ }\href
  {https://doi.org/10.1088/1475-7516/2014/02/019} {\bibfield  {journal}
  {\bibinfo  {journal} {JCAP}\ }\textbf {\bibinfo {volume} {02}},\ \bibinfo
  {pages} {019}},\ \Eprint {https://arxiv.org/abs/1309.5888} {arXiv:1309.5888
  [astro-ph.CO]} \BibitemShut {NoStop}%
\bibitem [{\citenamefont {De~Martino}\ \emph {et~al.}(2017)\citenamefont
  {De~Martino}, \citenamefont {Broadhurst}, \citenamefont {Henry~Tye},
  \citenamefont {Chiueh}, \citenamefont {Schive},\ and\ \citenamefont
  {Lazkoz}}]{DeMartino:2017qsa}%
  \BibitemOpen
  \bibfield  {author} {\bibinfo {author} {\bibfnamefont {I.}~\bibnamefont
  {De~Martino}}, \bibinfo {author} {\bibfnamefont {T.}~\bibnamefont
  {Broadhurst}}, \bibinfo {author} {\bibfnamefont {S.~H.}\ \bibnamefont
  {Henry~Tye}}, \bibinfo {author} {\bibfnamefont {T.}~\bibnamefont {Chiueh}},
  \bibinfo {author} {\bibfnamefont {H.-Y.}\ \bibnamefont {Schive}},\ and\
  \bibinfo {author} {\bibfnamefont {R.}~\bibnamefont {Lazkoz}},\ }\bibfield
  {title} {\bibinfo {title} {{Recognizing Axionic Dark Matter by Compton and de
  Broglie Scale Modulation of Pulsar Timing}},\ }\href
  {https://doi.org/10.1103/PhysRevLett.119.221103} {\bibfield  {journal}
  {\bibinfo  {journal} {Phys. Rev. Lett.}\ }\textbf {\bibinfo {volume} {119}},\
  \bibinfo {pages} {221103} (\bibinfo {year} {2017})},\ \Eprint
  {https://arxiv.org/abs/1705.04367} {arXiv:1705.04367 [astro-ph.CO]}
  \BibitemShut {NoStop}%
\bibitem [{\citenamefont {Stegmann}\ and\ \citenamefont
  {Vermeulen}(2024)}]{Stegmann:2023glt}%
  \BibitemOpen
  \bibfield  {author} {\bibinfo {author} {\bibfnamefont {J.}~\bibnamefont
  {Stegmann}}\ and\ \bibinfo {author} {\bibfnamefont {S.~M.}\ \bibnamefont
  {Vermeulen}},\ }\bibfield  {title} {\bibinfo {title} {{Detecting the
  heterodyning of gravitational waves}},\ }\href
  {https://doi.org/10.1088/1361-6382/ad682c} {\bibfield  {journal} {\bibinfo
  {journal} {Class. Quant. Grav.}\ }\textbf {\bibinfo {volume} {41}},\ \bibinfo
  {pages} {175012} (\bibinfo {year} {2024})},\ \Eprint
  {https://arxiv.org/abs/2301.02672} {arXiv:2301.02672 [gr-qc]} \BibitemShut
  {NoStop}%
\bibitem [{\citenamefont {Stegmann}\ \emph {et~al.}(2024)\citenamefont
  {Stegmann}, \citenamefont {Zwick}, \citenamefont {Vermeulen}, \citenamefont
  {Antonini},\ and\ \citenamefont {Mayer}}]{Stegmann:2023wzy}%
  \BibitemOpen
  \bibfield  {author} {\bibinfo {author} {\bibfnamefont {J.}~\bibnamefont
  {Stegmann}}, \bibinfo {author} {\bibfnamefont {L.}~\bibnamefont {Zwick}},
  \bibinfo {author} {\bibfnamefont {S.~M.}\ \bibnamefont {Vermeulen}}, \bibinfo
  {author} {\bibfnamefont {F.}~\bibnamefont {Antonini}},\ and\ \bibinfo
  {author} {\bibfnamefont {L.}~\bibnamefont {Mayer}},\ }\bibfield  {title}
  {\bibinfo {title} {{Imprints of massive black-hole binaries on neighbouring
  decihertz gravitational-wave sources}},\ }\href
  {https://doi.org/10.1038/s41550-024-02338-0} {\bibfield  {journal} {\bibinfo
  {journal} {Nature Astron.}\ }\textbf {\bibinfo {volume} {8}},\ \bibinfo
  {pages} {1321} (\bibinfo {year} {2024})},\ \Eprint
  {https://arxiv.org/abs/2311.06335} {arXiv:2311.06335 [astro-ph.HE]}
  \BibitemShut {NoStop}%
\bibitem [{\citenamefont {Bustamante-Rosell}\ \emph {et~al.}(2022)\citenamefont
  {Bustamante-Rosell}, \citenamefont {Meyers}, \citenamefont {Pearson},
  \citenamefont {Trendafilova},\ and\ \citenamefont
  {Zimmerman}}]{Bustamante-Rosell:2021daj}%
  \BibitemOpen
  \bibfield  {author} {\bibinfo {author} {\bibfnamefont {M.~J.}\ \bibnamefont
  {Bustamante-Rosell}}, \bibinfo {author} {\bibfnamefont {J.}~\bibnamefont
  {Meyers}}, \bibinfo {author} {\bibfnamefont {N.}~\bibnamefont {Pearson}},
  \bibinfo {author} {\bibfnamefont {C.}~\bibnamefont {Trendafilova}},\ and\
  \bibinfo {author} {\bibfnamefont {A.}~\bibnamefont {Zimmerman}},\ }\bibfield
  {title} {\bibinfo {title} {{Gravitational wave timing array}},\ }\href
  {https://doi.org/10.1103/PhysRevD.105.044005} {\bibfield  {journal} {\bibinfo
   {journal} {Phys. Rev. D}\ }\textbf {\bibinfo {volume} {105}},\ \bibinfo
  {pages} {044005} (\bibinfo {year} {2022})},\ \Eprint
  {https://arxiv.org/abs/2107.02788} {arXiv:2107.02788 [gr-qc]} \BibitemShut
  {NoStop}%
\bibitem [{\citenamefont {Wang}\ and\ \citenamefont
  {Zhong}(2023)}]{Wang:2023phr}%
  \BibitemOpen
  \bibfield  {author} {\bibinfo {author} {\bibfnamefont {K.}~\bibnamefont
  {Wang}}\ and\ \bibinfo {author} {\bibfnamefont {Y.}~\bibnamefont {Zhong}},\
  }\bibfield  {title} {\bibinfo {title} {{Frequency modulation of gravitational
  waves by ultralight scalar dark matter}},\ }\href
  {https://doi.org/10.1103/PhysRevD.108.123531} {\bibfield  {journal} {\bibinfo
   {journal} {Phys. Rev. D}\ }\textbf {\bibinfo {volume} {108}},\ \bibinfo
  {pages} {123531} (\bibinfo {year} {2023})},\ \Eprint
  {https://arxiv.org/abs/2306.10732} {arXiv:2306.10732 [astro-ph.CO]}
  \BibitemShut {NoStop}%
\bibitem [{\citenamefont {Brax}\ \emph {et~al.}(2024)\citenamefont {Brax},
  \citenamefont {Valageas}, \citenamefont {Burrage},\ and\ \citenamefont
  {Cembranos}}]{Brax:2024yqh}%
  \BibitemOpen
  \bibfield  {author} {\bibinfo {author} {\bibfnamefont {P.}~\bibnamefont
  {Brax}}, \bibinfo {author} {\bibfnamefont {P.}~\bibnamefont {Valageas}},
  \bibinfo {author} {\bibfnamefont {C.}~\bibnamefont {Burrage}},\ and\ \bibinfo
  {author} {\bibfnamefont {J.~A.~R.}\ \bibnamefont {Cembranos}},\ }\bibfield
  {title} {\bibinfo {title} {{Detecting dark matter oscillations with
  gravitational waveforms}},\ }\href
  {https://doi.org/10.1103/PhysRevD.110.083515} {\bibfield  {journal} {\bibinfo
   {journal} {Phys. Rev. D}\ }\textbf {\bibinfo {volume} {110}},\ \bibinfo
  {pages} {083515} (\bibinfo {year} {2024})},\ \Eprint
  {https://arxiv.org/abs/2402.04819} {arXiv:2402.04819 [astro-ph.CO]}
  \BibitemShut {NoStop}%
\bibitem [{\citenamefont {Arvanitaki}\ \emph {et~al.}(2010)\citenamefont
  {Arvanitaki}, \citenamefont {Dimopoulos}, \citenamefont {Dubovsky},
  \citenamefont {Kaloper},\ and\ \citenamefont
  {March-Russell}}]{Arvanitaki:2009fg}%
  \BibitemOpen
  \bibfield  {author} {\bibinfo {author} {\bibfnamefont {A.}~\bibnamefont
  {Arvanitaki}}, \bibinfo {author} {\bibfnamefont {S.}~\bibnamefont
  {Dimopoulos}}, \bibinfo {author} {\bibfnamefont {S.}~\bibnamefont
  {Dubovsky}}, \bibinfo {author} {\bibfnamefont {N.}~\bibnamefont {Kaloper}},\
  and\ \bibinfo {author} {\bibfnamefont {J.}~\bibnamefont {March-Russell}},\
  }\bibfield  {title} {\bibinfo {title} {{String Axiverse}},\ }\href
  {https://doi.org/10.1103/PhysRevD.81.123530} {\bibfield  {journal} {\bibinfo
  {journal} {Phys. Rev. D}\ }\textbf {\bibinfo {volume} {81}},\ \bibinfo
  {pages} {123530} (\bibinfo {year} {2010})},\ \Eprint
  {https://arxiv.org/abs/0905.4720} {arXiv:0905.4720 [hep-th]} \BibitemShut
  {NoStop}%
\bibitem [{\citenamefont {Marsh}(2016)}]{Marsh:2015xka}%
  \BibitemOpen
  \bibfield  {author} {\bibinfo {author} {\bibfnamefont {D.~J.~E.}\
  \bibnamefont {Marsh}},\ }\bibfield  {title} {\bibinfo {title} {{Axion
  Cosmology}},\ }\href {https://doi.org/10.1016/j.physrep.2016.06.005}
  {\bibfield  {journal} {\bibinfo  {journal} {Phys. Rept.}\ }\textbf {\bibinfo
  {volume} {643}},\ \bibinfo {pages} {1} (\bibinfo {year} {2016})},\ \Eprint
  {https://arxiv.org/abs/1510.07633} {arXiv:1510.07633 [astro-ph.CO]}
  \BibitemShut {NoStop}%
\bibitem [{\citenamefont {Blas}\ \emph {et~al.}(2017)\citenamefont {Blas},
  \citenamefont {Nacir},\ and\ \citenamefont {Sibiryakov}}]{Blas:2016ddr}%
  \BibitemOpen
  \bibfield  {author} {\bibinfo {author} {\bibfnamefont {D.}~\bibnamefont
  {Blas}}, \bibinfo {author} {\bibfnamefont {D.~L.}\ \bibnamefont {Nacir}},\
  and\ \bibinfo {author} {\bibfnamefont {S.}~\bibnamefont {Sibiryakov}},\
  }\bibfield  {title} {\bibinfo {title} {{Ultralight Dark Matter Resonates with
  Binary Pulsars}},\ }\href {https://doi.org/10.1103/PhysRevLett.118.261102}
  {\bibfield  {journal} {\bibinfo  {journal} {Phys. Rev. Lett.}\ }\textbf
  {\bibinfo {volume} {118}},\ \bibinfo {pages} {261102} (\bibinfo {year}
  {2017})},\ \Eprint {https://arxiv.org/abs/1612.06789} {arXiv:1612.06789
  [hep-ph]} \BibitemShut {NoStop}%
\bibitem [{\citenamefont {Blas}\ \emph {et~al.}(2020)\citenamefont {Blas},
  \citenamefont {L\'opez~Nacir},\ and\ \citenamefont
  {Sibiryakov}}]{Blas:2019hxz}%
  \BibitemOpen
  \bibfield  {author} {\bibinfo {author} {\bibfnamefont {D.}~\bibnamefont
  {Blas}}, \bibinfo {author} {\bibfnamefont {D.}~\bibnamefont
  {L\'opez~Nacir}},\ and\ \bibinfo {author} {\bibfnamefont {S.}~\bibnamefont
  {Sibiryakov}},\ }\bibfield  {title} {\bibinfo {title} {{Secular effects of
  ultralight dark matter on binary pulsars}},\ }\href
  {https://doi.org/10.1103/PhysRevD.101.063016} {\bibfield  {journal} {\bibinfo
   {journal} {Phys. Rev. D}\ }\textbf {\bibinfo {volume} {101}},\ \bibinfo
  {pages} {063016} (\bibinfo {year} {2020})},\ \Eprint
  {https://arxiv.org/abs/1910.08544} {arXiv:1910.08544 [gr-qc]} \BibitemShut
  {NoStop}%
\bibitem [{\citenamefont {Fierz}(1956)}]{Fierz:1956zz}%
  \BibitemOpen
  \bibfield  {author} {\bibinfo {author} {\bibfnamefont {M.}~\bibnamefont
  {Fierz}},\ }\bibfield  {title} {\bibinfo {title} {{On the physical
  interpretation of P.Jordan's extended theory of gravitation}},\ }\href@noop
  {} {\bibfield  {journal} {\bibinfo  {journal} {Helv. Phys. Acta}\ }\textbf
  {\bibinfo {volume} {29}},\ \bibinfo {pages} {128} (\bibinfo {year}
  {1956})}\BibitemShut {NoStop}%
\bibitem [{\citenamefont {Jordan}(1959)}]{Jordan:1959eg}%
  \BibitemOpen
  \bibfield  {author} {\bibinfo {author} {\bibfnamefont {P.}~\bibnamefont
  {Jordan}},\ }\bibfield  {title} {\bibinfo {title} {{The present state of
  Dirac's cosmological hypothesis}},\ }\href
  {https://doi.org/10.1007/BF01375155} {\bibfield  {journal} {\bibinfo
  {journal} {Z. Phys.}\ }\textbf {\bibinfo {volume} {157}},\ \bibinfo {pages}
  {112} (\bibinfo {year} {1959})}\BibitemShut {NoStop}%
\bibitem [{\citenamefont {Brans}\ and\ \citenamefont
  {Dicke}(1961)}]{Brans:1961sx}%
  \BibitemOpen
  \bibfield  {author} {\bibinfo {author} {\bibfnamefont {C.}~\bibnamefont
  {Brans}}\ and\ \bibinfo {author} {\bibfnamefont {R.~H.}\ \bibnamefont
  {Dicke}},\ }\bibfield  {title} {\bibinfo {title} {{Mach's principle and a
  relativistic theory of gravitation}},\ }\href
  {https://doi.org/10.1103/PhysRev.124.925} {\bibfield  {journal} {\bibinfo
  {journal} {Phys. Rev.}\ }\textbf {\bibinfo {volume} {124}},\ \bibinfo {pages}
  {925} (\bibinfo {year} {1961})}\BibitemShut {NoStop}%
\bibitem [{\citenamefont {Dicke}(1962)}]{Dicke:1961gz}%
  \BibitemOpen
  \bibfield  {author} {\bibinfo {author} {\bibfnamefont {R.~H.}\ \bibnamefont
  {Dicke}},\ }\bibfield  {title} {\bibinfo {title} {{Mach's principle and
  invariance under transformation of units}},\ }\href
  {https://doi.org/10.1103/PhysRev.125.2163} {\bibfield  {journal} {\bibinfo
  {journal} {Phys. Rev.}\ }\textbf {\bibinfo {volume} {125}},\ \bibinfo {pages}
  {2163} (\bibinfo {year} {1962})}\BibitemShut {NoStop}%
\bibitem [{\citenamefont {Damour}\ and\ \citenamefont
  {Esposito-Farese}(1992)}]{Damour:1992we}%
  \BibitemOpen
  \bibfield  {author} {\bibinfo {author} {\bibfnamefont {T.}~\bibnamefont
  {Damour}}\ and\ \bibinfo {author} {\bibfnamefont {G.}~\bibnamefont
  {Esposito-Farese}},\ }\bibfield  {title} {\bibinfo {title} {{Tensor
  multiscalar theories of gravitation}},\ }\href
  {https://doi.org/10.1088/0264-9381/9/9/015} {\bibfield  {journal} {\bibinfo
  {journal} {Class. Quant. Grav.}\ }\textbf {\bibinfo {volume} {9}},\ \bibinfo
  {pages} {2093} (\bibinfo {year} {1992})}\BibitemShut {NoStop}%
\bibitem [{\citenamefont {Damour}\ and\ \citenamefont
  {Esposito-Farese}(1993)}]{Damour:1993hw}%
  \BibitemOpen
  \bibfield  {author} {\bibinfo {author} {\bibfnamefont {T.}~\bibnamefont
  {Damour}}\ and\ \bibinfo {author} {\bibfnamefont {G.}~\bibnamefont
  {Esposito-Farese}},\ }\bibfield  {title} {\bibinfo {title} {{Nonperturbative
  strong field effects in tensor - scalar theories of gravitation}},\ }\href
  {https://doi.org/10.1103/PhysRevLett.70.2220} {\bibfield  {journal} {\bibinfo
   {journal} {Phys. Rev. Lett.}\ }\textbf {\bibinfo {volume} {70}},\ \bibinfo
  {pages} {2220} (\bibinfo {year} {1993})}\BibitemShut {NoStop}%
\bibitem [{\citenamefont {Damour}\ and\ \citenamefont
  {Donoghue}(2010)}]{Damour:2010rp}%
  \BibitemOpen
  \bibfield  {author} {\bibinfo {author} {\bibfnamefont {T.}~\bibnamefont
  {Damour}}\ and\ \bibinfo {author} {\bibfnamefont {J.~F.}\ \bibnamefont
  {Donoghue}},\ }\bibfield  {title} {\bibinfo {title} {{Equivalence Principle
  Violations and Couplings of a Light Dilaton}},\ }\href
  {https://doi.org/10.1103/PhysRevD.82.084033} {\bibfield  {journal} {\bibinfo
  {journal} {Phys. Rev. D}\ }\textbf {\bibinfo {volume} {82}},\ \bibinfo
  {pages} {084033} (\bibinfo {year} {2010})},\ \Eprint
  {https://arxiv.org/abs/1007.2792} {arXiv:1007.2792 [gr-qc]} \BibitemShut
  {NoStop}%
\bibitem [{\citenamefont {Smarra}\ \emph {et~al.}(2024)\citenamefont {Smarra}
  \emph {et~al.}}]{Smarra:2024kvv}%
  \BibitemOpen
  \bibfield  {author} {\bibinfo {author} {\bibfnamefont {C.}~\bibnamefont
  {Smarra}} \emph {et~al.},\ }\bibfield  {title} {\bibinfo {title}
  {{Constraints on conformal ultralight dark matter couplings from the European
  Pulsar Timing Array}},\ }\href {https://doi.org/10.1103/PhysRevD.110.043033}
  {\bibfield  {journal} {\bibinfo  {journal} {Phys. Rev. D}\ }\textbf {\bibinfo
  {volume} {110}},\ \bibinfo {pages} {043033} (\bibinfo {year} {2024})},\
  \Eprint {https://arxiv.org/abs/2405.01633} {arXiv:2405.01633 [astro-ph.HE]}
  \BibitemShut {NoStop}%
\bibitem [{\citenamefont {Kaup}(1968)}]{Kaup:1968}%
  \BibitemOpen
  \bibfield  {author} {\bibinfo {author} {\bibfnamefont {D.~J.}\ \bibnamefont
  {Kaup}},\ }\bibfield  {title} {\bibinfo {title} {Klein-gordon geon},\ }\href
  {https://doi.org/10.1103/PhysRev.172.1331} {\bibfield  {journal} {\bibinfo
  {journal} {Phys. Rev.}\ }\textbf {\bibinfo {volume} {172}},\ \bibinfo {pages}
  {1331} (\bibinfo {year} {1968})}\BibitemShut {NoStop}%
\bibitem [{\citenamefont {Ruffini}\ and\ \citenamefont
  {Bonazzola}(1969)}]{Ruffini:1969qy}%
  \BibitemOpen
  \bibfield  {author} {\bibinfo {author} {\bibfnamefont {R.}~\bibnamefont
  {Ruffini}}\ and\ \bibinfo {author} {\bibfnamefont {S.}~\bibnamefont
  {Bonazzola}},\ }\bibfield  {title} {\bibinfo {title} {{Systems of
  selfgravitating particles in general relativity and the concept of an
  equation of state}},\ }\href {https://doi.org/10.1103/PhysRev.187.1767}
  {\bibfield  {journal} {\bibinfo  {journal} {Phys. Rev.}\ }\textbf {\bibinfo
  {volume} {187}},\ \bibinfo {pages} {1767} (\bibinfo {year}
  {1969})}\BibitemShut {NoStop}%
\bibitem [{\citenamefont {Liebling}\ and\ \citenamefont
  {Palenzuela}(2023)}]{Liebling:2012fv}%
  \BibitemOpen
  \bibfield  {author} {\bibinfo {author} {\bibfnamefont {S.~L.}\ \bibnamefont
  {Liebling}}\ and\ \bibinfo {author} {\bibfnamefont {C.}~\bibnamefont
  {Palenzuela}},\ }\bibfield  {title} {\bibinfo {title} {{Dynamical boson
  stars}},\ }\href {https://doi.org/10.1007/s41114-023-00043-4} {\bibfield
  {journal} {\bibinfo  {journal} {Living Rev. Rel.}\ }\textbf {\bibinfo
  {volume} {26}},\ \bibinfo {pages} {1} (\bibinfo {year} {2023})},\ \Eprint
  {https://arxiv.org/abs/1202.5809} {arXiv:1202.5809 [gr-qc]} \BibitemShut
  {NoStop}%
\bibitem [{\citenamefont {Seidel}\ and\ \citenamefont
  {Suen}(1991)}]{Seidel:1991zh}%
  \BibitemOpen
  \bibfield  {author} {\bibinfo {author} {\bibfnamefont {E.}~\bibnamefont
  {Seidel}}\ and\ \bibinfo {author} {\bibfnamefont {W.~M.}\ \bibnamefont
  {Suen}},\ }\bibfield  {title} {\bibinfo {title} {{Oscillating soliton
  stars}},\ }\href {https://doi.org/10.1103/PhysRevLett.66.1659} {\bibfield
  {journal} {\bibinfo  {journal} {Phys. Rev. Lett.}\ }\textbf {\bibinfo
  {volume} {66}},\ \bibinfo {pages} {1659} (\bibinfo {year}
  {1991})}\BibitemShut {NoStop}%
\bibitem [{\citenamefont {Guzman}\ and\ \citenamefont
  {Urena-Lopez}(2004)}]{Guzman:2004wj}%
  \BibitemOpen
  \bibfield  {author} {\bibinfo {author} {\bibfnamefont {F.~S.}\ \bibnamefont
  {Guzman}}\ and\ \bibinfo {author} {\bibfnamefont {L.~A.}\ \bibnamefont
  {Urena-Lopez}},\ }\bibfield  {title} {\bibinfo {title} {{Evolution of the
  Schrodinger-Newton system for a selfgravitating scalar field}},\ }\href
  {https://doi.org/10.1103/PhysRevD.69.124033} {\bibfield  {journal} {\bibinfo
  {journal} {Phys. Rev. D}\ }\textbf {\bibinfo {volume} {69}},\ \bibinfo
  {pages} {124033} (\bibinfo {year} {2004})},\ \Eprint
  {https://arxiv.org/abs/gr-qc/0404014} {arXiv:gr-qc/0404014} \BibitemShut
  {NoStop}%
\bibitem [{\citenamefont {Visinelli}(2021)}]{Visinelli:2021uve}%
  \BibitemOpen
  \bibfield  {author} {\bibinfo {author} {\bibfnamefont {L.}~\bibnamefont
  {Visinelli}},\ }\bibfield  {title} {\bibinfo {title} {{Boson stars and
  oscillatons: A review}},\ }\href {https://doi.org/10.1142/S0218271821300068}
  {\bibfield  {journal} {\bibinfo  {journal} {Int. J. Mod. Phys. D}\ }\textbf
  {\bibinfo {volume} {30}},\ \bibinfo {pages} {2130006} (\bibinfo {year}
  {2021})},\ \Eprint {https://arxiv.org/abs/2109.05481} {arXiv:2109.05481
  [gr-qc]} \BibitemShut {NoStop}%
\bibitem [{\citenamefont {Schive}\ \emph
  {et~al.}(2014{\natexlab{b}})\citenamefont {Schive}, \citenamefont {Liao},
  \citenamefont {Woo}, \citenamefont {Wong}, \citenamefont {Chiueh},
  \citenamefont {Broadhurst},\ and\ \citenamefont {Hwang}}]{Schive:2014hza}%
  \BibitemOpen
  \bibfield  {author} {\bibinfo {author} {\bibfnamefont {H.-Y.}\ \bibnamefont
  {Schive}}, \bibinfo {author} {\bibfnamefont {M.-H.}\ \bibnamefont {Liao}},
  \bibinfo {author} {\bibfnamefont {T.-P.}\ \bibnamefont {Woo}}, \bibinfo
  {author} {\bibfnamefont {S.-K.}\ \bibnamefont {Wong}}, \bibinfo {author}
  {\bibfnamefont {T.}~\bibnamefont {Chiueh}}, \bibinfo {author} {\bibfnamefont
  {T.}~\bibnamefont {Broadhurst}},\ and\ \bibinfo {author} {\bibfnamefont
  {W.~Y.~P.}\ \bibnamefont {Hwang}},\ }\bibfield  {title} {\bibinfo {title}
  {{Understanding the Core-Halo Relation of Quantum Wave Dark Matter from 3D
  Simulations}},\ }\href {https://doi.org/10.1103/PhysRevLett.113.261302}
  {\bibfield  {journal} {\bibinfo  {journal} {Phys. Rev. Lett.}\ }\textbf
  {\bibinfo {volume} {113}},\ \bibinfo {pages} {261302} (\bibinfo {year}
  {2014}{\natexlab{b}})},\ \Eprint {https://arxiv.org/abs/1407.7762}
  {arXiv:1407.7762 [astro-ph.GA]} \BibitemShut {NoStop}%
\bibitem [{\citenamefont {Mocz}\ \emph {et~al.}(2017)\citenamefont {Mocz},
  \citenamefont {Vogelsberger}, \citenamefont {Robles}, \citenamefont {Zavala},
  \citenamefont {Boylan-Kolchin}, \citenamefont {Fialkov},\ and\ \citenamefont
  {Hernquist}}]{Mocz:2017wlg}%
  \BibitemOpen
  \bibfield  {author} {\bibinfo {author} {\bibfnamefont {P.}~\bibnamefont
  {Mocz}}, \bibinfo {author} {\bibfnamefont {M.}~\bibnamefont {Vogelsberger}},
  \bibinfo {author} {\bibfnamefont {V.~H.}\ \bibnamefont {Robles}}, \bibinfo
  {author} {\bibfnamefont {J.}~\bibnamefont {Zavala}}, \bibinfo {author}
  {\bibfnamefont {M.}~\bibnamefont {Boylan-Kolchin}}, \bibinfo {author}
  {\bibfnamefont {A.}~\bibnamefont {Fialkov}},\ and\ \bibinfo {author}
  {\bibfnamefont {L.}~\bibnamefont {Hernquist}},\ }\bibfield  {title} {\bibinfo
  {title} {{Galaxy formation with BECDM \textendash{} I. Turbulence and
  relaxation of idealized haloes}},\ }\href
  {https://doi.org/10.1093/mnras/stx1887} {\bibfield  {journal} {\bibinfo
  {journal} {Mon. Not. Roy. Astron. Soc.}\ }\textbf {\bibinfo {volume} {471}},\
  \bibinfo {pages} {4559} (\bibinfo {year} {2017})},\ \Eprint
  {https://arxiv.org/abs/1705.05845} {arXiv:1705.05845 [astro-ph.CO]}
  \BibitemShut {NoStop}%
\bibitem [{\citenamefont {Veltmaat}\ \emph {et~al.}(2018)\citenamefont
  {Veltmaat}, \citenamefont {Niemeyer},\ and\ \citenamefont
  {Schwabe}}]{Veltmaat:2018dfz}%
  \BibitemOpen
  \bibfield  {author} {\bibinfo {author} {\bibfnamefont {J.}~\bibnamefont
  {Veltmaat}}, \bibinfo {author} {\bibfnamefont {J.~C.}\ \bibnamefont
  {Niemeyer}},\ and\ \bibinfo {author} {\bibfnamefont {B.}~\bibnamefont
  {Schwabe}},\ }\bibfield  {title} {\bibinfo {title} {{Formation and structure
  of ultralight bosonic dark matter halos}},\ }\href
  {https://doi.org/10.1103/PhysRevD.98.043509} {\bibfield  {journal} {\bibinfo
  {journal} {Phys. Rev. D}\ }\textbf {\bibinfo {volume} {98}},\ \bibinfo
  {pages} {043509} (\bibinfo {year} {2018})},\ \Eprint
  {https://arxiv.org/abs/1804.09647} {arXiv:1804.09647 [astro-ph.CO]}
  \BibitemShut {NoStop}%
\bibitem [{\citenamefont {{Chan}}\ \emph {et~al.}(2018)\citenamefont {{Chan}},
  \citenamefont {{Schive}}, \citenamefont {{Woo}},\ and\ \citenamefont
  {{Chiueh}}}]{Chan:2018}%
  \BibitemOpen
  \bibfield  {author} {\bibinfo {author} {\bibfnamefont {J.~H.~H.}\
  \bibnamefont {{Chan}}}, \bibinfo {author} {\bibfnamefont {H.-Y.}\
  \bibnamefont {{Schive}}}, \bibinfo {author} {\bibfnamefont {T.-P.}\
  \bibnamefont {{Woo}}},\ and\ \bibinfo {author} {\bibfnamefont
  {T.}~\bibnamefont {{Chiueh}}},\ }\bibfield  {title} {\bibinfo {title} {{How
  do stars affect {\ensuremath{\psi}}DM haloes?}},\ }\href
  {https://doi.org/10.1093/mnras/sty900} {\bibfield  {journal} {\bibinfo
  {journal} {\mnras}\ }\textbf {\bibinfo {volume} {478}},\ \bibinfo {pages}
  {2686} (\bibinfo {year} {2018})},\ \Eprint {https://arxiv.org/abs/1712.01947}
  {arXiv:1712.01947 [astro-ph.GA]} \BibitemShut {NoStop}%
\bibitem [{\citenamefont {Mina}\ \emph {et~al.}(2022)\citenamefont {Mina},
  \citenamefont {Mota},\ and\ \citenamefont {Winther}}]{Mina:2020eik}%
  \BibitemOpen
  \bibfield  {author} {\bibinfo {author} {\bibfnamefont {M.}~\bibnamefont
  {Mina}}, \bibinfo {author} {\bibfnamefont {D.~F.}\ \bibnamefont {Mota}},\
  and\ \bibinfo {author} {\bibfnamefont {H.~A.}\ \bibnamefont {Winther}},\
  }\bibfield  {title} {\bibinfo {title} {{Solitons in the dark: First approach
  to non-linear structure formation with fuzzy dark matter}},\ }\href
  {https://doi.org/10.1051/0004-6361/202038876} {\bibfield  {journal} {\bibinfo
   {journal} {Astron. Astrophys.}\ }\textbf {\bibinfo {volume} {662}},\
  \bibinfo {pages} {A29} (\bibinfo {year} {2022})},\ \Eprint
  {https://arxiv.org/abs/2007.04119} {arXiv:2007.04119 [astro-ph.CO]}
  \BibitemShut {NoStop}%
\bibitem [{\citenamefont {Nori}\ and\ \citenamefont
  {Baldi}(2021)}]{Nori:2020jzx}%
  \BibitemOpen
  \bibfield  {author} {\bibinfo {author} {\bibfnamefont {M.}~\bibnamefont
  {Nori}}\ and\ \bibinfo {author} {\bibfnamefont {M.}~\bibnamefont {Baldi}},\
  }\bibfield  {title} {\bibinfo {title} {{Scaling relations of fuzzy dark
  matter haloes \textendash{} I. Individual systems in their cosmological
  environment}},\ }\href {https://doi.org/10.1093/mnras/staa3772} {\bibfield
  {journal} {\bibinfo  {journal} {Mon. Not. Roy. Astron. Soc.}\ }\textbf
  {\bibinfo {volume} {501}},\ \bibinfo {pages} {1539} (\bibinfo {year}
  {2021})},\ \Eprint {https://arxiv.org/abs/2007.01316} {arXiv:2007.01316
  [astro-ph.CO]} \BibitemShut {NoStop}%
\bibitem [{\citenamefont {Chen}\ \emph {et~al.}(2021)\citenamefont {Chen},
  \citenamefont {Du}, \citenamefont {Lentz}, \citenamefont {Marsh},\ and\
  \citenamefont {Niemeyer}}]{Chen:2020cef}%
  \BibitemOpen
  \bibfield  {author} {\bibinfo {author} {\bibfnamefont {J.}~\bibnamefont
  {Chen}}, \bibinfo {author} {\bibfnamefont {X.}~\bibnamefont {Du}}, \bibinfo
  {author} {\bibfnamefont {E.~W.}\ \bibnamefont {Lentz}}, \bibinfo {author}
  {\bibfnamefont {D.~J.~E.}\ \bibnamefont {Marsh}},\ and\ \bibinfo {author}
  {\bibfnamefont {J.~C.}\ \bibnamefont {Niemeyer}},\ }\bibfield  {title}
  {\bibinfo {title} {{New insights into the formation and growth of boson stars
  in dark matter halos}},\ }\href {https://doi.org/10.1103/PhysRevD.104.083022}
  {\bibfield  {journal} {\bibinfo  {journal} {Phys. Rev. D}\ }\textbf {\bibinfo
  {volume} {104}},\ \bibinfo {pages} {083022} (\bibinfo {year} {2021})},\
  \Eprint {https://arxiv.org/abs/2011.01333} {arXiv:2011.01333 [astro-ph.CO]}
  \BibitemShut {NoStop}%
\bibitem [{\citenamefont {Chan}\ \emph {et~al.}(2022)\citenamefont {Chan},
  \citenamefont {Ferreira}, \citenamefont {May}, \citenamefont {Hayashi},\ and\
  \citenamefont {Chiba}}]{Chan:2021bja}%
  \BibitemOpen
  \bibfield  {author} {\bibinfo {author} {\bibfnamefont {H.~Y.~J.}\
  \bibnamefont {Chan}}, \bibinfo {author} {\bibfnamefont {E.~G.~M.}\
  \bibnamefont {Ferreira}}, \bibinfo {author} {\bibfnamefont {S.}~\bibnamefont
  {May}}, \bibinfo {author} {\bibfnamefont {K.}~\bibnamefont {Hayashi}},\ and\
  \bibinfo {author} {\bibfnamefont {M.}~\bibnamefont {Chiba}},\ }\bibfield
  {title} {\bibinfo {title} {{The diversity of core\textendash{}halo structure
  in the fuzzy dark matter model}},\ }\href
  {https://doi.org/10.1093/mnras/stac063} {\bibfield  {journal} {\bibinfo
  {journal} {Mon. Not. Roy. Astron. Soc.}\ }\textbf {\bibinfo {volume} {511}},\
  \bibinfo {pages} {943} (\bibinfo {year} {2022})},\ \Eprint
  {https://arxiv.org/abs/2110.11882} {arXiv:2110.11882 [astro-ph.CO]}
  \BibitemShut {NoStop}%
\bibitem [{\citenamefont {Mocz}\ \emph {et~al.}(2023)\citenamefont {Mocz} \emph
  {et~al.}}]{Mocz:2023adf}%
  \BibitemOpen
  \bibfield  {author} {\bibinfo {author} {\bibfnamefont {P.}~\bibnamefont
  {Mocz}} \emph {et~al.},\ }\bibfield  {title} {\bibinfo {title} {{Cosmological
  structure formation and soliton phase transition in fuzzy dark matter with
  axion self-interactions}},\ }\href {https://doi.org/10.1093/mnras/stad694}
  {\bibfield  {journal} {\bibinfo  {journal} {Mon. Not. Roy. Astron. Soc.}\
  }\textbf {\bibinfo {volume} {521}},\ \bibinfo {pages} {2608} (\bibinfo {year}
  {2023})},\ \Eprint {https://arxiv.org/abs/2301.10266} {arXiv:2301.10266
  [astro-ph.CO]} \BibitemShut {NoStop}%
\bibitem [{\citenamefont {Painter}\ \emph {et~al.}(2024)\citenamefont
  {Painter}, \citenamefont {Boylan-Kolchin}, \citenamefont {Mocz},\ and\
  \citenamefont {Vogelsberger}}]{Painter:2024rnc}%
  \BibitemOpen
  \bibfield  {author} {\bibinfo {author} {\bibfnamefont {C.~A.}\ \bibnamefont
  {Painter}}, \bibinfo {author} {\bibfnamefont {M.}~\bibnamefont
  {Boylan-Kolchin}}, \bibinfo {author} {\bibfnamefont {P.}~\bibnamefont
  {Mocz}},\ and\ \bibinfo {author} {\bibfnamefont {M.}~\bibnamefont
  {Vogelsberger}},\ }\bibfield  {title} {\bibinfo {title} {{An attractive
  model: simulating fuzzy dark matter with attractive self-interactions}},\
  }\href {https://doi.org/10.1093/mnras/stae1912} {\bibfield  {journal}
  {\bibinfo  {journal} {Mon. Not. Roy. Astron. Soc.}\ }\textbf {\bibinfo
  {volume} {533}},\ \bibinfo {pages} {2454} (\bibinfo {year} {2024})},\ \Eprint
  {https://arxiv.org/abs/2402.16945} {arXiv:2402.16945 [astro-ph.CO]}
  \BibitemShut {NoStop}%
\bibitem [{\citenamefont {{Chavanis}}(2011)}]{Chavanis:2011a}%
  \BibitemOpen
  \bibfield  {author} {\bibinfo {author} {\bibfnamefont {P.-H.}\ \bibnamefont
  {{Chavanis}}},\ }\bibfield  {title} {\bibinfo {title} {{Mass-radius relation
  of Newtonian self-gravitating Bose-Einstein condensates with short-range
  interactions. I. Analytical results}},\ }\href
  {https://doi.org/10.1103/PhysRevD.84.043531} {\bibfield  {journal} {\bibinfo
  {journal} {\prd}\ }\textbf {\bibinfo {volume} {84}},\ \bibinfo {eid} {043531}
  (\bibinfo {year} {2011})},\ \Eprint {https://arxiv.org/abs/1103.2050}
  {arXiv:1103.2050 [astro-ph.CO]} \BibitemShut {NoStop}%
\bibitem [{\citenamefont {{Chavanis}}\ and\ \citenamefont
  {{Delfini}}(2011)}]{Chavanis:2011b}%
  \BibitemOpen
  \bibfield  {author} {\bibinfo {author} {\bibfnamefont {P.-H.}\ \bibnamefont
  {{Chavanis}}}\ and\ \bibinfo {author} {\bibfnamefont {L.}~\bibnamefont
  {{Delfini}}},\ }\bibfield  {title} {\bibinfo {title} {{Mass-radius relation
  of Newtonian self-gravitating Bose-Einstein condensates with short-range
  interactions. II. Numerical results}},\ }\href
  {https://doi.org/10.1103/PhysRevD.84.043532} {\bibfield  {journal} {\bibinfo
  {journal} {\prd}\ }\textbf {\bibinfo {volume} {84}},\ \bibinfo {eid} {043532}
  (\bibinfo {year} {2011})},\ \Eprint {https://arxiv.org/abs/1103.2054}
  {arXiv:1103.2054 [astro-ph.CO]} \BibitemShut {NoStop}%
\bibitem [{\citenamefont {Eby}\ \emph {et~al.}(2016)\citenamefont {Eby},
  \citenamefont {Kouvaris}, \citenamefont {Nielsen},\ and\ \citenamefont
  {Wijewardhana}}]{Eby:2015hsq}%
  \BibitemOpen
  \bibfield  {author} {\bibinfo {author} {\bibfnamefont {J.}~\bibnamefont
  {Eby}}, \bibinfo {author} {\bibfnamefont {C.}~\bibnamefont {Kouvaris}},
  \bibinfo {author} {\bibfnamefont {N.~G.}\ \bibnamefont {Nielsen}},\ and\
  \bibinfo {author} {\bibfnamefont {L.~C.~R.}\ \bibnamefont {Wijewardhana}},\
  }\bibfield  {title} {\bibinfo {title} {{Boson Stars from Self-Interacting
  Dark Matter}},\ }\href {https://doi.org/10.1007/JHEP02(2016)028} {\bibfield
  {journal} {\bibinfo  {journal} {JHEP}\ }\textbf {\bibinfo {volume} {02}},\
  \bibinfo {pages} {028}},\ \Eprint {https://arxiv.org/abs/1511.04474}
  {arXiv:1511.04474 [hep-ph]} \BibitemShut {NoStop}%
\bibitem [{\citenamefont {Braaten}\ \emph {et~al.}(2016)\citenamefont
  {Braaten}, \citenamefont {Mohapatra},\ and\ \citenamefont
  {Zhang}}]{Braaten:2015eeu}%
  \BibitemOpen
  \bibfield  {author} {\bibinfo {author} {\bibfnamefont {E.}~\bibnamefont
  {Braaten}}, \bibinfo {author} {\bibfnamefont {A.}~\bibnamefont {Mohapatra}},\
  and\ \bibinfo {author} {\bibfnamefont {H.}~\bibnamefont {Zhang}},\ }\bibfield
   {title} {\bibinfo {title} {{Dense Axion Stars}},\ }\href
  {https://doi.org/10.1103/PhysRevLett.117.121801} {\bibfield  {journal}
  {\bibinfo  {journal} {Phys. Rev. Lett.}\ }\textbf {\bibinfo {volume} {117}},\
  \bibinfo {pages} {121801} (\bibinfo {year} {2016})},\ \Eprint
  {https://arxiv.org/abs/1512.00108} {arXiv:1512.00108 [hep-ph]} \BibitemShut
  {NoStop}%
\bibitem [{\citenamefont {Chavanis}(2018)}]{Chavanis:2017loo}%
  \BibitemOpen
  \bibfield  {author} {\bibinfo {author} {\bibfnamefont {P.-H.}\ \bibnamefont
  {Chavanis}},\ }\bibfield  {title} {\bibinfo {title} {{Phase transitions
  between dilute and dense axion stars}},\ }\href
  {https://doi.org/10.1103/PhysRevD.98.023009} {\bibfield  {journal} {\bibinfo
  {journal} {Phys. Rev. D}\ }\textbf {\bibinfo {volume} {98}},\ \bibinfo
  {pages} {023009} (\bibinfo {year} {2018})},\ \Eprint
  {https://arxiv.org/abs/1710.06268} {arXiv:1710.06268 [gr-qc]} \BibitemShut
  {NoStop}%
\bibitem [{\citenamefont {Chavanis}(2016)}]{Chavanis:2016dab}%
  \BibitemOpen
  \bibfield  {author} {\bibinfo {author} {\bibfnamefont {P.-H.}\ \bibnamefont
  {Chavanis}},\ }\bibfield  {title} {\bibinfo {title} {{Collapse of a
  self-gravitating Bose-Einstein condensate with attractive
  self-interaction}},\ }\href {https://doi.org/10.1103/PhysRevD.94.083007}
  {\bibfield  {journal} {\bibinfo  {journal} {Phys. Rev. D}\ }\textbf {\bibinfo
  {volume} {94}},\ \bibinfo {pages} {083007} (\bibinfo {year} {2016})},\
  \Eprint {https://arxiv.org/abs/1604.05904} {arXiv:1604.05904 [astro-ph.CO]}
  \BibitemShut {NoStop}%
\bibitem [{\citenamefont {Helfer}\ \emph {et~al.}(2017)\citenamefont {Helfer},
  \citenamefont {Marsh}, \citenamefont {Clough}, \citenamefont {Fairbairn},
  \citenamefont {Lim},\ and\ \citenamefont {Becerril}}]{Helfer:2016ljl}%
  \BibitemOpen
  \bibfield  {author} {\bibinfo {author} {\bibfnamefont {T.}~\bibnamefont
  {Helfer}}, \bibinfo {author} {\bibfnamefont {D.~J.~E.}\ \bibnamefont
  {Marsh}}, \bibinfo {author} {\bibfnamefont {K.}~\bibnamefont {Clough}},
  \bibinfo {author} {\bibfnamefont {M.}~\bibnamefont {Fairbairn}}, \bibinfo
  {author} {\bibfnamefont {E.~A.}\ \bibnamefont {Lim}},\ and\ \bibinfo {author}
  {\bibfnamefont {R.}~\bibnamefont {Becerril}},\ }\bibfield  {title} {\bibinfo
  {title} {{Black hole formation from axion stars}},\ }\href
  {https://doi.org/10.1088/1475-7516/2017/03/055} {\bibfield  {journal}
  {\bibinfo  {journal} {JCAP}\ }\textbf {\bibinfo {volume} {03}},\ \bibinfo
  {pages} {055}},\ \Eprint {https://arxiv.org/abs/1609.04724} {arXiv:1609.04724
  [astro-ph.CO]} \BibitemShut {NoStop}%
\bibitem [{\citenamefont {Bar}\ \emph {et~al.}(2018)\citenamefont {Bar},
  \citenamefont {Blas}, \citenamefont {Blum},\ and\ \citenamefont
  {Sibiryakov}}]{Bar:2018acw}%
  \BibitemOpen
  \bibfield  {author} {\bibinfo {author} {\bibfnamefont {N.}~\bibnamefont
  {Bar}}, \bibinfo {author} {\bibfnamefont {D.}~\bibnamefont {Blas}}, \bibinfo
  {author} {\bibfnamefont {K.}~\bibnamefont {Blum}},\ and\ \bibinfo {author}
  {\bibfnamefont {S.}~\bibnamefont {Sibiryakov}},\ }\bibfield  {title}
  {\bibinfo {title} {{Galactic rotation curves versus ultralight dark matter:
  Implications of the soliton-host halo relation}},\ }\href
  {https://doi.org/10.1103/PhysRevD.98.083027} {\bibfield  {journal} {\bibinfo
  {journal} {Phys. Rev. D}\ }\textbf {\bibinfo {volume} {98}},\ \bibinfo
  {pages} {083027} (\bibinfo {year} {2018})},\ \Eprint
  {https://arxiv.org/abs/1805.00122} {arXiv:1805.00122 [astro-ph.CO]}
  \BibitemShut {NoStop}%
\bibitem [{\citenamefont {Rogers}\ and\ \citenamefont
  {Peiris}(2021)}]{Rogers:2020ltq}%
  \BibitemOpen
  \bibfield  {author} {\bibinfo {author} {\bibfnamefont {K.~K.}\ \bibnamefont
  {Rogers}}\ and\ \bibinfo {author} {\bibfnamefont {H.~V.}\ \bibnamefont
  {Peiris}},\ }\bibfield  {title} {\bibinfo {title} {{Strong Bound on Canonical
  Ultralight Axion Dark Matter from the Lyman-Alpha Forest}},\ }\href
  {https://doi.org/10.1103/PhysRevLett.126.071302} {\bibfield  {journal}
  {\bibinfo  {journal} {Phys. Rev. Lett.}\ }\textbf {\bibinfo {volume} {126}},\
  \bibinfo {pages} {071302} (\bibinfo {year} {2021})},\ \Eprint
  {https://arxiv.org/abs/2007.12705} {arXiv:2007.12705 [astro-ph.CO]}
  \BibitemShut {NoStop}%
\bibitem [{\citenamefont {van Dissel}\ \emph {et~al.}(2024)\citenamefont {van
  Dissel}, \citenamefont {Hertzberg},\ and\ \citenamefont
  {Shapiro}}]{vanDissel:2023vhu}%
  \BibitemOpen
  \bibfield  {author} {\bibinfo {author} {\bibfnamefont {F.}~\bibnamefont {van
  Dissel}}, \bibinfo {author} {\bibfnamefont {M.~P.}\ \bibnamefont
  {Hertzberg}},\ and\ \bibinfo {author} {\bibfnamefont {J.}~\bibnamefont
  {Shapiro}},\ }\bibfield  {title} {\bibinfo {title} {{Core and halo properties
  in multi-field wave dark matter}},\ }\href
  {https://doi.org/10.1088/1475-7516/2024/04/077} {\bibfield  {journal}
  {\bibinfo  {journal} {JCAP}\ }\textbf {\bibinfo {volume} {04}},\ \bibinfo
  {pages} {077}},\ \Eprint {https://arxiv.org/abs/2310.19762} {arXiv:2310.19762
  [astro-ph.CO]} \BibitemShut {NoStop}%
\bibitem [{\citenamefont {Mirasola}\ \emph {et~al.}(2024)\citenamefont
  {Mirasola}, \citenamefont {Musoke}, \citenamefont {Neyrinck}, \citenamefont
  {Prescod-Weinstein},\ and\ \citenamefont {Zagorac}}]{Mirasola:2024pmw}%
  \BibitemOpen
  \bibfield  {author} {\bibinfo {author} {\bibfnamefont {A.~E.}\ \bibnamefont
  {Mirasola}}, \bibinfo {author} {\bibfnamefont {N.}~\bibnamefont {Musoke}},
  \bibinfo {author} {\bibfnamefont {M.~C.}\ \bibnamefont {Neyrinck}}, \bibinfo
  {author} {\bibfnamefont {C.}~\bibnamefont {Prescod-Weinstein}},\ and\
  \bibinfo {author} {\bibfnamefont {J.~L.}\ \bibnamefont {Zagorac}},\
  }\bibfield  {title} {\bibinfo {title} {{The three phases of self-gravitating
  scalar field ground states}},\ }\href@noop {} {\  (\bibinfo {year} {2024})},\
  \Eprint {https://arxiv.org/abs/2410.02663} {arXiv:2410.02663 [astro-ph.CO]}
  \BibitemShut {NoStop}%
\bibitem [{\citenamefont {{Kormendy}}\ and\ \citenamefont
  {{Richstone}}(1995)}]{Kormendy:1995}%
  \BibitemOpen
  \bibfield  {author} {\bibinfo {author} {\bibfnamefont {J.}~\bibnamefont
  {{Kormendy}}}\ and\ \bibinfo {author} {\bibfnamefont {D.}~\bibnamefont
  {{Richstone}}},\ }\bibfield  {title} {\bibinfo {title} {{Inward Bound---The
  Search For Supermassive Black Holes In Galactic Nuclei}},\ }\href
  {https://doi.org/10.1146/annurev.aa.33.090195.003053} {\bibfield  {journal}
  {\bibinfo  {journal} {\araa}\ }\textbf {\bibinfo {volume} {33}},\ \bibinfo
  {pages} {581} (\bibinfo {year} {1995})}\BibitemShut {NoStop}%
\bibitem [{\citenamefont {Kormendy}\ and\ \citenamefont
  {Ho}(2013)}]{Kormendy:2013dxa}%
  \BibitemOpen
  \bibfield  {author} {\bibinfo {author} {\bibfnamefont {J.}~\bibnamefont
  {Kormendy}}\ and\ \bibinfo {author} {\bibfnamefont {L.~C.}\ \bibnamefont
  {Ho}},\ }\bibfield  {title} {\bibinfo {title} {{Coevolution (Or Not) of
  Supermassive Black Holes and Host Galaxies}},\ }\href
  {https://doi.org/10.1146/annurev-astro-082708-101811} {\bibfield  {journal}
  {\bibinfo  {journal} {Ann. Rev. Astron. Astrophys.}\ }\textbf {\bibinfo
  {volume} {51}},\ \bibinfo {pages} {511} (\bibinfo {year} {2013})},\ \Eprint
  {https://arxiv.org/abs/1304.7762} {arXiv:1304.7762 [astro-ph.CO]}
  \BibitemShut {NoStop}%
\bibitem [{\citenamefont {{Reines}}(2022)}]{Reines:2022}%
  \BibitemOpen
  \bibfield  {author} {\bibinfo {author} {\bibfnamefont {A.~E.}\ \bibnamefont
  {{Reines}}},\ }\bibfield  {title} {\bibinfo {title} {{Hunting for massive
  black holes in dwarf galaxies}},\ }\href
  {https://doi.org/10.1038/s41550-021-01556-0} {\bibfield  {journal} {\bibinfo
  {journal} {Nature Astronomy}\ }\textbf {\bibinfo {volume} {6}},\ \bibinfo
  {pages} {26} (\bibinfo {year} {2022})},\ \Eprint
  {https://arxiv.org/abs/2201.10569} {arXiv:2201.10569 [astro-ph.GA]}
  \BibitemShut {NoStop}%
\bibitem [{\citenamefont {Bar}\ \emph {et~al.}(2019)\citenamefont {Bar},
  \citenamefont {Blum}, \citenamefont {Lacroix},\ and\ \citenamefont
  {Panci}}]{Bar:2019pnz}%
  \BibitemOpen
  \bibfield  {author} {\bibinfo {author} {\bibfnamefont {N.}~\bibnamefont
  {Bar}}, \bibinfo {author} {\bibfnamefont {K.}~\bibnamefont {Blum}}, \bibinfo
  {author} {\bibfnamefont {T.}~\bibnamefont {Lacroix}},\ and\ \bibinfo {author}
  {\bibfnamefont {P.}~\bibnamefont {Panci}},\ }\bibfield  {title} {\bibinfo
  {title} {{Looking for ultralight dark matter near supermassive black
  holes}},\ }\href {https://doi.org/10.1088/1475-7516/2019/07/045} {\bibfield
  {journal} {\bibinfo  {journal} {JCAP}\ }\textbf {\bibinfo {volume} {07}},\
  \bibinfo {pages} {045}},\ \Eprint {https://arxiv.org/abs/1905.11745}
  {arXiv:1905.11745 [astro-ph.CO]} \BibitemShut {NoStop}%
\bibitem [{\citenamefont {Davies}\ and\ \citenamefont
  {Mocz}(2020)}]{Davies:2019wgi}%
  \BibitemOpen
  \bibfield  {author} {\bibinfo {author} {\bibfnamefont {E.~Y.}\ \bibnamefont
  {Davies}}\ and\ \bibinfo {author} {\bibfnamefont {P.}~\bibnamefont {Mocz}},\
  }\bibfield  {title} {\bibinfo {title} {{Fuzzy Dark Matter Soliton Cores
  around Supermassive Black Holes}},\ }\href
  {https://doi.org/10.1093/mnras/staa202} {\bibfield  {journal} {\bibinfo
  {journal} {Mon. Not. Roy. Astron. Soc.}\ }\textbf {\bibinfo {volume} {492}},\
  \bibinfo {pages} {5721} (\bibinfo {year} {2020})},\ \Eprint
  {https://arxiv.org/abs/1908.04790} {arXiv:1908.04790 [astro-ph.GA]}
  \BibitemShut {NoStop}%
\bibitem [{\citenamefont {Cardoso}\ \emph {et~al.}(2022)\citenamefont
  {Cardoso}, \citenamefont {Ikeda}, \citenamefont {Vicente},\ and\
  \citenamefont {Zilh\~ao}}]{Cardoso:2022nzc}%
  \BibitemOpen
  \bibfield  {author} {\bibinfo {author} {\bibfnamefont {V.}~\bibnamefont
  {Cardoso}}, \bibinfo {author} {\bibfnamefont {T.}~\bibnamefont {Ikeda}},
  \bibinfo {author} {\bibfnamefont {R.}~\bibnamefont {Vicente}},\ and\ \bibinfo
  {author} {\bibfnamefont {M.}~\bibnamefont {Zilh\~ao}},\ }\bibfield  {title}
  {\bibinfo {title} {{Parasitic black holes: The swallowing of a fuzzy dark
  matter soliton}},\ }\href {https://doi.org/10.1103/PhysRevD.106.L121302}
  {\bibfield  {journal} {\bibinfo  {journal} {Phys. Rev. D}\ }\textbf {\bibinfo
  {volume} {106}},\ \bibinfo {pages} {L121302} (\bibinfo {year} {2022})},\
  \Eprint {https://arxiv.org/abs/2207.09469} {arXiv:2207.09469 [gr-qc]}
  \BibitemShut {NoStop}%
\bibitem [{\citenamefont {Gillessen}\ \emph {et~al.}(2009)\citenamefont
  {Gillessen}, \citenamefont {Eisenhauer}, \citenamefont {Trippe},
  \citenamefont {Alexander}, \citenamefont {Genzel}, \citenamefont {Martins},\
  and\ \citenamefont {Ott}}]{Gillessen:2008qv}%
  \BibitemOpen
  \bibfield  {author} {\bibinfo {author} {\bibfnamefont {S.}~\bibnamefont
  {Gillessen}}, \bibinfo {author} {\bibfnamefont {F.}~\bibnamefont
  {Eisenhauer}}, \bibinfo {author} {\bibfnamefont {S.}~\bibnamefont {Trippe}},
  \bibinfo {author} {\bibfnamefont {T.}~\bibnamefont {Alexander}}, \bibinfo
  {author} {\bibfnamefont {R.}~\bibnamefont {Genzel}}, \bibinfo {author}
  {\bibfnamefont {F.}~\bibnamefont {Martins}},\ and\ \bibinfo {author}
  {\bibfnamefont {T.}~\bibnamefont {Ott}},\ }\bibfield  {title} {\bibinfo
  {title} {{Monitoring stellar orbits around the Massive Black Hole in the
  Galactic Center}},\ }\href {https://doi.org/10.1088/0004-637X/692/2/1075}
  {\bibfield  {journal} {\bibinfo  {journal} {Astrophys. J.}\ }\textbf
  {\bibinfo {volume} {692}},\ \bibinfo {pages} {1075} (\bibinfo {year}
  {2009})},\ \Eprint {https://arxiv.org/abs/0810.4674} {arXiv:0810.4674
  [astro-ph]} \BibitemShut {NoStop}%
\bibitem [{\citenamefont {Ghez}\ \emph {et~al.}(2008)\citenamefont {Ghez} \emph
  {et~al.}}]{Ghez:2008ms}%
  \BibitemOpen
  \bibfield  {author} {\bibinfo {author} {\bibfnamefont {A.~M.}\ \bibnamefont
  {Ghez}} \emph {et~al.},\ }\bibfield  {title} {\bibinfo {title} {{Measuring
  Distance and Properties of the Milky Way's Central Supermassive Black Hole
  with Stellar Orbits}},\ }\href {https://doi.org/10.1086/592738} {\bibfield
  {journal} {\bibinfo  {journal} {Astrophys. J.}\ }\textbf {\bibinfo {volume}
  {689}},\ \bibinfo {pages} {1044} (\bibinfo {year} {2008})},\ \Eprint
  {https://arxiv.org/abs/0808.2870} {arXiv:0808.2870 [astro-ph]} \BibitemShut
  {NoStop}%
\bibitem [{\citenamefont {Genzel}\ \emph {et~al.}(2010)\citenamefont {Genzel},
  \citenamefont {Eisenhauer},\ and\ \citenamefont {Gillessen}}]{Genzel:2010zy}%
  \BibitemOpen
  \bibfield  {author} {\bibinfo {author} {\bibfnamefont {R.}~\bibnamefont
  {Genzel}}, \bibinfo {author} {\bibfnamefont {F.}~\bibnamefont {Eisenhauer}},\
  and\ \bibinfo {author} {\bibfnamefont {S.}~\bibnamefont {Gillessen}},\
  }\bibfield  {title} {\bibinfo {title} {{The Galactic Center Massive Black
  Hole and Nuclear Star Cluster}},\ }\href
  {https://doi.org/10.1103/RevModPhys.82.3121} {\bibfield  {journal} {\bibinfo
  {journal} {Rev. Mod. Phys.}\ }\textbf {\bibinfo {volume} {82}},\ \bibinfo
  {pages} {3121} (\bibinfo {year} {2010})},\ \Eprint
  {https://arxiv.org/abs/1006.0064} {arXiv:1006.0064 [astro-ph.GA]}
  \BibitemShut {NoStop}%
\bibitem [{\citenamefont {Derevianko}(2018)}]{Derevianko:2016vpm}%
  \BibitemOpen
  \bibfield  {author} {\bibinfo {author} {\bibfnamefont {A.}~\bibnamefont
  {Derevianko}},\ }\bibfield  {title} {\bibinfo {title} {{Detecting dark-matter
  waves with a network of precision-measurement tools}},\ }\href
  {https://doi.org/10.1103/PhysRevA.97.042506} {\bibfield  {journal} {\bibinfo
  {journal} {Phys. Rev. A}\ }\textbf {\bibinfo {volume} {97}},\ \bibinfo
  {pages} {042506} (\bibinfo {year} {2018})},\ \Eprint
  {https://arxiv.org/abs/1605.09717} {arXiv:1605.09717 [physics.atom-ph]}
  \BibitemShut {NoStop}%
\bibitem [{\citenamefont {Centers}\ \emph {et~al.}(2021)\citenamefont {Centers}
  \emph {et~al.}}]{Centers:2019dyn}%
  \BibitemOpen
  \bibfield  {author} {\bibinfo {author} {\bibfnamefont {G.~P.}\ \bibnamefont
  {Centers}} \emph {et~al.},\ }\bibfield  {title} {\bibinfo {title}
  {{Stochastic fluctuations of bosonic dark matter}},\ }\href
  {https://doi.org/10.1038/s41467-021-27632-7} {\bibfield  {journal} {\bibinfo
  {journal} {Nature Commun.}\ }\textbf {\bibinfo {volume} {12}},\ \bibinfo
  {pages} {7321} (\bibinfo {year} {2021})},\ \Eprint
  {https://arxiv.org/abs/1905.13650} {arXiv:1905.13650 [astro-ph.CO]}
  \BibitemShut {NoStop}%
\bibitem [{\citenamefont {Navarro}\ \emph {et~al.}(1996)\citenamefont
  {Navarro}, \citenamefont {Frenk},\ and\ \citenamefont
  {White}}]{Navarro:1995iw}%
  \BibitemOpen
  \bibfield  {author} {\bibinfo {author} {\bibfnamefont {J.~F.}\ \bibnamefont
  {Navarro}}, \bibinfo {author} {\bibfnamefont {C.~S.}\ \bibnamefont {Frenk}},\
  and\ \bibinfo {author} {\bibfnamefont {S.~D.~M.}\ \bibnamefont {White}},\
  }\bibfield  {title} {\bibinfo {title} {{The Structure of cold dark matter
  halos}},\ }\href {https://doi.org/10.1086/177173} {\bibfield  {journal}
  {\bibinfo  {journal} {Astrophys. J.}\ }\textbf {\bibinfo {volume} {462}},\
  \bibinfo {pages} {563} (\bibinfo {year} {1996})},\ \Eprint
  {https://arxiv.org/abs/astro-ph/9508025} {arXiv:astro-ph/9508025}
  \BibitemShut {NoStop}%
\bibitem [{\citenamefont {Chatrchyan}\ \emph {et~al.}(2023)\citenamefont
  {Chatrchyan}, \citenamefont {Er\"oncel}, \citenamefont {Koschnitzke},\ and\
  \citenamefont {Servant}}]{Chatrchyan:2023cmz}%
  \BibitemOpen
  \bibfield  {author} {\bibinfo {author} {\bibfnamefont {A.}~\bibnamefont
  {Chatrchyan}}, \bibinfo {author} {\bibfnamefont {C.}~\bibnamefont
  {Er\"oncel}}, \bibinfo {author} {\bibfnamefont {M.}~\bibnamefont
  {Koschnitzke}},\ and\ \bibinfo {author} {\bibfnamefont {G.}~\bibnamefont
  {Servant}},\ }\bibfield  {title} {\bibinfo {title} {{ALP dark matter with
  non-periodic potentials: parametric resonance, halo formation and
  gravitational signatures}},\ }\href
  {https://doi.org/10.1088/1475-7516/2023/10/068} {\bibfield  {journal}
  {\bibinfo  {journal} {JCAP}\ }\textbf {\bibinfo {volume} {10}},\ \bibinfo
  {pages} {068}},\ \Eprint {https://arxiv.org/abs/2305.03756} {arXiv:2305.03756
  [hep-ph]} \BibitemShut {NoStop}%
\bibitem [{\citenamefont {Er\"oncel}\ and\ \citenamefont
  {Servant}(2023)}]{Eroncel:2022efc}%
  \BibitemOpen
  \bibfield  {author} {\bibinfo {author} {\bibfnamefont {C.}~\bibnamefont
  {Er\"oncel}}\ and\ \bibinfo {author} {\bibfnamefont {G.}~\bibnamefont
  {Servant}},\ }\bibfield  {title} {\bibinfo {title} {{ALP dark matter
  mini-clusters from kinetic fragmentation}},\ }\href
  {https://doi.org/10.1088/1475-7516/2023/01/009} {\bibfield  {journal}
  {\bibinfo  {journal} {JCAP}\ }\textbf {\bibinfo {volume} {01}},\ \bibinfo
  {pages} {009}},\ \Eprint {https://arxiv.org/abs/2207.10111} {arXiv:2207.10111
  [hep-ph]} \BibitemShut {NoStop}%
\bibitem [{\citenamefont {Er\"oncel}\ \emph {et~al.}(2022)\citenamefont
  {Er\"oncel}, \citenamefont {Sato}, \citenamefont {Servant},\ and\
  \citenamefont {S\o{}rensen}}]{Eroncel:2022vjg}%
  \BibitemOpen
  \bibfield  {author} {\bibinfo {author} {\bibfnamefont {C.}~\bibnamefont
  {Er\"oncel}}, \bibinfo {author} {\bibfnamefont {R.}~\bibnamefont {Sato}},
  \bibinfo {author} {\bibfnamefont {G.}~\bibnamefont {Servant}},\ and\ \bibinfo
  {author} {\bibfnamefont {P.}~\bibnamefont {S\o{}rensen}},\ }\bibfield
  {title} {\bibinfo {title} {{ALP dark matter from kinetic fragmentation:
  opening up the parameter window}},\ }\href
  {https://doi.org/10.1088/1475-7516/2022/10/053} {\bibfield  {journal}
  {\bibinfo  {journal} {JCAP}\ }\textbf {\bibinfo {volume} {10}},\ \bibinfo
  {pages} {053}},\ \Eprint {https://arxiv.org/abs/2206.14259} {arXiv:2206.14259
  [hep-ph]} \BibitemShut {NoStop}%
\bibitem [{\citenamefont {Prabhu}(2020)}]{Prabhu:2020pzm}%
  \BibitemOpen
  \bibfield  {author} {\bibinfo {author} {\bibfnamefont {A.}~\bibnamefont
  {Prabhu}},\ }\bibfield  {title} {\bibinfo {title} {{Optical Lensing by Axion
  Stars: Observational Prospects with Radio Astrometry}},\ }\href@noop {} {\
  (\bibinfo {year} {2020})},\ \Eprint {https://arxiv.org/abs/2006.10231}
  {arXiv:2006.10231 [astro-ph.CO]} \BibitemShut {NoStop}%
\bibitem [{\citenamefont {{Hobbs}}\ \emph {et~al.}(2010)\citenamefont
  {{Hobbs}}, \citenamefont {{Archibald}}, \citenamefont {{Arzoumanian}},
  \citenamefont {{Backer}}, \citenamefont {{Bailes}}, \citenamefont {{Bhat}},
  \citenamefont {{Burgay}}, \citenamefont {{Burke-Spolaor}}, \citenamefont
  {{Champion}}, \citenamefont {{Cognard}}, \citenamefont {{Coles}},
  \citenamefont {{Cordes}}, \citenamefont {{Demorest}}, \citenamefont
  {{Desvignes}}, \citenamefont {{Ferdman}}, \citenamefont {{Finn}},
  \citenamefont {{Freire}}, \citenamefont {{Gonzalez}}, \citenamefont
  {{Hessels}}, \citenamefont {{Hotan}}, \citenamefont {{Janssen}},
  \citenamefont {{Jenet}}, \citenamefont {{Jessner}}, \citenamefont {{Jordan}},
  \citenamefont {{Kaspi}}, \citenamefont {{Kramer}}, \citenamefont
  {{Kondratiev}}, \citenamefont {{Lazio}}, \citenamefont {{Lazaridis}},
  \citenamefont {{Lee}}, \citenamefont {{Levin}}, \citenamefont {{Lommen}},
  \citenamefont {{Lorimer}}, \citenamefont {{Lynch}}, \citenamefont {{Lyne}},
  \citenamefont {{Manchester}}, \citenamefont {{McLaughlin}}, \citenamefont
  {{Nice}}, \citenamefont {{Oslowski}}, \citenamefont {{Pilia}}, \citenamefont
  {{Possenti}}, \citenamefont {{Purver}}, \citenamefont {{Ransom}},
  \citenamefont {{Reynolds}}, \citenamefont {{Sanidas}}, \citenamefont
  {{Sarkissian}}, \citenamefont {{Sesana}}, \citenamefont {{Shannon}},
  \citenamefont {{Siemens}}, \citenamefont {{Stairs}}, \citenamefont
  {{Stappers}}, \citenamefont {{Stinebring}}, \citenamefont {{Theureau}},
  \citenamefont {{van Haasteren}}, \citenamefont {{van Straten}}, \citenamefont
  {{Verbiest}}, \citenamefont {{Yardley}},\ and\ \citenamefont
  {{You}}}]{2010CQGra..27h4013H}%
  \BibitemOpen
  \bibfield  {author} {\bibinfo {author} {\bibfnamefont {G.}~\bibnamefont
  {{Hobbs}}}, \bibinfo {author} {\bibfnamefont {A.}~\bibnamefont
  {{Archibald}}}, \bibinfo {author} {\bibfnamefont {Z.}~\bibnamefont
  {{Arzoumanian}}}, \bibinfo {author} {\bibfnamefont {D.}~\bibnamefont
  {{Backer}}}, \bibinfo {author} {\bibfnamefont {M.}~\bibnamefont {{Bailes}}},
  \bibinfo {author} {\bibfnamefont {N.~D.~R.}\ \bibnamefont {{Bhat}}}, \bibinfo
  {author} {\bibfnamefont {M.}~\bibnamefont {{Burgay}}}, \bibinfo {author}
  {\bibfnamefont {S.}~\bibnamefont {{Burke-Spolaor}}}, \bibinfo {author}
  {\bibfnamefont {D.}~\bibnamefont {{Champion}}}, \bibinfo {author}
  {\bibfnamefont {I.}~\bibnamefont {{Cognard}}}, \bibinfo {author}
  {\bibfnamefont {W.}~\bibnamefont {{Coles}}}, \bibinfo {author} {\bibfnamefont
  {J.}~\bibnamefont {{Cordes}}}, \bibinfo {author} {\bibfnamefont
  {P.}~\bibnamefont {{Demorest}}}, \bibinfo {author} {\bibfnamefont
  {G.}~\bibnamefont {{Desvignes}}}, \bibinfo {author} {\bibfnamefont {R.~D.}\
  \bibnamefont {{Ferdman}}}, \bibinfo {author} {\bibfnamefont {L.}~\bibnamefont
  {{Finn}}}, \bibinfo {author} {\bibfnamefont {P.}~\bibnamefont {{Freire}}},
  \bibinfo {author} {\bibfnamefont {M.}~\bibnamefont {{Gonzalez}}}, \bibinfo
  {author} {\bibfnamefont {J.}~\bibnamefont {{Hessels}}}, \bibinfo {author}
  {\bibfnamefont {A.}~\bibnamefont {{Hotan}}}, \bibinfo {author} {\bibfnamefont
  {G.}~\bibnamefont {{Janssen}}}, \bibinfo {author} {\bibfnamefont
  {F.}~\bibnamefont {{Jenet}}}, \bibinfo {author} {\bibfnamefont
  {A.}~\bibnamefont {{Jessner}}}, \bibinfo {author} {\bibfnamefont
  {C.}~\bibnamefont {{Jordan}}}, \bibinfo {author} {\bibfnamefont
  {V.}~\bibnamefont {{Kaspi}}}, \bibinfo {author} {\bibfnamefont
  {M.}~\bibnamefont {{Kramer}}}, \bibinfo {author} {\bibfnamefont
  {V.}~\bibnamefont {{Kondratiev}}}, \bibinfo {author} {\bibfnamefont
  {J.}~\bibnamefont {{Lazio}}}, \bibinfo {author} {\bibfnamefont
  {K.}~\bibnamefont {{Lazaridis}}}, \bibinfo {author} {\bibfnamefont {K.~J.}\
  \bibnamefont {{Lee}}}, \bibinfo {author} {\bibfnamefont {Y.}~\bibnamefont
  {{Levin}}}, \bibinfo {author} {\bibfnamefont {A.}~\bibnamefont {{Lommen}}},
  \bibinfo {author} {\bibfnamefont {D.}~\bibnamefont {{Lorimer}}}, \bibinfo
  {author} {\bibfnamefont {R.}~\bibnamefont {{Lynch}}}, \bibinfo {author}
  {\bibfnamefont {A.}~\bibnamefont {{Lyne}}}, \bibinfo {author} {\bibfnamefont
  {R.}~\bibnamefont {{Manchester}}}, \bibinfo {author} {\bibfnamefont
  {M.}~\bibnamefont {{McLaughlin}}}, \bibinfo {author} {\bibfnamefont
  {D.}~\bibnamefont {{Nice}}}, \bibinfo {author} {\bibfnamefont
  {S.}~\bibnamefont {{Oslowski}}}, \bibinfo {author} {\bibfnamefont
  {M.}~\bibnamefont {{Pilia}}}, \bibinfo {author} {\bibfnamefont
  {A.}~\bibnamefont {{Possenti}}}, \bibinfo {author} {\bibfnamefont
  {M.}~\bibnamefont {{Purver}}}, \bibinfo {author} {\bibfnamefont
  {S.}~\bibnamefont {{Ransom}}}, \bibinfo {author} {\bibfnamefont
  {J.}~\bibnamefont {{Reynolds}}}, \bibinfo {author} {\bibfnamefont
  {S.}~\bibnamefont {{Sanidas}}}, \bibinfo {author} {\bibfnamefont
  {J.}~\bibnamefont {{Sarkissian}}}, \bibinfo {author} {\bibfnamefont
  {A.}~\bibnamefont {{Sesana}}}, \bibinfo {author} {\bibfnamefont
  {R.}~\bibnamefont {{Shannon}}}, \bibinfo {author} {\bibfnamefont
  {X.}~\bibnamefont {{Siemens}}}, \bibinfo {author} {\bibfnamefont
  {I.}~\bibnamefont {{Stairs}}}, \bibinfo {author} {\bibfnamefont
  {B.}~\bibnamefont {{Stappers}}}, \bibinfo {author} {\bibfnamefont
  {D.}~\bibnamefont {{Stinebring}}}, \bibinfo {author} {\bibfnamefont
  {G.}~\bibnamefont {{Theureau}}}, \bibinfo {author} {\bibfnamefont
  {R.}~\bibnamefont {{van Haasteren}}}, \bibinfo {author} {\bibfnamefont
  {W.}~\bibnamefont {{van Straten}}}, \bibinfo {author} {\bibfnamefont
  {J.~P.~W.}\ \bibnamefont {{Verbiest}}}, \bibinfo {author} {\bibfnamefont
  {D.~R.~B.}\ \bibnamefont {{Yardley}}},\ and\ \bibinfo {author} {\bibfnamefont
  {X.~P.}\ \bibnamefont {{You}}},\ }\bibfield  {title} {\bibinfo {title} {{The
  International Pulsar Timing Array project: using pulsars as a gravitational
  wave detector}},\ }\href {https://doi.org/10.1088/0264-9381/27/8/084013}
  {\bibfield  {journal} {\bibinfo  {journal} {Classical and Quantum Gravity}\
  }\textbf {\bibinfo {volume} {27}},\ \bibinfo {eid} {084013} (\bibinfo {year}
  {2010})},\ \Eprint {https://arxiv.org/abs/0911.5206} {arXiv:0911.5206
  [astro-ph.SR]} \BibitemShut {NoStop}%
\bibitem [{\citenamefont {K\r{u}s}\ \emph {et~al.}(2024)\citenamefont
  {K\r{u}s}, \citenamefont {L\'opez~Nacir},\ and\ \citenamefont
  {Urban}}]{Kus:2024vpa}%
  \BibitemOpen
  \bibfield  {author} {\bibinfo {author} {\bibfnamefont {P.}~\bibnamefont
  {K\r{u}s}}, \bibinfo {author} {\bibfnamefont {D.}~\bibnamefont
  {L\'opez~Nacir}},\ and\ \bibinfo {author} {\bibfnamefont {F.~R.}\
  \bibnamefont {Urban}},\ }\bibfield  {title} {\bibinfo {title} {{Bayesian
  sensitivity of binary pulsars to ultra-light dark matter}},\ }\href@noop {}
  {\  (\bibinfo {year} {2024})},\ \Eprint {https://arxiv.org/abs/2402.04099}
  {arXiv:2402.04099 [astro-ph.HE]} \BibitemShut {NoStop}%
\bibitem [{\citenamefont {Sesana}\ \emph {et~al.}(2008)\citenamefont {Sesana},
  \citenamefont {Vecchio},\ and\ \citenamefont {Colacino}}]{Sesana:2008mz}%
  \BibitemOpen
  \bibfield  {author} {\bibinfo {author} {\bibfnamefont {A.}~\bibnamefont
  {Sesana}}, \bibinfo {author} {\bibfnamefont {A.}~\bibnamefont {Vecchio}},\
  and\ \bibinfo {author} {\bibfnamefont {C.~N.}\ \bibnamefont {Colacino}},\
  }\bibfield  {title} {\bibinfo {title} {{The stochastic gravitational-wave
  background from massive black hole binary systems: implications for
  observations with Pulsar Timing Arrays}},\ }\href
  {https://doi.org/10.1111/j.1365-2966.2008.13682.x} {\bibfield  {journal}
  {\bibinfo  {journal} {Mon. Not. Roy. Astron. Soc.}\ }\textbf {\bibinfo
  {volume} {390}},\ \bibinfo {pages} {192} (\bibinfo {year} {2008})},\ \Eprint
  {https://arxiv.org/abs/0804.4476} {arXiv:0804.4476 [astro-ph]} \BibitemShut
  {NoStop}%
\bibitem [{\citenamefont {Lower}\ \emph {et~al.}(2024)\citenamefont {Lower},
  \citenamefont {Dai}, \citenamefont {Johnston},\ and\ \citenamefont
  {Barr}}]{Lower:2024sdi}%
  \BibitemOpen
  \bibfield  {author} {\bibinfo {author} {\bibfnamefont {M.~E.}\ \bibnamefont
  {Lower}}, \bibinfo {author} {\bibfnamefont {S.}~\bibnamefont {Dai}}, \bibinfo
  {author} {\bibfnamefont {S.}~\bibnamefont {Johnston}},\ and\ \bibinfo
  {author} {\bibfnamefont {E.~D.}\ \bibnamefont {Barr}},\ }\bibfield  {title}
  {\bibinfo {title} {{A Millisecond Pulsar Binary Embedded in a Galactic Center
  Radio Filament}},\ }\href {https://doi.org/10.3847/2041-8213/ad4866}
  {\bibfield  {journal} {\bibinfo  {journal} {Astrophys. J. Lett.}\ }\textbf
  {\bibinfo {volume} {967}},\ \bibinfo {pages} {L16} (\bibinfo {year}
  {2024})},\ \Eprint {https://arxiv.org/abs/2404.09098} {arXiv:2404.09098
  [astro-ph.HE]} \BibitemShut {NoStop}%
\bibitem [{\citenamefont {Read}(2014)}]{Read:2014qva}%
  \BibitemOpen
  \bibfield  {author} {\bibinfo {author} {\bibfnamefont {J.~I.}\ \bibnamefont
  {Read}},\ }\bibfield  {title} {\bibinfo {title} {{The Local Dark Matter
  Density}},\ }\href {https://doi.org/10.1088/0954-3899/41/6/063101} {\bibfield
   {journal} {\bibinfo  {journal} {J. Phys. G}\ }\textbf {\bibinfo {volume}
  {41}},\ \bibinfo {pages} {063101} (\bibinfo {year} {2014})},\ \Eprint
  {https://arxiv.org/abs/1404.1938} {arXiv:1404.1938 [astro-ph.GA]}
  \BibitemShut {NoStop}%
\bibitem [{\citenamefont {{Bland-Hawthorn}}\ and\ \citenamefont
  {{Gerhard}}(2016)}]{Bland-Hawthorn:2016}%
  \BibitemOpen
  \bibfield  {author} {\bibinfo {author} {\bibfnamefont {J.}~\bibnamefont
  {{Bland-Hawthorn}}}\ and\ \bibinfo {author} {\bibfnamefont {O.}~\bibnamefont
  {{Gerhard}}},\ }\bibfield  {title} {\bibinfo {title} {{The Galaxy in Context:
  Structural, Kinematic, and Integrated Properties}},\ }\href
  {https://doi.org/10.1146/annurev-astro-081915-023441} {\bibfield  {journal}
  {\bibinfo  {journal} {\araa}\ }\textbf {\bibinfo {volume} {54}},\ \bibinfo
  {pages} {529} (\bibinfo {year} {2016})},\ \Eprint
  {https://arxiv.org/abs/1602.07702} {arXiv:1602.07702 [astro-ph.GA]}
  \BibitemShut {NoStop}%
\bibitem [{\citenamefont {Iocco}\ \emph {et~al.}(2015)\citenamefont {Iocco},
  \citenamefont {Pato},\ and\ \citenamefont {Bertone}}]{Iocco:2015xga}%
  \BibitemOpen
  \bibfield  {author} {\bibinfo {author} {\bibfnamefont {F.}~\bibnamefont
  {Iocco}}, \bibinfo {author} {\bibfnamefont {M.}~\bibnamefont {Pato}},\ and\
  \bibinfo {author} {\bibfnamefont {G.}~\bibnamefont {Bertone}},\ }\bibfield
  {title} {\bibinfo {title} {{Evidence for dark matter in the inner Milky
  Way}},\ }\href {https://doi.org/10.1038/nphys3237} {\bibfield  {journal}
  {\bibinfo  {journal} {Nature Phys.}\ }\textbf {\bibinfo {volume} {11}},\
  \bibinfo {pages} {245} (\bibinfo {year} {2015})},\ \Eprint
  {https://arxiv.org/abs/1502.03821} {arXiv:1502.03821 [astro-ph.GA]}
  \BibitemShut {NoStop}%
\bibitem [{\citenamefont {{Portail}}\ \emph {et~al.}(2017)\citenamefont
  {{Portail}}, \citenamefont {{Gerhard}}, \citenamefont {{Wegg}},\ and\
  \citenamefont {{Ness}}}]{Portail:2017}%
  \BibitemOpen
  \bibfield  {author} {\bibinfo {author} {\bibfnamefont {M.}~\bibnamefont
  {{Portail}}}, \bibinfo {author} {\bibfnamefont {O.}~\bibnamefont
  {{Gerhard}}}, \bibinfo {author} {\bibfnamefont {C.}~\bibnamefont {{Wegg}}},\
  and\ \bibinfo {author} {\bibfnamefont {M.}~\bibnamefont {{Ness}}},\
  }\bibfield  {title} {\bibinfo {title} {{Dynamical modelling of the galactic
  bulge and bar: the Milky Way's pattern speed, stellar and dark matter mass
  distribution}},\ }\href {https://doi.org/10.1093/mnras/stw2819} {\bibfield
  {journal} {\bibinfo  {journal} {\mnras}\ }\textbf {\bibinfo {volume} {465}},\
  \bibinfo {pages} {1621} (\bibinfo {year} {2017})},\ \Eprint
  {https://arxiv.org/abs/1608.07954} {arXiv:1608.07954 [astro-ph.GA]}
  \BibitemShut {NoStop}%
\bibitem [{\citenamefont {{Zoccali}}\ \emph {et~al.}(2014)\citenamefont
  {{Zoccali}}, \citenamefont {{Gonzalez}}, \citenamefont {{Vasquez}},
  \citenamefont {{Hill}}, \citenamefont {{Rejkuba}}, \citenamefont {{Valenti}},
  \citenamefont {{Renzini}}, \citenamefont {{Rojas-Arriagada}}, \citenamefont
  {{Martinez-Valpuesta}}, \citenamefont {{Babusiaux}}, \citenamefont {{Brown}},
  \citenamefont {{Minniti}},\ and\ \citenamefont {{McWilliam}}}]{Zoccali:2014}%
  \BibitemOpen
  \bibfield  {author} {\bibinfo {author} {\bibfnamefont {M.}~\bibnamefont
  {{Zoccali}}}, \bibinfo {author} {\bibfnamefont {O.~A.}\ \bibnamefont
  {{Gonzalez}}}, \bibinfo {author} {\bibfnamefont {S.}~\bibnamefont
  {{Vasquez}}}, \bibinfo {author} {\bibfnamefont {V.}~\bibnamefont {{Hill}}},
  \bibinfo {author} {\bibfnamefont {M.}~\bibnamefont {{Rejkuba}}}, \bibinfo
  {author} {\bibfnamefont {E.}~\bibnamefont {{Valenti}}}, \bibinfo {author}
  {\bibfnamefont {A.}~\bibnamefont {{Renzini}}}, \bibinfo {author}
  {\bibfnamefont {A.}~\bibnamefont {{Rojas-Arriagada}}}, \bibinfo {author}
  {\bibfnamefont {I.}~\bibnamefont {{Martinez-Valpuesta}}}, \bibinfo {author}
  {\bibfnamefont {C.}~\bibnamefont {{Babusiaux}}}, \bibinfo {author}
  {\bibfnamefont {T.}~\bibnamefont {{Brown}}}, \bibinfo {author} {\bibfnamefont
  {D.}~\bibnamefont {{Minniti}}},\ and\ \bibinfo {author} {\bibfnamefont
  {A.}~\bibnamefont {{McWilliam}}},\ }\bibfield  {title} {\bibinfo {title}
  {{The GIRAFFE Inner Bulge Survey (GIBS). I. Survey description and a
  kinematical map of the Milky Way bulge}},\ }\href
  {https://doi.org/10.1051/0004-6361/201323120} {\bibfield  {journal} {\bibinfo
   {journal} {\aap}\ }\textbf {\bibinfo {volume} {562}},\ \bibinfo {eid} {A66}
  (\bibinfo {year} {2014})},\ \Eprint {https://arxiv.org/abs/1401.4878}
  {arXiv:1401.4878 [astro-ph.GA]} \BibitemShut {NoStop}%
\bibitem [{\citenamefont {{Launhardt}}\ \emph {et~al.}(2002)\citenamefont
  {{Launhardt}}, \citenamefont {{Zylka}},\ and\ \citenamefont
  {{Mezger}}}]{Launhardt:2002}%
  \BibitemOpen
  \bibfield  {author} {\bibinfo {author} {\bibfnamefont {R.}~\bibnamefont
  {{Launhardt}}}, \bibinfo {author} {\bibfnamefont {R.}~\bibnamefont
  {{Zylka}}},\ and\ \bibinfo {author} {\bibfnamefont {P.~G.}\ \bibnamefont
  {{Mezger}}},\ }\bibfield  {title} {\bibinfo {title} {{The nuclear bulge of
  the Galaxy. III. Large-scale physical characteristics of stars and
  interstellar matter}},\ }\href {https://doi.org/10.1051/0004-6361:20020017}
  {\bibfield  {journal} {\bibinfo  {journal} {\aap}\ }\textbf {\bibinfo
  {volume} {384}},\ \bibinfo {pages} {112} (\bibinfo {year} {2002})},\ \Eprint
  {https://arxiv.org/abs/astro-ph/0201294} {arXiv:astro-ph/0201294 [astro-ph]}
  \BibitemShut {NoStop}%
\bibitem [{\citenamefont {{Sch{\"o}nrich}}\ \emph {et~al.}(2015)\citenamefont
  {{Sch{\"o}nrich}}, \citenamefont {{Aumer}},\ and\ \citenamefont
  {{Sale}}}]{Schonrich:2015}%
  \BibitemOpen
  \bibfield  {author} {\bibinfo {author} {\bibfnamefont {R.}~\bibnamefont
  {{Sch{\"o}nrich}}}, \bibinfo {author} {\bibfnamefont {M.}~\bibnamefont
  {{Aumer}}},\ and\ \bibinfo {author} {\bibfnamefont {S.~E.}\ \bibnamefont
  {{Sale}}},\ }\bibfield  {title} {\bibinfo {title} {{Kinematic Detection of
  the Galactic Nuclear Disk}},\ }\href
  {https://doi.org/10.1088/2041-8205/812/2/L21} {\bibfield  {journal} {\bibinfo
   {journal} {\apjl}\ }\textbf {\bibinfo {volume} {812}},\ \bibinfo {eid} {L21}
  (\bibinfo {year} {2015})},\ \Eprint {https://arxiv.org/abs/1507.02695}
  {arXiv:1507.02695 [astro-ph.GA]} \BibitemShut {NoStop}%
\bibitem [{\citenamefont {{Wegg}}\ \emph {et~al.}(2017)\citenamefont {{Wegg}},
  \citenamefont {{Gerhard}},\ and\ \citenamefont {{Portail}}}]{Wegg:2017}%
  \BibitemOpen
  \bibfield  {author} {\bibinfo {author} {\bibfnamefont {C.}~\bibnamefont
  {{Wegg}}}, \bibinfo {author} {\bibfnamefont {O.}~\bibnamefont {{Gerhard}}},\
  and\ \bibinfo {author} {\bibfnamefont {M.}~\bibnamefont {{Portail}}},\
  }\bibfield  {title} {\bibinfo {title} {{The Initial Mass Function of the
  Inner Galaxy Measured from OGLE-III Microlensing Timescales}},\ }\href
  {https://doi.org/10.3847/2041-8213/aa794e} {\bibfield  {journal} {\bibinfo
  {journal} {\apjl}\ }\textbf {\bibinfo {volume} {843}},\ \bibinfo {eid} {L5}
  (\bibinfo {year} {2017})},\ \Eprint {https://arxiv.org/abs/1706.04193}
  {arXiv:1706.04193 [astro-ph.GA]} \BibitemShut {NoStop}%
\bibitem [{\citenamefont {De~Martino}\ \emph {et~al.}(2020)\citenamefont
  {De~Martino}, \citenamefont {Broadhurst}, \citenamefont {Tye}, \citenamefont
  {Chiueh},\ and\ \citenamefont {Schive}}]{DeMartino:2018zkx}%
  \BibitemOpen
  \bibfield  {author} {\bibinfo {author} {\bibfnamefont {I.}~\bibnamefont
  {De~Martino}}, \bibinfo {author} {\bibfnamefont {T.}~\bibnamefont
  {Broadhurst}}, \bibinfo {author} {\bibfnamefont {S.~H.~H.}\ \bibnamefont
  {Tye}}, \bibinfo {author} {\bibfnamefont {T.}~\bibnamefont {Chiueh}},\ and\
  \bibinfo {author} {\bibfnamefont {H.-Y.}\ \bibnamefont {Schive}},\ }\bibfield
   {title} {\bibinfo {title} {{Dynamical Evidence of a Solitonic Core of
  $10^{9}M_\odot$ in the Milky Way}},\ }\href
  {https://doi.org/10.1016/j.dark.2020.100503} {\bibfield  {journal} {\bibinfo
  {journal} {Phys. Dark Univ.}\ }\textbf {\bibinfo {volume} {28}},\ \bibinfo
  {pages} {100503} (\bibinfo {year} {2020})},\ \Eprint
  {https://arxiv.org/abs/1807.08153} {arXiv:1807.08153 [astro-ph.GA]}
  \BibitemShut {NoStop}%
\bibitem [{\citenamefont {Li}\ \emph {et~al.}(2020)\citenamefont {Li},
  \citenamefont {Shen},\ and\ \citenamefont {Schive}}]{Li:2020qva}%
  \BibitemOpen
  \bibfield  {author} {\bibinfo {author} {\bibfnamefont {Z.}~\bibnamefont
  {Li}}, \bibinfo {author} {\bibfnamefont {J.}~\bibnamefont {Shen}},\ and\
  \bibinfo {author} {\bibfnamefont {H.-Y.}\ \bibnamefont {Schive}},\ }\bibfield
   {title} {\bibinfo {title} {{Testing the Prediction of Fuzzy Dark Matter
  Theory in the Milky Way Center}}\ }\href
  {https://doi.org/10.3847/1538-4357/ab6598} {10.3847/1538-4357/ab6598}
  (\bibinfo {year} {2020}),\ \Eprint {https://arxiv.org/abs/2001.00318}
  {arXiv:2001.00318 [astro-ph.GA]} \BibitemShut {NoStop}%
\bibitem [{\citenamefont {Ellis}\ \emph {et~al.}(2012)\citenamefont {Ellis},
  \citenamefont {Jenet},\ and\ \citenamefont {McLaughlin}}]{Ellis:2012in}%
  \BibitemOpen
  \bibfield  {author} {\bibinfo {author} {\bibfnamefont {J.~A.}\ \bibnamefont
  {Ellis}}, \bibinfo {author} {\bibfnamefont {F.~A.}\ \bibnamefont {Jenet}},\
  and\ \bibinfo {author} {\bibfnamefont {M.~A.}\ \bibnamefont {McLaughlin}},\
  }\bibfield  {title} {\bibinfo {title} {{Practical Methods for Continuous
  Gravitational Wave Detection using Pulsar Timing Data}},\ }\href
  {https://doi.org/10.1088/0004-637X/753/2/96} {\bibfield  {journal} {\bibinfo
  {journal} {Astrophys. J.}\ }\textbf {\bibinfo {volume} {753}},\ \bibinfo
  {pages} {96} (\bibinfo {year} {2012})},\ \Eprint
  {https://arxiv.org/abs/1202.0808} {arXiv:1202.0808 [astro-ph.IM]}
  \BibitemShut {NoStop}%
\bibitem [{\citenamefont {Siemens}\ \emph {et~al.}(2013)\citenamefont
  {Siemens}, \citenamefont {Ellis}, \citenamefont {Jenet},\ and\ \citenamefont
  {Romano}}]{Siemens:2013zla}%
  \BibitemOpen
  \bibfield  {author} {\bibinfo {author} {\bibfnamefont {X.}~\bibnamefont
  {Siemens}}, \bibinfo {author} {\bibfnamefont {J.}~\bibnamefont {Ellis}},
  \bibinfo {author} {\bibfnamefont {F.}~\bibnamefont {Jenet}},\ and\ \bibinfo
  {author} {\bibfnamefont {J.~D.}\ \bibnamefont {Romano}},\ }\bibfield  {title}
  {\bibinfo {title} {{The stochastic background: scaling laws and time to
  detection for pulsar timing arrays}},\ }\href
  {https://doi.org/10.1088/0264-9381/30/22/224015} {\bibfield  {journal}
  {\bibinfo  {journal} {Class. Quant. Grav.}\ }\textbf {\bibinfo {volume}
  {30}},\ \bibinfo {pages} {224015} (\bibinfo {year} {2013})},\ \Eprint
  {https://arxiv.org/abs/1305.3196} {arXiv:1305.3196 [astro-ph.IM]}
  \BibitemShut {NoStop}%
\bibitem [{\citenamefont {Perrodin}\ and\ \citenamefont
  {Sesana}(2018)}]{Perrodin:2017bxr}%
  \BibitemOpen
  \bibfield  {author} {\bibinfo {author} {\bibfnamefont {D.}~\bibnamefont
  {Perrodin}}\ and\ \bibinfo {author} {\bibfnamefont {A.}~\bibnamefont
  {Sesana}},\ }\bibfield  {title} {\bibinfo {title} {{Radio pulsars: testing
  gravity and detecting gravitational waves}},\ }\href
  {https://doi.org/10.1007/978-3-319-97616-7_3} {\bibfield  {journal} {\bibinfo
   {journal} {Astrophys. Space Sci. Libr.}\ }\textbf {\bibinfo {volume}
  {457}},\ \bibinfo {pages} {95} (\bibinfo {year} {2018})},\ \Eprint
  {https://arxiv.org/abs/1709.02816} {arXiv:1709.02816 [astro-ph.HE]}
  \BibitemShut {NoStop}%
\bibitem [{\citenamefont {Colpi}\ \emph {et~al.}(2024)\citenamefont {Colpi}
  \emph {et~al.}}]{Colpi:2024xhw}%
  \BibitemOpen
  \bibfield  {author} {\bibinfo {author} {\bibfnamefont {M.}~\bibnamefont
  {Colpi}} \emph {et~al.},\ }\bibfield  {title} {\bibinfo {title} {{LISA
  Definition Study Report}},\ }\href@noop {} {\  (\bibinfo {year} {2024})},\
  \Eprint {https://arxiv.org/abs/2402.07571} {arXiv:2402.07571 [astro-ph.CO]}
  \BibitemShut {NoStop}%
\bibitem [{\citenamefont {Gong}\ \emph {et~al.}(2021)\citenamefont {Gong},
  \citenamefont {Luo},\ and\ \citenamefont {Wang}}]{Gong:2021gvw}%
  \BibitemOpen
  \bibfield  {author} {\bibinfo {author} {\bibfnamefont {Y.}~\bibnamefont
  {Gong}}, \bibinfo {author} {\bibfnamefont {J.}~\bibnamefont {Luo}},\ and\
  \bibinfo {author} {\bibfnamefont {B.}~\bibnamefont {Wang}},\ }\bibfield
  {title} {\bibinfo {title} {{Concepts and status of Chinese space
  gravitational wave detection projects}},\ }\href
  {https://doi.org/10.1038/s41550-021-01480-3} {\bibfield  {journal} {\bibinfo
  {journal} {Nature Astron.}\ }\textbf {\bibinfo {volume} {5}},\ \bibinfo
  {pages} {881} (\bibinfo {year} {2021})},\ \Eprint
  {https://arxiv.org/abs/2109.07442} {arXiv:2109.07442 [astro-ph.IM]}
  \BibitemShut {NoStop}%
\bibitem [{\citenamefont {Korol}\ \emph {et~al.}(2019)\citenamefont {Korol},
  \citenamefont {Rossi},\ and\ \citenamefont {Barausse}}]{Korol:2018wep}%
  \BibitemOpen
  \bibfield  {author} {\bibinfo {author} {\bibfnamefont {V.}~\bibnamefont
  {Korol}}, \bibinfo {author} {\bibfnamefont {E.~M.}\ \bibnamefont {Rossi}},\
  and\ \bibinfo {author} {\bibfnamefont {E.}~\bibnamefont {Barausse}},\
  }\bibfield  {title} {\bibinfo {title} {{A multimessenger study of the Milky
  Way\textquoteright{}s stellar disc and bulge with LISA, Gaia, and LSST}},\
  }\href {https://doi.org/10.1093/mnras/sty3440} {\bibfield  {journal}
  {\bibinfo  {journal} {Mon. Not. Roy. Astron. Soc.}\ }\textbf {\bibinfo
  {volume} {483}},\ \bibinfo {pages} {5518} (\bibinfo {year} {2019})},\ \Eprint
  {https://arxiv.org/abs/1806.03306} {arXiv:1806.03306 [astro-ph.GA]}
  \BibitemShut {NoStop}%
\bibitem [{\citenamefont {Korol}\ \emph {et~al.}(2020)\citenamefont {Korol}
  \emph {et~al.}}]{Korol:2020lpq}%
  \BibitemOpen
  \bibfield  {author} {\bibinfo {author} {\bibfnamefont {V.}~\bibnamefont
  {Korol}} \emph {et~al.},\ }\bibfield  {title} {\bibinfo {title} {{Populations
  of double white dwarfs in Milky Way satellites and their detectability with
  LISA}},\ }\href {https://doi.org/10.1051/0004-6361/202037764} {\bibfield
  {journal} {\bibinfo  {journal} {Astron. Astrophys.}\ }\textbf {\bibinfo
  {volume} {638}},\ \bibinfo {pages} {A153} (\bibinfo {year} {2020})},\ \Eprint
  {https://arxiv.org/abs/2002.10462} {arXiv:2002.10462 [astro-ph.GA]}
  \BibitemShut {NoStop}%
\bibitem [{\citenamefont {Korol}\ \emph {et~al.}(2022)\citenamefont {Korol},
  \citenamefont {Hallakoun}, \citenamefont {Toonen},\ and\ \citenamefont
  {Karnesis}}]{Korol:2021pun}%
  \BibitemOpen
  \bibfield  {author} {\bibinfo {author} {\bibfnamefont {V.}~\bibnamefont
  {Korol}}, \bibinfo {author} {\bibfnamefont {N.}~\bibnamefont {Hallakoun}},
  \bibinfo {author} {\bibfnamefont {S.}~\bibnamefont {Toonen}},\ and\ \bibinfo
  {author} {\bibfnamefont {N.}~\bibnamefont {Karnesis}},\ }\bibfield  {title}
  {\bibinfo {title} {{Observationally driven Galactic double white dwarf
  population for LISA}},\ }\href {https://doi.org/10.1093/mnras/stac415}
  {\bibfield  {journal} {\bibinfo  {journal} {Mon. Not. Roy. Astron. Soc.}\
  }\textbf {\bibinfo {volume} {511}},\ \bibinfo {pages} {5936} (\bibinfo {year}
  {2022})},\ \Eprint {https://arxiv.org/abs/2109.10972} {arXiv:2109.10972
  [astro-ph.HE]} \BibitemShut {NoStop}%
\bibitem [{\citenamefont {Strokov}\ and\ \citenamefont
  {Berti}(2023)}]{Strokov:2023ypy}%
  \BibitemOpen
  \bibfield  {author} {\bibinfo {author} {\bibfnamefont {V.}~\bibnamefont
  {Strokov}}\ and\ \bibinfo {author} {\bibfnamefont {E.}~\bibnamefont
  {Berti}},\ }\bibfield  {title} {\bibinfo {title} {{Quasimonochromatic LISA
  Sources in the Frequency Domain}},\ }\href@noop {} {\  (\bibinfo {year}
  {2023})},\ \Eprint {https://arxiv.org/abs/2312.00121} {arXiv:2312.00121
  [gr-qc]} \BibitemShut {NoStop}%
\bibitem [{\citenamefont {Ebadi}\ \emph {et~al.}(2024)\citenamefont {Ebadi},
  \citenamefont {Strokov}, \citenamefont {Tanin}, \citenamefont {Berti},\ and\
  \citenamefont {Walsworth}}]{Ebadi:2024oaq}%
  \BibitemOpen
  \bibfield  {author} {\bibinfo {author} {\bibfnamefont {R.}~\bibnamefont
  {Ebadi}}, \bibinfo {author} {\bibfnamefont {V.}~\bibnamefont {Strokov}},
  \bibinfo {author} {\bibfnamefont {E.~H.}\ \bibnamefont {Tanin}}, \bibinfo
  {author} {\bibfnamefont {E.}~\bibnamefont {Berti}},\ and\ \bibinfo {author}
  {\bibfnamefont {R.~L.}\ \bibnamefont {Walsworth}},\ }\bibfield  {title}
  {\bibinfo {title} {{LISA double white dwarf binaries as Galactic
  accelerometers}},\ }\href@noop {} {\  (\bibinfo {year} {2024})},\ \Eprint
  {https://arxiv.org/abs/2405.13109} {arXiv:2405.13109 [gr-qc]} \BibitemShut
  {NoStop}%
\bibitem [{\citenamefont {{Maoz}}\ \emph {et~al.}(2018)\citenamefont {{Maoz}},
  \citenamefont {{Hallakoun}},\ and\ \citenamefont {{Badenes}}}]{Maoz:2018}%
  \BibitemOpen
  \bibfield  {author} {\bibinfo {author} {\bibfnamefont {D.}~\bibnamefont
  {{Maoz}}}, \bibinfo {author} {\bibfnamefont {N.}~\bibnamefont
  {{Hallakoun}}},\ and\ \bibinfo {author} {\bibfnamefont {C.}~\bibnamefont
  {{Badenes}}},\ }\bibfield  {title} {\bibinfo {title} {{The separation
  distribution and merger rate of double white dwarfs: improved constraints}},\
  }\href {https://doi.org/10.1093/mnras/sty339} {\bibfield  {journal} {\bibinfo
   {journal} {\mnras}\ }\textbf {\bibinfo {volume} {476}},\ \bibinfo {pages}
  {2584} (\bibinfo {year} {2018})},\ \Eprint {https://arxiv.org/abs/1801.04275}
  {arXiv:1801.04275 [astro-ph.SR]} \BibitemShut {NoStop}%
\bibitem [{\citenamefont {Sesana}\ \emph {et~al.}(2021)\citenamefont {Sesana}
  \emph {et~al.}}]{Sesana:2019vho}%
  \BibitemOpen
  \bibfield  {author} {\bibinfo {author} {\bibfnamefont {A.}~\bibnamefont
  {Sesana}} \emph {et~al.},\ }\bibfield  {title} {\bibinfo {title} {{Unveiling
  the gravitational universe at $\mu$-Hz frequencies}},\ }\href
  {https://doi.org/10.1007/s10686-021-09709-9} {\bibfield  {journal} {\bibinfo
  {journal} {Exper. Astron.}\ }\textbf {\bibinfo {volume} {51}},\ \bibinfo
  {pages} {1333} (\bibinfo {year} {2021})},\ \Eprint
  {https://arxiv.org/abs/1908.11391} {arXiv:1908.11391 [astro-ph.IM]}
  \BibitemShut {NoStop}%
\bibitem [{\citenamefont {Wilhelm}\ \emph {et~al.}(2020)\citenamefont
  {Wilhelm}, \citenamefont {Korol}, \citenamefont {Rossi},\ and\ \citenamefont
  {D’Onghia}}]{Wilhelm_2020}%
  \BibitemOpen
  \bibfield  {author} {\bibinfo {author} {\bibfnamefont {M.~J.~C.}\
  \bibnamefont {Wilhelm}}, \bibinfo {author} {\bibfnamefont {V.}~\bibnamefont
  {Korol}}, \bibinfo {author} {\bibfnamefont {E.~M.}\ \bibnamefont {Rossi}},\
  and\ \bibinfo {author} {\bibfnamefont {E.}~\bibnamefont {D’Onghia}},\
  }\bibfield  {title} {\bibinfo {title} {The milky way’s bar structural
  properties from gravitational waves},\ }\href
  {https://doi.org/10.1093/mnras/staa3457} {\bibfield  {journal} {\bibinfo
  {journal} {Monthly Notices of the Royal Astronomical Society}\ }\textbf
  {\bibinfo {volume} {500}},\ \bibinfo {pages} {4958–4971} (\bibinfo {year}
  {2020})}\BibitemShut {NoStop}%
\bibitem [{\citenamefont {Korol}\ \emph {et~al.}(2023)\citenamefont {Korol},
  \citenamefont {Igoshev}, \citenamefont {Toonen}, \citenamefont {Karnesis},
  \citenamefont {Moore}, \citenamefont {Finch},\ and\ \citenamefont
  {Klein}}]{Korol:2023jfz}%
  \BibitemOpen
  \bibfield  {author} {\bibinfo {author} {\bibfnamefont {V.}~\bibnamefont
  {Korol}}, \bibinfo {author} {\bibfnamefont {A.~P.}\ \bibnamefont {Igoshev}},
  \bibinfo {author} {\bibfnamefont {S.}~\bibnamefont {Toonen}}, \bibinfo
  {author} {\bibfnamefont {N.}~\bibnamefont {Karnesis}}, \bibinfo {author}
  {\bibfnamefont {C.~J.}\ \bibnamefont {Moore}}, \bibinfo {author}
  {\bibfnamefont {E.}~\bibnamefont {Finch}},\ and\ \bibinfo {author}
  {\bibfnamefont {A.}~\bibnamefont {Klein}},\ }\bibfield  {title} {\bibinfo
  {title} {{Neutron Star - White Dwarf Binaries: Probing Formation Pathways and
  Natal Kicks with LISA}},\ }\href@noop {} {\  (\bibinfo {year} {2023})},\
  \Eprint {https://arxiv.org/abs/2310.06559} {arXiv:2310.06559 [astro-ph.HE]}
  \BibitemShut {NoStop}%
\bibitem [{\citenamefont {Nelemans}\ \emph {et~al.}(2001)\citenamefont
  {Nelemans}, \citenamefont {Yungelson},\ and\ \citenamefont
  {Portegies~Zwart}}]{Nelemans:2001hp}%
  \BibitemOpen
  \bibfield  {author} {\bibinfo {author} {\bibfnamefont {G.}~\bibnamefont
  {Nelemans}}, \bibinfo {author} {\bibfnamefont {L.~R.}\ \bibnamefont
  {Yungelson}},\ and\ \bibinfo {author} {\bibfnamefont {S.~F.}\ \bibnamefont
  {Portegies~Zwart}},\ }\bibfield  {title} {\bibinfo {title} {{The
  gravitational wave signal from the galactic disk population of binaries
  containing two compact objects}},\ }\href
  {https://doi.org/10.1051/0004-6361:20010683} {\bibfield  {journal} {\bibinfo
  {journal} {Astron. Astrophys.}\ }\textbf {\bibinfo {volume} {375}},\ \bibinfo
  {pages} {890} (\bibinfo {year} {2001})},\ \Eprint
  {https://arxiv.org/abs/astro-ph/0105221} {arXiv:astro-ph/0105221}
  \BibitemShut {NoStop}%
\bibitem [{\citenamefont {Amaro-Seoane}(2019)}]{Amaro-Seoane:2019umn}%
  \BibitemOpen
  \bibfield  {author} {\bibinfo {author} {\bibfnamefont {P.}~\bibnamefont
  {Amaro-Seoane}},\ }\bibfield  {title} {\bibinfo {title} {{Extremely large
  mass-ratio inspirals}},\ }\href {https://doi.org/10.1103/PhysRevD.99.123025}
  {\bibfield  {journal} {\bibinfo  {journal} {Phys. Rev. D}\ }\textbf {\bibinfo
  {volume} {99}},\ \bibinfo {pages} {123025} (\bibinfo {year} {2019})},\
  \Eprint {https://arxiv.org/abs/1903.10871} {arXiv:1903.10871 [astro-ph.GA]}
  \BibitemShut {NoStop}%
\bibitem [{\citenamefont {Gourgoulhon}\ \emph {et~al.}(2019)\citenamefont
  {Gourgoulhon}, \citenamefont {Le~Tiec}, \citenamefont {Vincent},\ and\
  \citenamefont {Warburton}}]{Gourgoulhon:2019iyu}%
  \BibitemOpen
  \bibfield  {author} {\bibinfo {author} {\bibfnamefont {E.}~\bibnamefont
  {Gourgoulhon}}, \bibinfo {author} {\bibfnamefont {A.}~\bibnamefont
  {Le~Tiec}}, \bibinfo {author} {\bibfnamefont {F.~H.}\ \bibnamefont
  {Vincent}},\ and\ \bibinfo {author} {\bibfnamefont {N.}~\bibnamefont
  {Warburton}},\ }\bibfield  {title} {\bibinfo {title} {{Gravitational waves
  from bodies orbiting the Galactic Center black hole and their detectability
  by LISA}},\ }\href {https://doi.org/10.1051/0004-6361/201935406} {\bibfield
  {journal} {\bibinfo  {journal} {Astron. Astrophys.}\ }\textbf {\bibinfo
  {volume} {627}},\ \bibinfo {pages} {A92} (\bibinfo {year} {2019})},\ \Eprint
  {https://arxiv.org/abs/1903.02049} {arXiv:1903.02049 [gr-qc]} \BibitemShut
  {NoStop}%
\bibitem [{\citenamefont {V\'azquez-Aceves}\ \emph {et~al.}(2022)\citenamefont
  {V\'azquez-Aceves}, \citenamefont {Zwick}, \citenamefont {Bortolas},
  \citenamefont {Capelo}, \citenamefont {Amaro-Seoane}, \citenamefont {Mayer},\
  and\ \citenamefont {Chen}}]{Vazquez-Aceves:2021xwl}%
  \BibitemOpen
  \bibfield  {author} {\bibinfo {author} {\bibfnamefont {V.}~\bibnamefont
  {V\'azquez-Aceves}}, \bibinfo {author} {\bibfnamefont {L.}~\bibnamefont
  {Zwick}}, \bibinfo {author} {\bibfnamefont {E.}~\bibnamefont {Bortolas}},
  \bibinfo {author} {\bibfnamefont {P.~R.}\ \bibnamefont {Capelo}}, \bibinfo
  {author} {\bibfnamefont {P.}~\bibnamefont {Amaro-Seoane}}, \bibinfo {author}
  {\bibfnamefont {L.}~\bibnamefont {Mayer}},\ and\ \bibinfo {author}
  {\bibfnamefont {X.}~\bibnamefont {Chen}},\ }\bibfield  {title} {\bibinfo
  {title} {{Revised event rates for extreme and extremely large mass-ratio
  inspirals}},\ }\href {https://doi.org/10.1093/mnras/stab3485} {\bibfield
  {journal} {\bibinfo  {journal} {Mon. Not. Roy. Astron. Soc.}\ }\textbf
  {\bibinfo {volume} {510}},\ \bibinfo {pages} {2379} (\bibinfo {year}
  {2022})},\ \Eprint {https://arxiv.org/abs/2108.00135} {arXiv:2108.00135
  [astro-ph.GA]} \BibitemShut {NoStop}%
\bibitem [{\citenamefont {Piccinni}(2022)}]{Piccinni:2022vsd}%
  \BibitemOpen
  \bibfield  {author} {\bibinfo {author} {\bibfnamefont {O.~J.}\ \bibnamefont
  {Piccinni}},\ }\bibfield  {title} {\bibinfo {title} {{Status and Perspectives
  of Continuous Gravitational Wave Searches}},\ }\href
  {https://doi.org/10.3390/galaxies10030072} {\bibfield  {journal} {\bibinfo
  {journal} {Galaxies}\ }\textbf {\bibinfo {volume} {10}},\ \bibinfo {pages}
  {72} (\bibinfo {year} {2022})},\ \Eprint {https://arxiv.org/abs/2202.01088}
  {arXiv:2202.01088 [gr-qc]} \BibitemShut {NoStop}%
\bibitem [{\citenamefont {Gittins}(2024)}]{Gittins:2024zbg}%
  \BibitemOpen
  \bibfield  {author} {\bibinfo {author} {\bibfnamefont {F.}~\bibnamefont
  {Gittins}},\ }\bibfield  {title} {\bibinfo {title} {{Gravitational waves from
  neutron-star mountains}},\ }\href {https://doi.org/10.1088/1361-6382/ad1c35}
  {\bibfield  {journal} {\bibinfo  {journal} {Class. Quant. Grav.}\ }\textbf
  {\bibinfo {volume} {41}},\ \bibinfo {pages} {043001} (\bibinfo {year}
  {2024})},\ \Eprint {https://arxiv.org/abs/2401.01670} {arXiv:2401.01670
  [gr-qc]} \BibitemShut {NoStop}%
\bibitem [{\citenamefont {Cie\'slar}\ \emph {et~al.}(2021)\citenamefont
  {Cie\'slar}, \citenamefont {Bulik}, \citenamefont {Cury\l{}o}, \citenamefont
  {Sieniawska}, \citenamefont {Singh},\ and\ \citenamefont
  {Bejger}}]{Cieslar:2021viw}%
  \BibitemOpen
  \bibfield  {author} {\bibinfo {author} {\bibfnamefont {M.}~\bibnamefont
  {Cie\'slar}}, \bibinfo {author} {\bibfnamefont {T.}~\bibnamefont {Bulik}},
  \bibinfo {author} {\bibfnamefont {M.}~\bibnamefont {Cury\l{}o}}, \bibinfo
  {author} {\bibfnamefont {M.}~\bibnamefont {Sieniawska}}, \bibinfo {author}
  {\bibfnamefont {N.}~\bibnamefont {Singh}},\ and\ \bibinfo {author}
  {\bibfnamefont {M.}~\bibnamefont {Bejger}},\ }\bibfield  {title} {\bibinfo
  {title} {{Detectability of continuous gravitational waves from isolated
  neutron stars in the Milky Way - The population synthesis approach}},\ }\href
  {https://doi.org/10.1051/0004-6361/202039503} {\bibfield  {journal} {\bibinfo
   {journal} {Astron. Astrophys.}\ }\textbf {\bibinfo {volume} {649}},\
  \bibinfo {pages} {A92} (\bibinfo {year} {2021})},\ \Eprint
  {https://arxiv.org/abs/2102.08854} {arXiv:2102.08854 [gr-qc]} \BibitemShut
  {NoStop}%
\bibitem [{\citenamefont {Pagliaro}\ \emph {et~al.}(2023)\citenamefont
  {Pagliaro}, \citenamefont {Papa}, \citenamefont {Ming}, \citenamefont {Lian},
  \citenamefont {Tsuna}, \citenamefont {Maraston},\ and\ \citenamefont
  {Thomas}}]{Pagliaro:2023bvi}%
  \BibitemOpen
  \bibfield  {author} {\bibinfo {author} {\bibfnamefont {G.}~\bibnamefont
  {Pagliaro}}, \bibinfo {author} {\bibfnamefont {M.~A.}\ \bibnamefont {Papa}},
  \bibinfo {author} {\bibfnamefont {J.}~\bibnamefont {Ming}}, \bibinfo {author}
  {\bibfnamefont {J.}~\bibnamefont {Lian}}, \bibinfo {author} {\bibfnamefont
  {D.}~\bibnamefont {Tsuna}}, \bibinfo {author} {\bibfnamefont
  {C.}~\bibnamefont {Maraston}},\ and\ \bibinfo {author} {\bibfnamefont
  {D.}~\bibnamefont {Thomas}},\ }\bibfield  {title} {\bibinfo {title}
  {{Continuous Gravitational Waves from Galactic Neutron Stars: Demography,
  Detectability, and Prospects}},\ }\href
  {https://doi.org/10.3847/1538-4357/acd76f} {\bibfield  {journal} {\bibinfo
  {journal} {Astrophys. J.}\ }\textbf {\bibinfo {volume} {952}},\ \bibinfo
  {pages} {123} (\bibinfo {year} {2023})},\ \Eprint
  {https://arxiv.org/abs/2303.04714} {arXiv:2303.04714 [gr-qc]} \BibitemShut
  {NoStop}%
\bibitem [{\citenamefont {Branchesi}\ \emph {et~al.}(2023)\citenamefont
  {Branchesi} \emph {et~al.}}]{Branchesi:2023mws}%
  \BibitemOpen
  \bibfield  {author} {\bibinfo {author} {\bibfnamefont {M.}~\bibnamefont
  {Branchesi}} \emph {et~al.},\ }\bibfield  {title} {\bibinfo {title} {{Science
  with the Einstein Telescope: a comparison of different designs}},\ }\href
  {https://doi.org/10.1088/1475-7516/2023/07/068} {\bibfield  {journal}
  {\bibinfo  {journal} {JCAP}\ }\textbf {\bibinfo {volume} {07}},\ \bibinfo
  {pages} {068}},\ \Eprint {https://arxiv.org/abs/2303.15923} {arXiv:2303.15923
  [gr-qc]} \BibitemShut {NoStop}%
\bibitem [{\citenamefont {Maggiore}\ \emph
  {et~al.}(2020{\natexlab{a}})\citenamefont {Maggiore} \emph
  {et~al.}}]{Maggiore:2019uih}%
  \BibitemOpen
  \bibfield  {author} {\bibinfo {author} {\bibfnamefont {M.}~\bibnamefont
  {Maggiore}} \emph {et~al.},\ }\bibfield  {title} {\bibinfo {title} {{Science
  Case for the Einstein Telescope}},\ }\href
  {https://doi.org/10.1088/1475-7516/2020/03/050} {\bibfield  {journal}
  {\bibinfo  {journal} {JCAP}\ }\textbf {\bibinfo {volume} {03}},\ \bibinfo
  {pages} {050}},\ \Eprint {https://arxiv.org/abs/1912.02622} {arXiv:1912.02622
  [astro-ph.CO]} \BibitemShut {NoStop}%
\bibitem [{\citenamefont {Maggiore}\ \emph
  {et~al.}(2020{\natexlab{b}})\citenamefont {Maggiore}, \citenamefont {Broeck},
  \citenamefont {Bartolo}, \citenamefont {Belgacem}, \citenamefont {Bertacca},
  \citenamefont {Bizouard}, \citenamefont {Branchesi}, \citenamefont {Clesse},
  \citenamefont {Foffa}, \citenamefont {García-Bellido}, \citenamefont
  {Grimm}, \citenamefont {Harms}, \citenamefont {Hinderer}, \citenamefont
  {Matarrese}, \citenamefont {Palomba}, \citenamefont {Peloso}, \citenamefont
  {Ricciardone},\ and\ \citenamefont {Sakellariadou}}]{Maggiore_2020}%
  \BibitemOpen
  \bibfield  {author} {\bibinfo {author} {\bibfnamefont {M.}~\bibnamefont
  {Maggiore}}, \bibinfo {author} {\bibfnamefont {C.~V.~D.}\ \bibnamefont
  {Broeck}}, \bibinfo {author} {\bibfnamefont {N.}~\bibnamefont {Bartolo}},
  \bibinfo {author} {\bibfnamefont {E.}~\bibnamefont {Belgacem}}, \bibinfo
  {author} {\bibfnamefont {D.}~\bibnamefont {Bertacca}}, \bibinfo {author}
  {\bibfnamefont {M.~A.}\ \bibnamefont {Bizouard}}, \bibinfo {author}
  {\bibfnamefont {M.}~\bibnamefont {Branchesi}}, \bibinfo {author}
  {\bibfnamefont {S.}~\bibnamefont {Clesse}}, \bibinfo {author} {\bibfnamefont
  {S.}~\bibnamefont {Foffa}}, \bibinfo {author} {\bibfnamefont
  {J.}~\bibnamefont {García-Bellido}}, \bibinfo {author} {\bibfnamefont
  {S.}~\bibnamefont {Grimm}}, \bibinfo {author} {\bibfnamefont
  {J.}~\bibnamefont {Harms}}, \bibinfo {author} {\bibfnamefont
  {T.}~\bibnamefont {Hinderer}}, \bibinfo {author} {\bibfnamefont
  {S.}~\bibnamefont {Matarrese}}, \bibinfo {author} {\bibfnamefont
  {C.}~\bibnamefont {Palomba}}, \bibinfo {author} {\bibfnamefont
  {M.}~\bibnamefont {Peloso}}, \bibinfo {author} {\bibfnamefont
  {A.}~\bibnamefont {Ricciardone}},\ and\ \bibinfo {author} {\bibfnamefont
  {M.}~\bibnamefont {Sakellariadou}},\ }\bibfield  {title} {\bibinfo {title}
  {Science case for the einstein telescope},\ }\href
  {https://doi.org/10.1088/1475-7516/2020/03/050} {\bibfield  {journal}
  {\bibinfo  {journal} {Journal of Cosmology and Astroparticle Physics}\
  }\textbf {\bibinfo {volume} {2020}}\bibinfo  {number} { (03)},\ \bibinfo
  {pages} {050}}\BibitemShut {NoStop}%
\bibitem [{\citenamefont {Reitze}\ \emph {et~al.}(2019)\citenamefont {Reitze}
  \emph {et~al.}}]{Reitze:2019iox}%
  \BibitemOpen
\bibfield  {number} {  }\bibfield  {author} {\bibinfo {author} {\bibfnamefont
  {D.}~\bibnamefont {Reitze}} \emph {et~al.},\ }\bibfield  {title} {\bibinfo
  {title} {{Cosmic Explorer: The U.S. Contribution to Gravitational-Wave
  Astronomy beyond LIGO}},\ }\href@noop {} {\bibfield  {journal} {\bibinfo
  {journal} {Bull. Am. Astron. Soc.}\ }\textbf {\bibinfo {volume} {51}},\
  \bibinfo {pages} {035} (\bibinfo {year} {2019})},\ \Eprint
  {https://arxiv.org/abs/1907.04833} {arXiv:1907.04833 [astro-ph.IM]}
  \BibitemShut {NoStop}%
\bibitem [{\citenamefont {Savastano}\ \emph {et~al.}(2024)\citenamefont
  {Savastano}, \citenamefont {Vernizzi},\ and\ \citenamefont
  {Zumalac\'arregui}}]{Savastano:2022jjv}%
  \BibitemOpen
  \bibfield  {author} {\bibinfo {author} {\bibfnamefont {S.}~\bibnamefont
  {Savastano}}, \bibinfo {author} {\bibfnamefont {F.}~\bibnamefont
  {Vernizzi}},\ and\ \bibinfo {author} {\bibfnamefont {M.}~\bibnamefont
  {Zumalac\'arregui}},\ }\bibfield  {title} {\bibinfo {title} {{Through the
  lens of Sgr A*: Identifying and resolving strongly lensed continuous
  gravitational waves beyond the Einstein radius}},\ }\href
  {https://doi.org/10.1103/PhysRevD.109.024064} {\bibfield  {journal} {\bibinfo
   {journal} {Phys. Rev. D}\ }\textbf {\bibinfo {volume} {109}},\ \bibinfo
  {pages} {024064} (\bibinfo {year} {2024})},\ \Eprint
  {https://arxiv.org/abs/2212.14697} {arXiv:2212.14697 [gr-qc]} \BibitemShut
  {NoStop}%
\bibitem [{\citenamefont {{Bertotti}}\ \emph {et~al.}(2003)\citenamefont
  {{Bertotti}}, \citenamefont {{Iess}},\ and\ \citenamefont
  {{Tortora}}}]{2003Natur.425..374B}%
  \BibitemOpen
  \bibfield  {author} {\bibinfo {author} {\bibfnamefont {B.}~\bibnamefont
  {{Bertotti}}}, \bibinfo {author} {\bibfnamefont {L.}~\bibnamefont {{Iess}}},\
  and\ \bibinfo {author} {\bibfnamefont {P.}~\bibnamefont {{Tortora}}},\
  }\bibfield  {title} {\bibinfo {title} {{A test of general relativity using
  radio links with the Cassini spacecraft}},\ }\href
  {https://doi.org/10.1038/nature01997} {\bibfield  {journal} {\bibinfo
  {journal} {\nat}\ }\textbf {\bibinfo {volume} {425}},\ \bibinfo {pages} {374}
  (\bibinfo {year} {2003})}\BibitemShut {NoStop}%
\bibitem [{\citenamefont {Armstrong}\ \emph {et~al.}(2003)\citenamefont
  {Armstrong}, \citenamefont {Iess}, \citenamefont {Tortora},\ and\
  \citenamefont {Bertotti}}]{Armstrong:2003ay}%
  \BibitemOpen
  \bibfield  {author} {\bibinfo {author} {\bibfnamefont {J.~W.}\ \bibnamefont
  {Armstrong}}, \bibinfo {author} {\bibfnamefont {L.}~\bibnamefont {Iess}},
  \bibinfo {author} {\bibfnamefont {P.}~\bibnamefont {Tortora}},\ and\ \bibinfo
  {author} {\bibfnamefont {B.}~\bibnamefont {Bertotti}},\ }\bibfield  {title}
  {\bibinfo {title} {{Stochastic gravitational wave background: Upper limits in
  the 10**-6-Hz 10**-3-Hz band}},\ }\href {https://doi.org/10.1086/379505}
  {\bibfield  {journal} {\bibinfo  {journal} {Astrophys. J.}\ }\textbf
  {\bibinfo {volume} {599}},\ \bibinfo {pages} {806} (\bibinfo {year}
  {2003})}\BibitemShut {NoStop}%
\bibitem [{\citenamefont {Ferreira}(2021)}]{Ferreira:2020fam}%
  \BibitemOpen
  \bibfield  {author} {\bibinfo {author} {\bibfnamefont {E.~G.~M.}\
  \bibnamefont {Ferreira}},\ }\bibfield  {title} {\bibinfo {title}
  {{Ultra-light dark matter}},\ }\href
  {https://doi.org/10.1007/s00159-021-00135-6} {\bibfield  {journal} {\bibinfo
  {journal} {Astron. Astrophys. Rev.}\ }\textbf {\bibinfo {volume} {29}},\
  \bibinfo {pages} {7} (\bibinfo {year} {2021})},\ \Eprint
  {https://arxiv.org/abs/2005.03254} {arXiv:2005.03254 [astro-ph.CO]}
  \BibitemShut {NoStop}%
\bibitem [{\citenamefont {Kawamura}\ \emph {et~al.}(2011)\citenamefont
  {Kawamura} \emph {et~al.}}]{Kawamura:2011zz}%
  \BibitemOpen
  \bibfield  {author} {\bibinfo {author} {\bibfnamefont {S.}~\bibnamefont
  {Kawamura}} \emph {et~al.},\ }\bibfield  {title} {\bibinfo {title} {{The
  Japanese space gravitational wave antenna: DECIGO}},\ }\href
  {https://doi.org/10.1088/0264-9381/28/9/094011} {\bibfield  {journal}
  {\bibinfo  {journal} {Class. Quant. Grav.}\ }\textbf {\bibinfo {volume}
  {28}},\ \bibinfo {pages} {094011} (\bibinfo {year} {2011})}\BibitemShut
  {NoStop}%
\bibitem [{\citenamefont {Yagi}\ and\ \citenamefont
  {Seto}(2011)}]{Yagi:2011wg}%
  \BibitemOpen
  \bibfield  {author} {\bibinfo {author} {\bibfnamefont {K.}~\bibnamefont
  {Yagi}}\ and\ \bibinfo {author} {\bibfnamefont {N.}~\bibnamefont {Seto}},\
  }\bibfield  {title} {\bibinfo {title} {{Detector configuration of DECIGO/BBO
  and identification of cosmological neutron-star binaries}},\ }\href
  {https://doi.org/10.1103/PhysRevD.83.044011} {\bibfield  {journal} {\bibinfo
  {journal} {Phys. Rev. D}\ }\textbf {\bibinfo {volume} {83}},\ \bibinfo
  {pages} {044011} (\bibinfo {year} {2011})},\ \bibinfo {note} {[Erratum:
  Phys.Rev.D 95, 109901 (2017)]},\ \Eprint {https://arxiv.org/abs/1101.3940}
  {arXiv:1101.3940 [astro-ph.CO]} \BibitemShut {NoStop}%
\bibitem [{\citenamefont {Kawamura}\ \emph {et~al.}(2021)\citenamefont
  {Kawamura} \emph {et~al.}}]{Kawamura:2020pcg}%
  \BibitemOpen
  \bibfield  {author} {\bibinfo {author} {\bibfnamefont {S.}~\bibnamefont
  {Kawamura}} \emph {et~al.},\ }\bibfield  {title} {\bibinfo {title} {{Current
  status of space gravitational wave antenna DECIGO and B-DECIGO}},\ }\href
  {https://doi.org/10.1093/ptep/ptab019} {\bibfield  {journal} {\bibinfo
  {journal} {PTEP}\ }\textbf {\bibinfo {volume} {2021}},\ \bibinfo {pages}
  {05A105} (\bibinfo {year} {2021})},\ \Eprint
  {https://arxiv.org/abs/2006.13545} {arXiv:2006.13545 [gr-qc]} \BibitemShut
  {NoStop}%
\bibitem [{\citenamefont {Abbott}\ \emph
  {et~al.}(2017{\natexlab{a}})\citenamefont {Abbott} \emph
  {et~al.}}]{LIGOScientific:2017vox}%
  \BibitemOpen
  \bibfield  {author} {\bibinfo {author} {\bibfnamefont {B.~. P.~.}\
  \bibnamefont {Abbott}} \emph {et~al.} (\bibinfo {collaboration} {LIGO
  Scientific, Virgo}),\ }\bibfield  {title} {\bibinfo {title} {{GW170608:
  Observation of a 19-solar-mass Binary Black Hole Coalescence}},\ }\href
  {https://doi.org/10.3847/2041-8213/aa9f0c} {\bibfield  {journal} {\bibinfo
  {journal} {Astrophys. J. Lett.}\ }\textbf {\bibinfo {volume} {851}},\
  \bibinfo {pages} {L35} (\bibinfo {year} {2017}{\natexlab{a}})},\ \Eprint
  {https://arxiv.org/abs/1711.05578} {arXiv:1711.05578 [astro-ph.HE]}
  \BibitemShut {NoStop}%
\bibitem [{\citenamefont {Abbott}\ \emph
  {et~al.}(2017{\natexlab{b}})\citenamefont {Abbott} \emph
  {et~al.}}]{LIGOScientific:2017vwq}%
  \BibitemOpen
  \bibfield  {author} {\bibinfo {author} {\bibfnamefont {B.~P.}\ \bibnamefont
  {Abbott}} \emph {et~al.} (\bibinfo {collaboration} {LIGO Scientific,
  Virgo}),\ }\bibfield  {title} {\bibinfo {title} {{GW170817: Observation of
  Gravitational Waves from a Binary Neutron Star Inspiral}},\ }\href
  {https://doi.org/10.1103/PhysRevLett.119.161101} {\bibfield  {journal}
  {\bibinfo  {journal} {Phys. Rev. Lett.}\ }\textbf {\bibinfo {volume} {119}},\
  \bibinfo {pages} {161101} (\bibinfo {year} {2017}{\natexlab{b}})},\ \Eprint
  {https://arxiv.org/abs/1710.05832} {arXiv:1710.05832 [gr-qc]} \BibitemShut
  {NoStop}%
\bibitem [{\citenamefont {Robson}\ \emph {et~al.}(2019)\citenamefont {Robson},
  \citenamefont {Cornish},\ and\ \citenamefont {Liu}}]{Robson:2018ifk}%
  \BibitemOpen
  \bibfield  {author} {\bibinfo {author} {\bibfnamefont {T.}~\bibnamefont
  {Robson}}, \bibinfo {author} {\bibfnamefont {N.~J.}\ \bibnamefont
  {Cornish}},\ and\ \bibinfo {author} {\bibfnamefont {C.}~\bibnamefont {Liu}},\
  }\bibfield  {title} {\bibinfo {title} {{The construction and use of LISA
  sensitivity curves}},\ }\href {https://doi.org/10.1088/1361-6382/ab1101}
  {\bibfield  {journal} {\bibinfo  {journal} {Class. Quant. Grav.}\ }\textbf
  {\bibinfo {volume} {36}},\ \bibinfo {pages} {105011} (\bibinfo {year}
  {2019})},\ \Eprint {https://arxiv.org/abs/1803.01944} {arXiv:1803.01944
  [astro-ph.HE]} \BibitemShut {NoStop}%
\bibitem [{\citenamefont {Bandara}\ \emph {et~al.}(2009)\citenamefont
  {Bandara}, \citenamefont {Crampton},\ and\ \citenamefont
  {Simard}}]{Bandara_2009}%
  \BibitemOpen
  \bibfield  {author} {\bibinfo {author} {\bibfnamefont {K.}~\bibnamefont
  {Bandara}}, \bibinfo {author} {\bibfnamefont {D.}~\bibnamefont {Crampton}},\
  and\ \bibinfo {author} {\bibfnamefont {L.}~\bibnamefont {Simard}},\
  }\bibfield  {title} {\bibinfo {title} {A relationship between supermassive
  black hole mass and the total gravitational mass of the host galaxy},\ }\href
  {https://doi.org/10.1088/0004-637x/704/2/1135} {\bibfield  {journal}
  {\bibinfo  {journal} {The Astrophysical Journal}\ }\textbf {\bibinfo {volume}
  {704}},\ \bibinfo {pages} {1135–1145} (\bibinfo {year} {2009})}\BibitemShut
  {NoStop}%
\bibitem [{\citenamefont {Kim}\ and\ \citenamefont
  {Mitridate}(2024)}]{Kim:2023kyy}%
  \BibitemOpen
  \bibfield  {author} {\bibinfo {author} {\bibfnamefont {H.}~\bibnamefont
  {Kim}}\ and\ \bibinfo {author} {\bibfnamefont {A.}~\bibnamefont
  {Mitridate}},\ }\bibfield  {title} {\bibinfo {title} {{Stochastic ultralight
  dark matter fluctuations in pulsar timing arrays}},\ }\href
  {https://doi.org/10.1103/PhysRevD.109.055017} {\bibfield  {journal} {\bibinfo
   {journal} {Phys. Rev. D}\ }\textbf {\bibinfo {volume} {109}},\ \bibinfo
  {pages} {055017} (\bibinfo {year} {2024})},\ \Eprint
  {https://arxiv.org/abs/2312.12225} {arXiv:2312.12225 [hep-ph]} \BibitemShut
  {NoStop}%
\bibitem [{\citenamefont {Kim}(2023)}]{Kim:2023pkx}%
  \BibitemOpen
  \bibfield  {author} {\bibinfo {author} {\bibfnamefont {H.}~\bibnamefont
  {Kim}},\ }\bibfield  {title} {\bibinfo {title} {{Gravitational interaction of
  ultralight dark matter with interferometers}},\ }\href
  {https://doi.org/10.1088/1475-7516/2023/12/018} {\bibfield  {journal}
  {\bibinfo  {journal} {JCAP}\ }\textbf {\bibinfo {volume} {12}},\ \bibinfo
  {pages} {018}},\ \Eprint {https://arxiv.org/abs/2306.13348} {arXiv:2306.13348
  [hep-ph]} \BibitemShut {NoStop}%
\bibitem [{\citenamefont {Yao}\ and\ \citenamefont
  {Tang}(2024)}]{yao2024probing}%
  \BibitemOpen
  \bibfield  {author} {\bibinfo {author} {\bibfnamefont {Y.-H.}\ \bibnamefont
  {Yao}}\ and\ \bibinfo {author} {\bibfnamefont {Y.}~\bibnamefont {Tang}},\
  }\href@noop {} {\bibinfo {title} {Probing stochastic ultralight dark matter
  with space-based gravitational-wave interferometers}} (\bibinfo {year}
  {2024}),\ \Eprint {https://arxiv.org/abs/2404.01494} {arXiv:2404.01494
  [hep-ph]} \BibitemShut {NoStop}%
\bibitem [{\citenamefont {Dev}\ \emph {et~al.}(2017)\citenamefont {Dev},
  \citenamefont {Lindner},\ and\ \citenamefont {Ohmer}}]{Dev:2016hxv}%
  \BibitemOpen
  \bibfield  {author} {\bibinfo {author} {\bibfnamefont {P.~S.~B.}\
  \bibnamefont {Dev}}, \bibinfo {author} {\bibfnamefont {M.}~\bibnamefont
  {Lindner}},\ and\ \bibinfo {author} {\bibfnamefont {S.}~\bibnamefont
  {Ohmer}},\ }\bibfield  {title} {\bibinfo {title} {{Gravitational waves as a
  new probe of Bose\textendash{}Einstein condensate Dark Matter}},\ }\href
  {https://doi.org/10.1016/j.physletb.2017.08.043} {\bibfield  {journal}
  {\bibinfo  {journal} {Phys. Lett. B}\ }\textbf {\bibinfo {volume} {773}},\
  \bibinfo {pages} {219} (\bibinfo {year} {2017})},\ \Eprint
  {https://arxiv.org/abs/1609.03939} {arXiv:1609.03939 [hep-ph]} \BibitemShut
  {NoStop}%
\bibitem [{\citenamefont {Banerjee}\ \emph {et~al.}(2023)\citenamefont
  {Banerjee}, \citenamefont {Bera},\ and\ \citenamefont
  {Mota}}]{Banerjee:2022zii}%
  \BibitemOpen
  \bibfield  {author} {\bibinfo {author} {\bibfnamefont {S.}~\bibnamefont
  {Banerjee}}, \bibinfo {author} {\bibfnamefont {S.}~\bibnamefont {Bera}},\
  and\ \bibinfo {author} {\bibfnamefont {D.~F.}\ \bibnamefont {Mota}},\
  }\bibfield  {title} {\bibinfo {title} {{Prospects of probing dark matter
  condensates with gravitational waves}},\ }\href
  {https://doi.org/10.1088/1475-7516/2023/03/041} {\bibfield  {journal}
  {\bibinfo  {journal} {JCAP}\ }\textbf {\bibinfo {volume} {03}},\ \bibinfo
  {pages} {041}},\ \Eprint {https://arxiv.org/abs/2211.13988} {arXiv:2211.13988
  [gr-qc]} \BibitemShut {NoStop}%
\bibitem [{\citenamefont {Annulli}\ \emph
  {et~al.}(2020{\natexlab{a}})\citenamefont {Annulli}, \citenamefont
  {Cardoso},\ and\ \citenamefont {Vicente}}]{Annulli:2020ilw}%
  \BibitemOpen
  \bibfield  {author} {\bibinfo {author} {\bibfnamefont {L.}~\bibnamefont
  {Annulli}}, \bibinfo {author} {\bibfnamefont {V.}~\bibnamefont {Cardoso}},\
  and\ \bibinfo {author} {\bibfnamefont {R.}~\bibnamefont {Vicente}},\
  }\bibfield  {title} {\bibinfo {title} {{Stirred and shaken: Dynamical
  behavior of boson stars and dark matter cores}},\ }\href
  {https://doi.org/10.1016/j.physletb.2020.135944} {\bibfield  {journal}
  {\bibinfo  {journal} {Phys. Lett. B}\ }\textbf {\bibinfo {volume} {811}},\
  \bibinfo {pages} {135944} (\bibinfo {year} {2020}{\natexlab{a}})},\ \Eprint
  {https://arxiv.org/abs/2007.03700} {arXiv:2007.03700 [astro-ph.HE]}
  \BibitemShut {NoStop}%
\bibitem [{\citenamefont {Annulli}\ \emph
  {et~al.}(2020{\natexlab{b}})\citenamefont {Annulli}, \citenamefont
  {Cardoso},\ and\ \citenamefont {Vicente}}]{Annulli:2020lyc}%
  \BibitemOpen
  \bibfield  {author} {\bibinfo {author} {\bibfnamefont {L.}~\bibnamefont
  {Annulli}}, \bibinfo {author} {\bibfnamefont {V.}~\bibnamefont {Cardoso}},\
  and\ \bibinfo {author} {\bibfnamefont {R.}~\bibnamefont {Vicente}},\
  }\bibfield  {title} {\bibinfo {title} {{Response of ultralight dark matter to
  supermassive black holes and binaries}},\ }\href
  {https://doi.org/10.1103/PhysRevD.102.063022} {\bibfield  {journal} {\bibinfo
   {journal} {Phys. Rev. D}\ }\textbf {\bibinfo {volume} {102}},\ \bibinfo
  {pages} {063022} (\bibinfo {year} {2020}{\natexlab{b}})},\ \Eprint
  {https://arxiv.org/abs/2009.00012} {arXiv:2009.00012 [gr-qc]} \BibitemShut
  {NoStop}%
\bibitem [{\citenamefont {Baumann}\ \emph {et~al.}(2022)\citenamefont
  {Baumann}, \citenamefont {Bertone}, \citenamefont {Stout},\ and\
  \citenamefont {Tomaselli}}]{Baumann:2021fkf}%
  \BibitemOpen
  \bibfield  {author} {\bibinfo {author} {\bibfnamefont {D.}~\bibnamefont
  {Baumann}}, \bibinfo {author} {\bibfnamefont {G.}~\bibnamefont {Bertone}},
  \bibinfo {author} {\bibfnamefont {J.}~\bibnamefont {Stout}},\ and\ \bibinfo
  {author} {\bibfnamefont {G.~M.}\ \bibnamefont {Tomaselli}},\ }\bibfield
  {title} {\bibinfo {title} {{Ionization of gravitational atoms}},\ }\href
  {https://doi.org/10.1103/PhysRevD.105.115036} {\bibfield  {journal} {\bibinfo
   {journal} {Phys. Rev. D}\ }\textbf {\bibinfo {volume} {105}},\ \bibinfo
  {pages} {115036} (\bibinfo {year} {2022})},\ \Eprint
  {https://arxiv.org/abs/2112.14777} {arXiv:2112.14777 [gr-qc]} \BibitemShut
  {NoStop}%
\bibitem [{\citenamefont {Traykova}\ \emph {et~al.}(2021)\citenamefont
  {Traykova}, \citenamefont {Clough}, \citenamefont {Helfer}, \citenamefont
  {Berti}, \citenamefont {Ferreira},\ and\ \citenamefont
  {Hui}}]{Traykova:2021dua}%
  \BibitemOpen
  \bibfield  {author} {\bibinfo {author} {\bibfnamefont {D.}~\bibnamefont
  {Traykova}}, \bibinfo {author} {\bibfnamefont {K.}~\bibnamefont {Clough}},
  \bibinfo {author} {\bibfnamefont {T.}~\bibnamefont {Helfer}}, \bibinfo
  {author} {\bibfnamefont {E.}~\bibnamefont {Berti}}, \bibinfo {author}
  {\bibfnamefont {P.~G.}\ \bibnamefont {Ferreira}},\ and\ \bibinfo {author}
  {\bibfnamefont {L.}~\bibnamefont {Hui}},\ }\bibfield  {title} {\bibinfo
  {title} {{Dynamical friction from scalar dark matter in the relativistic
  regime}},\ }\href {https://doi.org/10.1103/PhysRevD.104.103014} {\bibfield
  {journal} {\bibinfo  {journal} {Phys. Rev. D}\ }\textbf {\bibinfo {volume}
  {104}},\ \bibinfo {pages} {103014} (\bibinfo {year} {2021})},\ \Eprint
  {https://arxiv.org/abs/2106.08280} {arXiv:2106.08280 [gr-qc]} \BibitemShut
  {NoStop}%
\bibitem [{\citenamefont {Vicente}\ and\ \citenamefont
  {Cardoso}(2022)}]{Vicente:2022ivh}%
  \BibitemOpen
  \bibfield  {author} {\bibinfo {author} {\bibfnamefont {R.}~\bibnamefont
  {Vicente}}\ and\ \bibinfo {author} {\bibfnamefont {V.}~\bibnamefont
  {Cardoso}},\ }\bibfield  {title} {\bibinfo {title} {{Dynamical friction of
  black holes in ultralight dark matter}},\ }\href
  {https://doi.org/10.1103/PhysRevD.105.083008} {\bibfield  {journal} {\bibinfo
   {journal} {Phys. Rev. D}\ }\textbf {\bibinfo {volume} {105}},\ \bibinfo
  {pages} {083008} (\bibinfo {year} {2022})},\ \Eprint
  {https://arxiv.org/abs/2201.08854} {arXiv:2201.08854 [gr-qc]} \BibitemShut
  {NoStop}%
\bibitem [{\citenamefont {Bamber}\ \emph {et~al.}(2023)\citenamefont {Bamber},
  \citenamefont {Aurrekoetxea}, \citenamefont {Clough},\ and\ \citenamefont
  {Ferreira}}]{Bamber:2022pbs}%
  \BibitemOpen
  \bibfield  {author} {\bibinfo {author} {\bibfnamefont {J.}~\bibnamefont
  {Bamber}}, \bibinfo {author} {\bibfnamefont {J.~C.}\ \bibnamefont
  {Aurrekoetxea}}, \bibinfo {author} {\bibfnamefont {K.}~\bibnamefont
  {Clough}},\ and\ \bibinfo {author} {\bibfnamefont {P.~G.}\ \bibnamefont
  {Ferreira}},\ }\bibfield  {title} {\bibinfo {title} {{Black hole merger
  simulations in wave dark matter environments}},\ }\href
  {https://doi.org/10.1103/PhysRevD.107.024035} {\bibfield  {journal} {\bibinfo
   {journal} {Phys. Rev. D}\ }\textbf {\bibinfo {volume} {107}},\ \bibinfo
  {pages} {024035} (\bibinfo {year} {2023})},\ \Eprint
  {https://arxiv.org/abs/2210.09254} {arXiv:2210.09254 [gr-qc]} \BibitemShut
  {NoStop}%
\bibitem [{\citenamefont {Traykova}\ \emph {et~al.}(2023)\citenamefont
  {Traykova}, \citenamefont {Vicente}, \citenamefont {Clough}, \citenamefont
  {Helfer}, \citenamefont {Berti}, \citenamefont {Ferreira},\ and\
  \citenamefont {Hui}}]{Traykova:2023qyv}%
  \BibitemOpen
  \bibfield  {author} {\bibinfo {author} {\bibfnamefont {D.}~\bibnamefont
  {Traykova}}, \bibinfo {author} {\bibfnamefont {R.}~\bibnamefont {Vicente}},
  \bibinfo {author} {\bibfnamefont {K.}~\bibnamefont {Clough}}, \bibinfo
  {author} {\bibfnamefont {T.}~\bibnamefont {Helfer}}, \bibinfo {author}
  {\bibfnamefont {E.}~\bibnamefont {Berti}}, \bibinfo {author} {\bibfnamefont
  {P.~G.}\ \bibnamefont {Ferreira}},\ and\ \bibinfo {author} {\bibfnamefont
  {L.}~\bibnamefont {Hui}},\ }\bibfield  {title} {\bibinfo {title}
  {{Relativistic drag forces on black holes from scalar dark matter clouds of
  all sizes}},\ }\href {https://doi.org/10.1103/PhysRevD.108.L121502}
  {\bibfield  {journal} {\bibinfo  {journal} {Phys. Rev. D}\ }\textbf {\bibinfo
  {volume} {108}},\ \bibinfo {pages} {L121502} (\bibinfo {year} {2023})},\
  \Eprint {https://arxiv.org/abs/2305.10492} {arXiv:2305.10492 [gr-qc]}
  \BibitemShut {NoStop}%
\bibitem [{\citenamefont {Tomaselli}\ \emph {et~al.}(2023)\citenamefont
  {Tomaselli}, \citenamefont {Spieksma},\ and\ \citenamefont
  {Bertone}}]{Tomaselli:2023ysb}%
  \BibitemOpen
  \bibfield  {author} {\bibinfo {author} {\bibfnamefont {G.~M.}\ \bibnamefont
  {Tomaselli}}, \bibinfo {author} {\bibfnamefont {T.~F.~M.}\ \bibnamefont
  {Spieksma}},\ and\ \bibinfo {author} {\bibfnamefont {G.}~\bibnamefont
  {Bertone}},\ }\bibfield  {title} {\bibinfo {title} {{Dynamical friction in
  gravitational atoms}},\ }\href
  {https://doi.org/10.1088/1475-7516/2023/07/070} {\bibfield  {journal}
  {\bibinfo  {journal} {JCAP}\ }\textbf {\bibinfo {volume} {07}},\ \bibinfo
  {pages} {070}},\ \Eprint {https://arxiv.org/abs/2305.15460} {arXiv:2305.15460
  [gr-qc]} \BibitemShut {NoStop}%
\bibitem [{\citenamefont {Aurrekoetxea}\ \emph {et~al.}(2024)\citenamefont
  {Aurrekoetxea}, \citenamefont {Clough}, \citenamefont {Bamber},\ and\
  \citenamefont {Ferreira}}]{Aurrekoetxea:2023jwk}%
  \BibitemOpen
  \bibfield  {author} {\bibinfo {author} {\bibfnamefont {J.~C.}\ \bibnamefont
  {Aurrekoetxea}}, \bibinfo {author} {\bibfnamefont {K.}~\bibnamefont
  {Clough}}, \bibinfo {author} {\bibfnamefont {J.}~\bibnamefont {Bamber}},\
  and\ \bibinfo {author} {\bibfnamefont {P.~G.}\ \bibnamefont {Ferreira}},\
  }\bibfield  {title} {\bibinfo {title} {{Effect of Wave Dark Matter on Equal
  Mass Black Hole Mergers}},\ }\href
  {https://doi.org/10.1103/PhysRevLett.132.211401} {\bibfield  {journal}
  {\bibinfo  {journal} {Phys. Rev. Lett.}\ }\textbf {\bibinfo {volume} {132}},\
  \bibinfo {pages} {211401} (\bibinfo {year} {2024})},\ \Eprint
  {https://arxiv.org/abs/2311.18156} {arXiv:2311.18156 [gr-qc]} \BibitemShut
  {NoStop}%
\bibitem [{\citenamefont {Brito}\ and\ \citenamefont
  {Shah}(2023)}]{Brito:2023pyl}%
  \BibitemOpen
  \bibfield  {author} {\bibinfo {author} {\bibfnamefont {R.}~\bibnamefont
  {Brito}}\ and\ \bibinfo {author} {\bibfnamefont {S.}~\bibnamefont {Shah}},\
  }\bibfield  {title} {\bibinfo {title} {{Extreme mass-ratio inspirals into
  black holes surrounded by scalar clouds}},\ }\href
  {https://doi.org/10.1103/PhysRevD.108.084019} {\bibfield  {journal} {\bibinfo
   {journal} {Phys. Rev. D}\ }\textbf {\bibinfo {volume} {108}},\ \bibinfo
  {pages} {084019} (\bibinfo {year} {2023})},\ \Eprint
  {https://arxiv.org/abs/2307.16093} {arXiv:2307.16093 [gr-qc]} \BibitemShut
  {NoStop}%
\bibitem [{\citenamefont {Duque}\ \emph {et~al.}(2023)\citenamefont {Duque},
  \citenamefont {Macedo}, \citenamefont {Vicente},\ and\ \citenamefont
  {Cardoso}}]{Duque:2023cac}%
  \BibitemOpen
  \bibfield  {author} {\bibinfo {author} {\bibfnamefont {F.}~\bibnamefont
  {Duque}}, \bibinfo {author} {\bibfnamefont {C.~F.~B.}\ \bibnamefont
  {Macedo}}, \bibinfo {author} {\bibfnamefont {R.}~\bibnamefont {Vicente}},\
  and\ \bibinfo {author} {\bibfnamefont {V.}~\bibnamefont {Cardoso}},\
  }\bibfield  {title} {\bibinfo {title} {{Axion Weak Leaks: extreme mass-ratio
  inspirals in ultra-light dark matter}},\ }\href@noop {} {\  (\bibinfo {year}
  {2023})},\ \Eprint {https://arxiv.org/abs/2312.06767} {arXiv:2312.06767
  [gr-qc]} \BibitemShut {NoStop}%
\bibitem [{\citenamefont {Zwick}\ \emph {et~al.}(2024)\citenamefont {Zwick},
  \citenamefont {Soyuer}, \citenamefont {D'Orazio}, \citenamefont {O'Neill},
  \citenamefont {Derdzinski}, \citenamefont {Saha}, \citenamefont {Blas},
  \citenamefont {Jenkins},\ and\ \citenamefont {Kelley}}]{Zwick:2024hag}%
  \BibitemOpen
  \bibfield  {author} {\bibinfo {author} {\bibfnamefont {L.}~\bibnamefont
  {Zwick}}, \bibinfo {author} {\bibfnamefont {D.}~\bibnamefont {Soyuer}},
  \bibinfo {author} {\bibfnamefont {D.~J.}\ \bibnamefont {D'Orazio}}, \bibinfo
  {author} {\bibfnamefont {D.}~\bibnamefont {O'Neill}}, \bibinfo {author}
  {\bibfnamefont {A.}~\bibnamefont {Derdzinski}}, \bibinfo {author}
  {\bibfnamefont {P.}~\bibnamefont {Saha}}, \bibinfo {author} {\bibfnamefont
  {D.}~\bibnamefont {Blas}}, \bibinfo {author} {\bibfnamefont {A.~C.}\
  \bibnamefont {Jenkins}},\ and\ \bibinfo {author} {\bibfnamefont {L.~Z.}\
  \bibnamefont {Kelley}},\ }\bibfield  {title} {\bibinfo {title} {{Bridging the
  micro-Hz gravitational wave gap via Doppler tracking with the Uranus Orbiter
  and Probe Mission: Massive black hole binaries, early universe signals and
  ultra-light dark matter}},\ }\href@noop {} {\  (\bibinfo {year} {2024})},\
  \Eprint {https://arxiv.org/abs/2406.02306} {arXiv:2406.02306 [astro-ph.HE]}
  \BibitemShut {NoStop}%
\bibitem [{\citenamefont {Kim}\ and\ \citenamefont {Yang}(2024)}]{Kim:2024rgf}%
  \BibitemOpen
  \bibfield  {author} {\bibinfo {author} {\bibfnamefont {J.~H.}\ \bibnamefont
  {Kim}}\ and\ \bibinfo {author} {\bibfnamefont {X.-Y.}\ \bibnamefont {Yang}},\
  }\bibfield  {title} {\bibinfo {title} {{Gravitational Wave Duet by Resonating
  Binary Black Holes with Axion-Like Particles}},\ }\href@noop {} {\  (\bibinfo
  {year} {2024})},\ \Eprint {https://arxiv.org/abs/2407.14604}
  {arXiv:2407.14604 [astro-ph.CO]} \BibitemShut {NoStop}%
\bibitem [{\citenamefont {Smarra}\ \emph {et~al.}(2023)\citenamefont {Smarra}
  \emph {et~al.}}]{EuropeanPulsarTimingArray:2023egv}%
  \BibitemOpen
  \bibfield  {author} {\bibinfo {author} {\bibfnamefont {C.}~\bibnamefont
  {Smarra}} \emph {et~al.} (\bibinfo {collaboration} {European Pulsar Timing
  Array}),\ }\bibfield  {title} {\bibinfo {title} {{Second Data Release from
  the European Pulsar Timing Array: Challenging the Ultralight Dark Matter
  Paradigm}},\ }\href {https://doi.org/10.1103/PhysRevLett.131.171001}
  {\bibfield  {journal} {\bibinfo  {journal} {Phys. Rev. Lett.}\ }\textbf
  {\bibinfo {volume} {131}},\ \bibinfo {pages} {171001} (\bibinfo {year}
  {2023})},\ \Eprint {https://arxiv.org/abs/2306.16228} {arXiv:2306.16228
  [astro-ph.HE]} \BibitemShut {NoStop}%
\bibitem [{\citenamefont {Flanagan}\ and\ \citenamefont
  {Hughes}(2005)}]{Flanagan:2005yc}%
  \BibitemOpen
  \bibfield  {author} {\bibinfo {author} {\bibfnamefont {E.~E.}\ \bibnamefont
  {Flanagan}}\ and\ \bibinfo {author} {\bibfnamefont {S.~A.}\ \bibnamefont
  {Hughes}},\ }\bibfield  {title} {\bibinfo {title} {{The Basics of
  gravitational wave theory}},\ }\href
  {https://doi.org/10.1088/1367-2630/7/1/204} {\bibfield  {journal} {\bibinfo
  {journal} {New J. Phys.}\ }\textbf {\bibinfo {volume} {7}},\ \bibinfo {pages}
  {204} (\bibinfo {year} {2005})},\ \Eprint
  {https://arxiv.org/abs/gr-qc/0501041} {arXiv:gr-qc/0501041} \BibitemShut
  {NoStop}%
\bibitem [{\citenamefont {Maggiore}(2007)}]{Maggiore:2007ulw}%
  \BibitemOpen
  \bibfield  {author} {\bibinfo {author} {\bibfnamefont {M.}~\bibnamefont
  {Maggiore}},\ }\href
  {https://doi.org/10.1093/acprof:oso/9780198570745.001.0001} {\emph {\bibinfo
  {title} {{Gravitational Waves. Vol. 1: Theory and Experiments}}}}\ (\bibinfo
  {publisher} {Oxford University Press},\ \bibinfo {year} {2007})\BibitemShut
  {NoStop}%
\bibitem [{\citenamefont {Dodelson}(2003)}]{Dodelson:2003ft}%
  \BibitemOpen
  \bibfield  {author} {\bibinfo {author} {\bibfnamefont {S.}~\bibnamefont
  {Dodelson}},\ }\href@noop {} {\emph {\bibinfo {title} {{Modern Cosmology}}}}\
  (\bibinfo  {publisher} {Academic Press},\ \bibinfo {address} {Amsterdam},\
  \bibinfo {year} {2003})\BibitemShut {NoStop}%
\bibitem [{\citenamefont {Laguna}\ \emph {et~al.}(2010)\citenamefont {Laguna},
  \citenamefont {Larson}, \citenamefont {Spergel},\ and\ \citenamefont
  {Yunes}}]{Laguna:2009re}%
  \BibitemOpen
  \bibfield  {author} {\bibinfo {author} {\bibfnamefont {P.}~\bibnamefont
  {Laguna}}, \bibinfo {author} {\bibfnamefont {S.~L.}\ \bibnamefont {Larson}},
  \bibinfo {author} {\bibfnamefont {D.}~\bibnamefont {Spergel}},\ and\ \bibinfo
  {author} {\bibfnamefont {N.}~\bibnamefont {Yunes}},\ }\bibfield  {title}
  {\bibinfo {title} {{Integrated Sachs-Wolfe Effect for Gravitational
  Radiation}},\ }\href {https://doi.org/10.1088/2041-8205/715/1/L12} {\bibfield
   {journal} {\bibinfo  {journal} {Astrophys. J. Lett.}\ }\textbf {\bibinfo
  {volume} {715}},\ \bibinfo {pages} {L12} (\bibinfo {year} {2010})},\ \Eprint
  {https://arxiv.org/abs/0905.1908} {arXiv:0905.1908 [gr-qc]} \BibitemShut
  {NoStop}%
\bibitem [{\citenamefont {Podolsky}\ and\ \citenamefont
  {Svitek}(2004)}]{Podolsky:2003bm}%
  \BibitemOpen
  \bibfield  {author} {\bibinfo {author} {\bibfnamefont {J.}~\bibnamefont
  {Podolsky}}\ and\ \bibinfo {author} {\bibfnamefont {O.}~\bibnamefont
  {Svitek}},\ }\bibfield  {title} {\bibinfo {title} {{Some high frequency
  gravitational waves related to exact radiative space-times}},\ }\href
  {https://doi.org/10.1023/B:GERG.0000010483.02257.90} {\bibfield  {journal}
  {\bibinfo  {journal} {Gen. Rel. Grav.}\ }\textbf {\bibinfo {volume} {36}},\
  \bibinfo {pages} {387} (\bibinfo {year} {2004})},\ \Eprint
  {https://arxiv.org/abs/gr-qc/0310084} {arXiv:gr-qc/0310084} \BibitemShut
  {NoStop}%
\bibitem [{\citenamefont {Isaacson}(1968)}]{Isaacson:1968hbi}%
  \BibitemOpen
  \bibfield  {author} {\bibinfo {author} {\bibfnamefont {R.~A.}\ \bibnamefont
  {Isaacson}},\ }\bibfield  {title} {\bibinfo {title} {{Gravitational Radiation
  in the Limit of High Frequency. I. The Linear Approximation and Geometrical
  Optics}},\ }\href {https://doi.org/10.1103/PhysRev.166.1263} {\bibfield
  {journal} {\bibinfo  {journal} {Phys. Rev.}\ }\textbf {\bibinfo {volume}
  {166}},\ \bibinfo {pages} {1263} (\bibinfo {year} {1968})}\BibitemShut
  {NoStop}%
\bibitem [{\citenamefont {Cutler}(1998)}]{Cutler:1997ta}%
  \BibitemOpen
  \bibfield  {author} {\bibinfo {author} {\bibfnamefont {C.}~\bibnamefont
  {Cutler}},\ }\bibfield  {title} {\bibinfo {title} {{Angular resolution of the
  LISA gravitational wave detector}},\ }\href
  {https://doi.org/10.1103/PhysRevD.57.7089} {\bibfield  {journal} {\bibinfo
  {journal} {Phys. Rev. D}\ }\textbf {\bibinfo {volume} {57}},\ \bibinfo
  {pages} {7089} (\bibinfo {year} {1998})},\ \Eprint
  {https://arxiv.org/abs/gr-qc/9703068} {arXiv:gr-qc/9703068} \BibitemShut
  {NoStop}%
\bibitem [{\citenamefont {Cutler}\ and\ \citenamefont
  {Flanagan}(1994)}]{Cutler:1994ys}%
  \BibitemOpen
  \bibfield  {author} {\bibinfo {author} {\bibfnamefont {C.}~\bibnamefont
  {Cutler}}\ and\ \bibinfo {author} {\bibfnamefont {E.~E.}\ \bibnamefont
  {Flanagan}},\ }\bibfield  {title} {\bibinfo {title} {{Gravitational waves
  from merging compact binaries: How accurately can one extract the binary's
  parameters from the inspiral wave form?}},\ }\href
  {https://doi.org/10.1103/PhysRevD.49.2658} {\bibfield  {journal} {\bibinfo
  {journal} {Phys. Rev. D}\ }\textbf {\bibinfo {volume} {49}},\ \bibinfo
  {pages} {2658} (\bibinfo {year} {1994})},\ \Eprint
  {https://arxiv.org/abs/gr-qc/9402014} {arXiv:gr-qc/9402014} \BibitemShut
  {NoStop}%
\bibitem [{\citenamefont {Takahashi}\ and\ \citenamefont
  {Seto}(2002)}]{Takahashi:2002ky}%
  \BibitemOpen
  \bibfield  {author} {\bibinfo {author} {\bibfnamefont {R.}~\bibnamefont
  {Takahashi}}\ and\ \bibinfo {author} {\bibfnamefont {N.}~\bibnamefont
  {Seto}},\ }\bibfield  {title} {\bibinfo {title} {{Parameter estimation for
  galactic binaries by LISA}},\ }\href {https://doi.org/10.1086/341483}
  {\bibfield  {journal} {\bibinfo  {journal} {Astrophys. J.}\ }\textbf
  {\bibinfo {volume} {575}},\ \bibinfo {pages} {1030} (\bibinfo {year}
  {2002})},\ \Eprint {https://arxiv.org/abs/astro-ph/0204487}
  {arXiv:astro-ph/0204487} \BibitemShut {NoStop}%
\bibitem [{\citenamefont {Vallisneri}(2008)}]{Vallisneri:2007ev}%
  \BibitemOpen
  \bibfield  {author} {\bibinfo {author} {\bibfnamefont {M.}~\bibnamefont
  {Vallisneri}},\ }\bibfield  {title} {\bibinfo {title} {{Use and abuse of the
  Fisher information matrix in the assessment of gravitational-wave
  parameter-estimation prospects}},\ }\href
  {https://doi.org/10.1103/PhysRevD.77.042001} {\bibfield  {journal} {\bibinfo
  {journal} {Phys. Rev. D}\ }\textbf {\bibinfo {volume} {77}},\ \bibinfo
  {pages} {042001} (\bibinfo {year} {2008})},\ \Eprint
  {https://arxiv.org/abs/gr-qc/0703086} {arXiv:gr-qc/0703086} \BibitemShut
  {NoStop}%
\bibitem [{\citenamefont {Torres-Orjuela}\ \emph {et~al.}(2024)\citenamefont
  {Torres-Orjuela}, \citenamefont {Huang}, \citenamefont {Liang}, \citenamefont
  {Liu}, \citenamefont {Wang}, \citenamefont {Ye}, \citenamefont {Hu},\ and\
  \citenamefont {Mei}}]{Torres-Orjuela:2023hfd}%
  \BibitemOpen
  \bibfield  {author} {\bibinfo {author} {\bibfnamefont {A.}~\bibnamefont
  {Torres-Orjuela}}, \bibinfo {author} {\bibfnamefont {S.-J.}\ \bibnamefont
  {Huang}}, \bibinfo {author} {\bibfnamefont {Z.-C.}\ \bibnamefont {Liang}},
  \bibinfo {author} {\bibfnamefont {S.}~\bibnamefont {Liu}}, \bibinfo {author}
  {\bibfnamefont {H.-T.}\ \bibnamefont {Wang}}, \bibinfo {author}
  {\bibfnamefont {C.-Q.}\ \bibnamefont {Ye}}, \bibinfo {author} {\bibfnamefont
  {Y.-M.}\ \bibnamefont {Hu}},\ and\ \bibinfo {author} {\bibfnamefont
  {J.}~\bibnamefont {Mei}},\ }\bibfield  {title} {\bibinfo {title} {{Detection
  of astrophysical gravitational wave sources by TianQin and LISA}},\ }\href
  {https://doi.org/10.1007/s11433-023-2308-x} {\bibfield  {journal} {\bibinfo
  {journal} {Sci. China Phys. Mech. Astron.}\ }\textbf {\bibinfo {volume}
  {67}},\ \bibinfo {pages} {259511} (\bibinfo {year} {2024})},\ \Eprint
  {https://arxiv.org/abs/2307.16628} {arXiv:2307.16628 [gr-qc]} \BibitemShut
  {NoStop}%
\bibitem [{\citenamefont {Liu}\ \emph {et~al.}(2023)\citenamefont {Liu},
  \citenamefont {Ruan},\ and\ \citenamefont {Guo}}]{Liu:2023qap}%
  \BibitemOpen
  \bibfield  {author} {\bibinfo {author} {\bibfnamefont {C.}~\bibnamefont
  {Liu}}, \bibinfo {author} {\bibfnamefont {W.-H.}\ \bibnamefont {Ruan}},\ and\
  \bibinfo {author} {\bibfnamefont {Z.-K.}\ \bibnamefont {Guo}},\ }\bibfield
  {title} {\bibinfo {title} {{Confusion noise from Galactic binaries for
  Taiji}},\ }\href {https://doi.org/10.1103/PhysRevD.107.064021} {\bibfield
  {journal} {\bibinfo  {journal} {Phys. Rev. D}\ }\textbf {\bibinfo {volume}
  {107}},\ \bibinfo {pages} {064021} (\bibinfo {year} {2023})},\ \Eprint
  {https://arxiv.org/abs/2301.02821} {arXiv:2301.02821 [astro-ph.IM]}
  \BibitemShut {NoStop}%
\bibitem [{\citenamefont {Pi\'orkowska-Kurpas}\ \emph
  {et~al.}(2021)\citenamefont {Pi\'orkowska-Kurpas}, \citenamefont {Hou},
  \citenamefont {Biesiada}, \citenamefont {Ding}, \citenamefont {Cao},
  \citenamefont {Fan}, \citenamefont {Kawamura},\ and\ \citenamefont
  {Zhu}}]{Piorkowska-Kurpas:2020rfy}%
  \BibitemOpen
  \bibfield  {author} {\bibinfo {author} {\bibfnamefont {A.}~\bibnamefont
  {Pi\'orkowska-Kurpas}}, \bibinfo {author} {\bibfnamefont {S.}~\bibnamefont
  {Hou}}, \bibinfo {author} {\bibfnamefont {M.}~\bibnamefont {Biesiada}},
  \bibinfo {author} {\bibfnamefont {X.}~\bibnamefont {Ding}}, \bibinfo {author}
  {\bibfnamefont {S.}~\bibnamefont {Cao}}, \bibinfo {author} {\bibfnamefont
  {X.}~\bibnamefont {Fan}}, \bibinfo {author} {\bibfnamefont {S.}~\bibnamefont
  {Kawamura}},\ and\ \bibinfo {author} {\bibfnamefont {Z.-H.}\ \bibnamefont
  {Zhu}},\ }\bibfield  {title} {\bibinfo {title} {{Inspiraling Double Compact
  Object Detection and Lensing Rate: Forecast for DECIGO and B-DECIGO}},\
  }\href {https://doi.org/10.3847/1538-4357/abd482} {\bibfield  {journal}
  {\bibinfo  {journal} {Astrophys. J.}\ }\textbf {\bibinfo {volume} {908}},\
  \bibinfo {pages} {196} (\bibinfo {year} {2021})},\ \Eprint
  {https://arxiv.org/abs/2005.08727} {arXiv:2005.08727 [astro-ph.HE]}
  \BibitemShut {NoStop}%
\end{thebibliography}%

\end{document}